\documentstyle[12pt,epsfig,subfigure]{article}
\catcode`\@=11   
\@addtoreset{equation}{section}

\newcommand{\ccaption}[2]{
    \begin{center}
    \parbox{0.85\textwidth}{
      \caption[#1]{\small\it {#2}}}                   
    \end{center}    }            

\def\ijmp#1#2#3{{\it Int. Jour. Mod. Phys. }{\bf #1} (19#2) #3}
\def\pl#1#2#3{{\it Phys. Lett. }{\bf #1} (19#2) #3}
\def\zp#1#2#3{{\it Z. Phys. }{\bf #1} (19#2) #3}
\def\prl#1#2#3{{\it Phys. Rev. Lett. }{\bf #1} (19#2) #3}

\def\prep#1#2#3{{\it Phys. Rep. }{\bf #1} (19#2) #3}
\def\pr#1#2#3{{\it Phys. Rev. }{\bf #1} (19#2) #3}
\def\np#1#2#3{{\it Nucl. Phys. }{\bf #1} (19#2) #3}
\def\mpl#1#2#3{{\it Mod. Phys. Lett. }{\bf #1} (19#2) #3}
\def\rnc#1#2#3{{\it Riv. Nuovo Cim. }{\bf #1} (19#2) #3}
\def\arnps#1#2#3{{\it Annu. Rev. Nucl. Part. Sci. }{\bf #1} (19#2) #3}

\def\jetp#1#2#3{{\it JETP Lett. }{\bf #1} (19#2) #3}
\def\app#1#2#3{{\it Acta Phys. Polon. }{\bf #1} (19#2) #3}
\relax                          
\def    \chipm          {\mbox{$\tilde\chi^{\pm}$}}
\def    \chipma        {\mbox{$\tilde\chi^{\pm}_1$}}
\def    \chipa        {\mbox{$\tilde\chi^+_1$}}

\def    \chipmb        {\mbox{$\tilde\chi^{\pm}_2$}}
\def    \chipmi        {\mbox{$\tilde\chi^{\pm}_i$}}
\def    \chio          {\mbox{$\tilde\chi^0$}}
\def    \chioa         {\mbox{$\tilde\chi^0_1$}}
\def    \chiob         {\mbox{$\tilde\chi^0_2$}}
\def    \chioi         {\mbox{$\tilde\chi^0_i$}}

\def    \stop           {\mbox{$\tilde{t}$}}
\def\te{\tilde e}
\def\tnu{\tilde\nu}
\def    \be             {\begin{equation}}
\def    \ee             {\end{equation}}
\def    \ba             {\begin{eqnarray}}
\def    \ea             {\end{eqnarray}}
\def    \nn             {\nonumber}
\def    \=              {\;=\;}
\def    \frac           #1#2{{#1 \over #2}}

\def    \to             {\rightarrow }

\def    \rs             {\mbox{$\sqrt{s}$}}
\def    \tanb           {\mbox{$\tan \beta$}}

\def    \ipb            {\mbox{{\rm pb}$^{-1}$}}
\def    \fb             {{\rm fb}}
\def    \pb             {{\rm pb}}
\def    \gev            {{\rm GeV}}
\def    \w              {\mbox{$W$}}
\def    \z              {\mbox{$Z$}}

\def    \b              {\mbox{$b$}}

\def    \t              {\mbox{$t$}}

\def    \snu            {\mbox{$\tilde{\nu}_e$}}
\def    \stop           {\mbox{$\tilde{t}$}}
\def    \stopone        {\mbox{$\tilde{t}_1$}}

\def    \smu            {\mbox{$\tilde{\mu}$}}
\def    \sele           {\mbox{$\tilde{e}$}}
\def    \stau           {\mbox{$\tilde{\tau}$}}
\def    \slep           {\mbox{$\tilde{\ell}$}}
\def    \mw             {\mbox{$m_W$}}
\def    \mz             {\mbox{$m_Z$}}
\def    \mt             {\mbox{$m_t$}}

\def    \msnu           {\mbox{$m_{\tilde{\nu}_e}$}}
\def    \msnue          {\mbox{$m_{\tilde{\nu}_e}$}}
\def    \mslep          {\mbox{$m_{\tilde{\ell}}$}}
\def    \mchipm         {\mbox{$m_{\tilde\chi^\pm}$}}
\def    \mchipma       {\mbox{$m_{\tilde\chi^{\pm}_1}$}}
\def    \mchipmb       {\mbox{$m_{\tilde\chi^{\pm}_2}$}}
\def    \mchioa        {\mbox{$m_{\tilde\chi^0_1}$}}
\def    \mchioi        {\mbox{$m_{\tilde\chi^0_i}$}}
\def    \mchioj        {\mbox{$m_{\tilde\chi^0_j}$}}
\def    \mchiob        {\mbox{$m_{\tilde\chi^0_2}$}}
\def    \mchio         {\mbox{$m_{\chi^0}$}}

\def    \mstopone       {\mbox{$m_{\tilde{t}_1}$}}
\def    \msmu           {\mbox{$m_{\tilde{\mu}}$}}
\def    \msele          {\mbox{$m_{\tilde{e}}$}}

\def    \epem           {\mbox{$e^+e^-$}}

\def    \missEt      {\ifmmode{/\mkern-11mu E_T}\else{${/\mkern-11mu E_T}$}\fi}
\def    \missE          {\ifmmode{/\mkern-11mu E}\else{${/\mkern-11mu E}$}\fi}
\def \ll {\mbox{$\ell \ell$}} 
\def \jjl {\mbox{$jj\ell$}}
\def \jjjj {\mbox{$4j$}}
\def \pt   {\mbox{$p_T$}}
\def \gamgam {\mbox{$\gamma \gamma$}}
\def \qqg {\mbox{$q \bar{q} \, \gamma$}}
\def \WW {\mbox{$W^+W^-$}}
\def\beq             {\begin{equation}}
\def\eeq             {\end{equation}}
\def\beqd            {\begin{displaymath}}
\def\eeqd            {\end{displaymath}}
\def\baa             {\begin{array}}
\def\eaa             {\end{array}}
\def\beqaa           {\begin{eqnarray}}
\def\eeqaa           {\end{eqnarray}}
\def\beqaad          {\begin{eqnarray*}}
\def\eeqaad          {\end{eqnarray*}}
\def\btabu           {\begin{tabular}}
\def\etabu           {\end{tabular}}
\def\bfig            {\begin{figure}}
\def\efig            {\end{figure}}
\def\bce             {\begin{center}}
\def\ece             {\end{center}}

\def\noi             {\noindent}
\def\nn              {\nonumber}

\def\ti              {\tilde}
\def\q               {\bar}

\def\b               {\beta}

\def\t               {\theta}
\def\s               {\sigma}
\def\x               {\chi}

\def\eeto            {e^+ e^- \to}

\def\sq              {\ti q}
\def\st              {\ti t}

\def\stst            {\ti t_1\,\bar{\st}_1}

\def\sb              {\ti b}

\def\sbsb            {\ti b_1\,\bar{\sb}_1}

\def\ch              {\ti \x^\pm}

\def\nt              {\ti \x^0}

\def\tW              {\t_W}
\def\sth             {\sin\t}
\def\cth             {\cos\t}
\def\sthq            {\sin^2\t}
\def\cthq            {\cos^2\t}

\def\ra              {\rightarrow}

\newcommand{\gsim}{\;\raisebox{-0.9ex}
           {$\textstyle\stackrel{\textstyle >}{\sim}$}\;}
\newcommand{\lsim}{\;\raisebox{-0.9ex}{$\textstyle\stackrel{\textstyle<}
           {\sim}$}\;}
\def\ltap{\;\raisebox{-.5ex}{\rlap{$\sim$}} \raisebox{.5ex}{$<$}\;}
\def\gtap{\;\raisebox{-.5ex}{\rlap{$\sim$}} \raisebox{.5ex}{$>$}\;}
\newcommand{\sss}{\scriptscriptstyle}
\newcommand{\rar}{\rightarrow}
\newcommand{\tgb}{\tan\beta} 
\newcommand{\Z}{Z^{\sss 0}}

\def\n#1{\tilde{\chi}^{\sss 0}_{#1}}  
\def\cp#1{\tilde{\chi}^{\sss +}_{#1}}
\def\cm#1{\tilde{\chi}^{\sss -}_{#1}}

\def\cc   {\epem  \rar  \cp{1}\cm{1}} 
\def\nno   {\epem  \rar  \n{1}\n{2}}         
\def\nng  {\epem  \rar  \n{1}\n{1}\gamma}

\def\te{\tilde e}
\def\tnu{\tilde\nu}
\def\esl{E\llap/}

\def\gs{{g''}}
\def\eps{{\epsilon}}
\def\s{s_\theta}
\def\cthe{c_\theta}
\def    \dd             {\displaystyle}
\def    \be             {\begin{equation}}
\def    \ee             {\end{equation}}
\def    \bea            {\begin{eqnarray}}
\def    \eea            {\end{eqnarray}}
\def    \Ag             {\mbox{$ A_{\lambda{\bar\lambda}}^\gamma $}}
\def    \Az             {\mbox{$ A_{\lambda{\bar\lambda}}^Z $}}
\def    \dAg            {\mbox{$ \delta A_{\lambda{\bar\lambda}}^\gamma $}}
\def    \dAz            {\mbox{$ \delta A_{\lambda{\bar\lambda}}^Z $}}
\def    \Mg             {\mbox{$ {\tilde{\cal M}}^\gamma $}}
\def    \Mz             {\mbox{$ {\tilde{\cal M}}^Z $}}
\def    \Mnu            {\mbox{$ {\tilde{\cal M}}^\nu $}}
\def    \thetab         {\bar\theta}
\newcommand{\bfsp}{{\mbox{\boldmath$\sigma_+$}}}
\newcommand{\bfsm}{{\mbox{\boldmath$\sigma_-$}}}
\def\ie{ {\it i.e.} }
\def\eg{ {\it e.g.} }
\newcommand{\ls}{\ell^*}

\newcommand{\lsls}{\ell^*\overline{\ell^*}}
\newcommand{\es}{{\mathrm e^*}}
\newcommand{\ese}{{\mathrm e^*e}}
\newcommand{\eses}{{\mathrm e^*\overline{e^*}}}

\newcommand{\msm}{\mu^*\mu}
\newcommand{\msms}{\mu^*\overline{\mu^*}}

\newcommand{\tst}{\tau^*\tau}
\newcommand{\tsts}{\tau^*\overline{\tau^*}}

\newcommand{\W}{\mbox{\rm W}}
\begin{document}
{\flushright{
        \begin{minipage}{6cm}
        hep-ph/9602207\hfill \\
        \end{minipage}        }
 
}

\begin{center}
{\large \bf SEARCHES FOR NEW PHYSICS}
\end{center}                                       
\begin{center}
{\it Conveners}: G.F.~Giudice, M.L.~Mangano, G.~Ridolfi, and R.~R\"uckl
\end{center}          
\begin{quote}
{\it Working group}: S.~Ambrosanio, 
S.~Asai, G.~Azuelos, H.~Baer, 
A.~Bartl, W.~Bernreuther,
M.~Besan\c{c}on, 
G.~Bhattacharyya,
M.~Brhlik, L.M.~Bryant, G.~Burkart, 
M.~Carena, R.~Casalbuoni, P.~Chankowski, D.~Choudhury, A.~Culatti, 
A.~Deandrea, 
W.~de~Boer, G.~Carlino, S.~De Curtis, G.~Degrassi, C.~Dionisi, 
A.~Djouadi, D.~Dominici, H.~Dreiner, H.~Eberl,
L.~Favart, M.~Felcini,  F.~Feruglio, H.~Fraas, F.~Franke, R.~Gatto,
S.~Giagu, M.~Grazzini, J.F.~Grivaz, 
J.J.~Hernandez,
S.~Katsanevas, S.~Komamiya, S.~Kraml, P.~Le~Coultre, S.~Lola, 
W.~Majerotto, A.~Masiero                                         
T.~Medcalf, B.~Mele, G.~Montagna, R.~Munroe, S.~Navas, F.~Nessi-Tedaldi,       
O.~Nicrosini,           
S.~Orito, F.~Piccinini, S.~Pokorski, W.~Porod, 
P.~Rebecchi, F.~Richard, S.~Rigolin,
S.~Rosier-Lees, S.~Shevchenko, V.~Shoutko,
A.~Shvorob, L.~Silvestrini, A.~Sopczak,                   
R.~Tafirout, X.~Tata, J.~Toth
A.~Trombini, C.~Vander~Velde, R.~van~Kooten, 
A.~Vicini, J.H.~von~Wimmersperg-Toeller,
C.~Wagner, K.~Yoshimura.
\end{quote}                                                   
\tableofcontents                                                                           
\newpage

\section{Supersymmetry}
\subsection{Introduction}
\label{susyint}
Supersymmetry (SUSY) \cite{wessz,susyrev} represents the best motivated known
extension of the Standard Model (SM). It offers an elegant solution to the
naturalness problem of the Higgs sector \cite{natur}, it is consistent with
present experimental data, and it predicts new particles to be discovered in
this generation of collider experiments. LEP2 has great potential for
discovering SUSY, as a consistent solution of the naturalness problem suggests
that some of the new weakly interacting particles are within the LEP2 energy
range~\cite{lim}. 
      
In the minimal SUSY model (MSSM), each SM particle has one SUSY partner.
The partners of the gauge bosons (gauginos) have spin 1/2, the partners of
fermions (sfermions) have spin 0 and the partners of the Higgs fields
(higgsinos) have spin 1/2. 
The Higgs sector is enlarged to contain two complex weak
doublets. After electroweak (EW) symmetry breaking, the partners of the Higgs and
$SU_2 \times U_1$ gauge bosons mix, and the physical mass eigenstates are given
by two Dirac fermions with one unit of electric charge (the charginos \chipmi,
$i=1,2$ with $\mchipma < \mchipmb$) and four neutral Majorana fermions (the
neutralinos, \chioi, $i=1,...,4$ with $\mchioi < \mchioj$ if $i<j$). This
sector is described by three parameters: the $SU_2$ gaugino mass $M_2$, the
Higgsino mass $\mu$, and $\tan \beta$, the ratio of the two Higgs vacuum
expectation values (VEVs). The $U_1$ gaugino mass is given by the unification
condition $M_1 =\frac{5}{3} \tan^2\theta_W M_2$, a relation valid in most GUT
SUSY models. Relations between the three parameters $M_2$, $\mu$, $\tan\beta$
and the  masses and couplings of charginos and neutralinos can be found in 
ref.~\cite{susyrev}.             
                    
Each quark and lepton has two scalar partners, one for each chirality state. We
will refer to these as left and right squarks ($\tilde q$) and sleptons
($\tilde \ell$). The scalar mass eigenstates are given by mixtures of
these two left and right states. The squark and slepton mixing angles and
masses are in general free parameters of the theory, and we will treat them as
such. However, in sect.~\ref{susybaer}, 
we will analyze the more restricted case in which all
squark and slepton masses are universal at the grand unification scale and then
are evolved to low energies according to the renormalization group equations.
This case, suggested by a certain class of supergravity theories, allows a
simple comparison among the discovery potentials of different experimental
searches, because all SUSY particle masses and interactions are described by
only four free parameters. 

The requirement of baryon and lepton number conservation in renormalizable
interactions implies the existence
of a discrete symmetry in the MSSM. Such symmetry, called R-parity,
distinguishes between ordinary and SUSY particles and implies that:
{\it i)} the SUSY particles can only be pair produced and {\it ii)} the
lightest SUSY particle (LSP) is stable. An appealing feature of models with
universal SUSY breaking terms at the GUT scale is that the neutralino turns out
to be the LSP in almost all of the parameter space consistent with radiative
EW symmetry breaking. This is welcome because from cosmological
arguments on particle relic abundance \cite{jelli} 
it follows that the LSP must be a neutral
particle. In most of our analysis we will therefore assume that the LSP is a
neutralino, although we will also comment on the case in which the LSP is a
sneutrino, the only other neutral SUSY particle in the MSSM.

Searches at LEP1 have ruled out charged SUSY particles lighter than about
$\mz/2$ \cite{pdg}. For particular values of the SUSY parameters, the $\tilde
t$ can decouple from the $Z^0$, and in this case the limit on the stop mass is
slightly reduced \cite{opalstop,alephstop}.
From measurements of the invisible $Z^0$ width, a sneutrino lighter than about
$\mz/2$ is also ruled out. The experimental limits on neutralinos \cite{neutr}
strongly depend on the SUSY parameters which determine their couplings to the
$Z^0$. Tevatron bounds \cite{teva} on gluinos and squarks
exclude the possibility of producing these particles at LEP2, with the notable
exception of a light $\tilde t$. Information about the possible existence of
light SUSY particles can also be extracted from global fits of EW
observables, and it will be discussed in sect.~\ref{susyrb}. 

In sects.~\ref{susycharginos}--\ref{susyneutralinos} we present a study of the
expected signals and detection efficiencies respectively for production of
charginos, sleptons, stops, and neutralinos at LEP2. 
As alternatives to the MSSM, we will also consider two extensions which have
some theoretical interest. In sect.~\ref{susyfranke} we discuss how neutralino
searches are modified by the introduction of a new singlet superfield. Finally
the rather different experimental signatures of SUSY in the presence of
$R$-parity violation are studied in sect.~\ref{rparitysection}.
                     
\subsection{Supersymmetry and $R_b$}
\label{susyrb}            
The SM (and the MSSM with all superpartners heavy) is in excellent 
agreement with most of the electroweak measurements
\cite{EWWG2,FITS,ELLIS,LANER,ABC,MY_MSSM} except for $R_b$ and $R_c$ 
\cite{BRUSS}. The present experimental results, $R_b=0.2219\pm0.0017$, 
$R_c=0.1540\pm0.0074$, are respectively $+3.7\sigma$ and $-2.5\sigma$
away from the SM predictions $R^{SM}_b=0.2156$, $R^{SM}_c=0.172$
(for $m_t=180$ GeV) \cite{EWWG2}. 
If $R_c$ is fixed to its SM value, the fits give 
$R_b=0.2205\pm0.0016$~\cite{EWWG2}.
                      
In the MSSM it is possible to obtain larger $R_b$ \cite{BF,SOLA,KANE}
without altering the rest of the EW observables \cite{MY_MSSM}. 
Indeed, the latter are sensitive    
essentially only to additional sources of the custodial $SU_V(2)$ breaking
in the ``oblique'' corrections, \ie mainly to the left  slepton
and squark masses. The dependence on the right sfermion masses 
enters only through the left--right mixing. Since the breaking of the 
custodial $SU_V(2)$ in the gaugino and Higgs sectors is weak, 
the MSSM with  heavy enough left sfermions does not significantly alter  the SM
predictions for those observables~\cite{ABC,MY_MSSM}. The case of
$R_b$ is different.                                    
Its value can be larger than in the SM: for low $\tan\beta$ $R_b$ receives
important corrections from
loops involving right stops and higgsino-like charginos.
For large $\tan\beta$ loops involving the $CP-$odd Higgs contribute as well.

\begin{figure}                                                           
\centerline{
\epsfig{figure=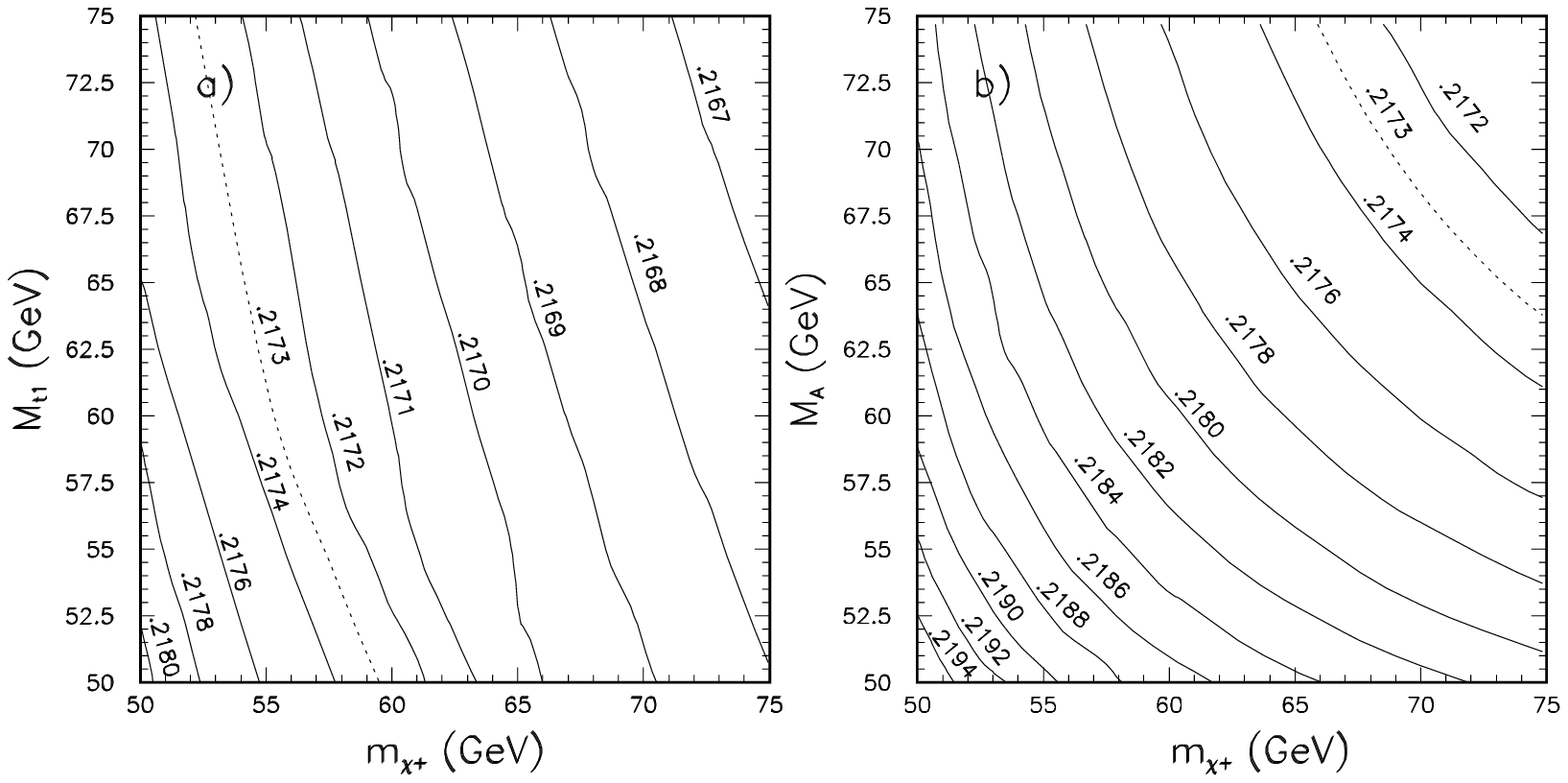,width=\textwidth,angle=0}}                                       
\ccaption{}{\label{fig:limits}                   
Contours of constant $R_b$ for $m_t=180$ GeV
\noindent a) in the chargino -- lighter stop mass plane
for $\tan\beta=1.6$; \noindent b) in the chargino -- $CP-$odd Higgs 
boson mass plane for large $\tan\beta=50$.}
\end{figure}

In fig.~\ref{fig:limits} we show the contours of constant $R_b$ in the 
(\mstopone,\mchipm) plane for low $\tan\beta$ and in the
($M_A$,\mchipm) plane for large $\tan\beta$. 
For each point in the plane of fig.~\ref{fig:limits} 
the plotted $R_b$ is the largest value obtainable by varying the remaining
SUSY parameters under the the following assumptions \cite{MY_MSSM}:
{\it i)} the overall $\chi^2$
in a fit to 14 electroweak observables is within $\Delta\chi^2<1$ from the
best fit in that particular point;
{\it ii)} the predicted $BR(b\rightarrow s\gamma)$ \cite{BSG} is in the range 
$(1-4)\times10^{-4}$; {\it iii)} $BR(t\rightarrow bW)>50$\%;
{\it iv)} $\Gamma(Z^0\rightarrow\chi^0_1\chi^0_1)<4$ MeV; {\it v)} the Higgs 
mass is larger than 50 GeV.                                          
                           
Although with the constraints from LEP1 
one cannot reach the central experimental value of $R_b$,
supersymmetric corrections can significantly reduce the discrepancy
\cite{MY_MSSM}. For instance, for low 
$\tan\beta$ and $M_{\tilde t_1}= m_{\chi^+}=50$ GeV one gets $R_b=0.2182$
and for large $\tan\beta$ and $M_A=55$ GeV, $m_{\chi^+}=50$ GeV
one gets $R_b=0.2196$ and a significant improvement in the overall $\chi^2$
w.r.t. the SM case. The supersymmetric $R_b$ prediction remains  within  the
95\%~C.L. range ($R_b>0.2173$) in the region bounded by the dashed contours.
The values of $\tan\beta$ chosen in fig.~\ref{fig:limits} are consistent with
perturbative Yukawa couplings up to the Grand Unification scale and with
constraints from $BR(b\rightarrow c\tau\nu)$~\cite{BCTN}.
                                             
LEP1.5 data could push the lower limit for charginos up to 
$\mchipm>65$ GeV provided that $\mchipm-\mchioa>10$ GeV. Notice
however that the regions of SUSY parameter space which provide the best $R_b$
values do not satisfy this constraint. 
As in our fit we find very small mass splittings $\mchipm-\mchioa$ the chargino
may well escape detection and therefore LEP1.5 measurements may not
significantly restrict the SUSY corrections to $R_b$.
However SUSY can bring the prediction for $R_b$ within the 95\%~CL range from
the data only if at least some SUSY particle is within the reach of LEP2.
The region in parameter space selected by the best fits is interesting from the
theoretical point of view \cite{IR,LARTAN}. 
The hierarchy $|\mu| << M_{2}$ (i.e. higgsino--like lightest
neutralino and chargino) is inconsistent with the mechanism of
radiative electroweak symmetry breaking and universal boundary conditions
for the scalar masses at the GUT scale 
However, it is predicted in models with certain pattern of 
non--universal boundary conditions \cite{OP2}.
                                                                     
\begin{figure}
\centerline{\epsfig{figure=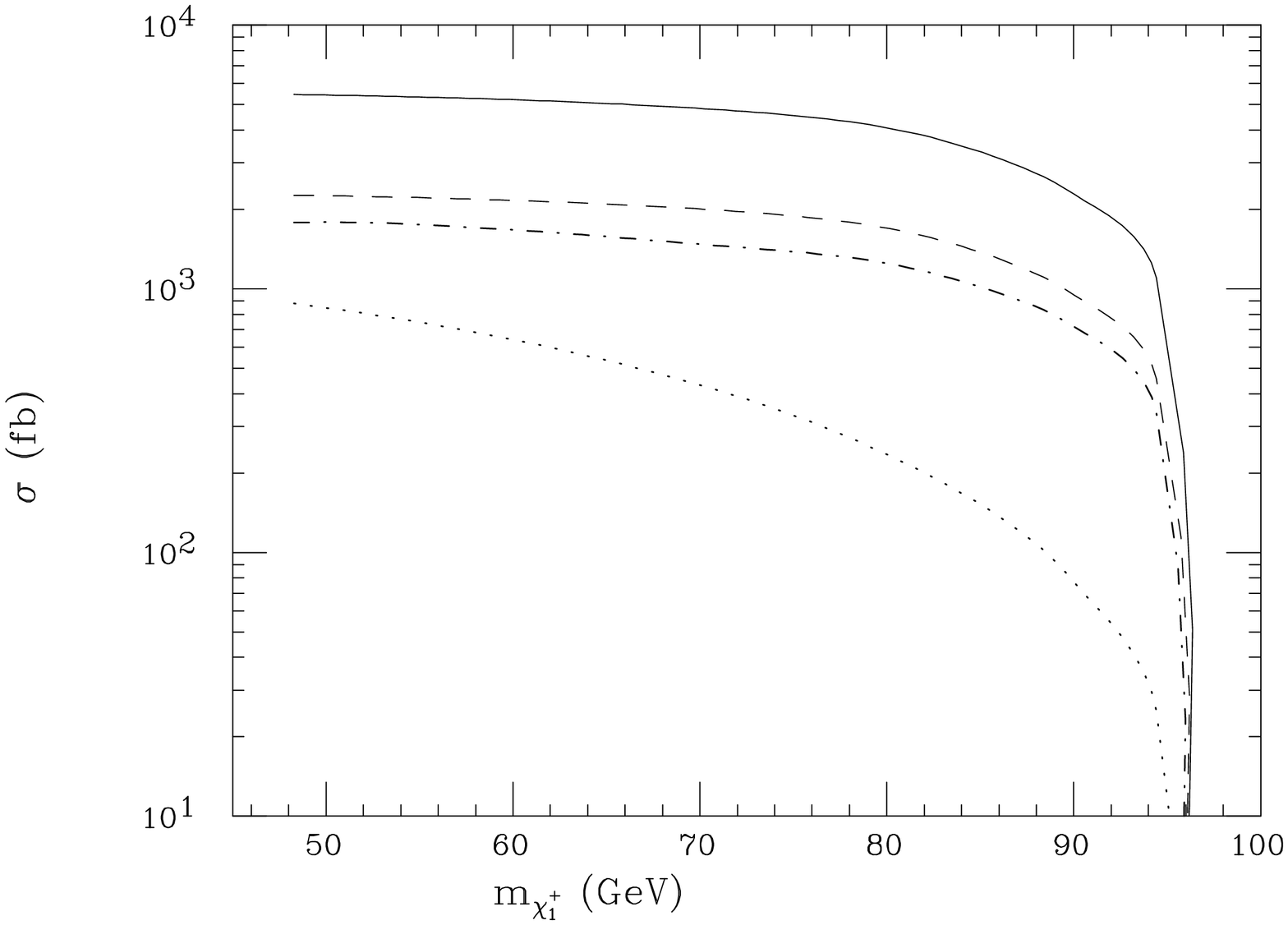,width=0.5\textwidth,angle=90}}
\vspace{-1cm}                                                       
\ccaption{}{ \label{fchplot}                            
Chargino production cross sections at LEP2, $\sqrt{s}=190$ \gev, as a function
of \mchipma. We show the ranges obtained by varying $M_2$, $\mu$, \tanb\ and
\msnue\
throughout the parameter space, requiring $\msnue>45$ GeV. 
The solid and dotted lines correspond to the maximum and minimum
production rates. The dashed (dash-dotted) line corresponds to the minimum
cross section if \msnu=2 TeV (\msnu=200 GeV).}
\end{figure}                            

\subsection{Chargino}
\label{susycharginos}
The MSSM contains two chargino mass eigenstates, \chipma\ and \chipmb. The
lighter one, \chipma, is a candidate for being the lightest SUSY charged
particle. In \epem\ collisions, charginos are pair produced via $\gamma$ and
$Z$ exchange in the $s$-channel, and via \snu--exchange in the $t$-channel.
This latter contribution is totally suppressed at LEP1.
However at LEP2 energies, the destructive interference of the two
contributions can lead to a considerable reduction of the production cross
section if the \snu\ is sufficiently light. In fig.~\ref{fchplot}\ we present 
the minimum and maximum
$\sigma(\epem\to\chi^+_1\chi^-_1$) as a function of \mchipma, obtained by
varying the SUSY parameters
\tanb, \msnu, $M_2$, and $\mu$. As
one can see, while the cross section is generally large, of the order of
several pb, for special values of the above parameters it can be reduced to
less than one pb, even away from the kinematic limit. Notice that the effects
of destructive interference become less dramatic as soon as $\msnue>200$ GeV.
                                                              
The main \chipma\ decay mode is expected to be:
\be
\chipma\to \chioa \, f \bar{f}'
\label{chardec}
\ee
with $f$ and $\bar{f}'$ being fermions of the same weak isospin doublet and
\chioa\ the lightest neutralino, which is assumed to be the lightest SUSY
particle. The chargino decay can occur via virtual exchange of \w, sfermions or
charged Higgs boson. If $H^+$ and all sfermions are very massive ($m_{H^+},
m_{\tilde f} \gg \mw$), the BR are the same as those of the \w. If the
slepton masses are significantly smaller than the squark masses and of the
order of \mw, the \chipma\ leptonic BR are enhanced. Suppression of the
hadronic modes due to phase space can also take place when the mass difference
between the chargino and the lightest
neutralino is small. Moreover, for some values of the
SUSY parameters the second lightest neutralino, \chiob, can be lighter than the
lightest chargino and therefore the decay $\chipma\to \, \chiob \, f \bar{f}'$ be
possible. This can reduce the BR of decay mode (\ref{chardec}) and give rise
to cascade events with more leptons and more jets. The possible outcomes in
terms of cross sections, decay modes and branching ratios as a function of the
SUSY parameters have been extensively studied in the
literature~\cite{bartl-feng}. 

If $\tilde{\nu}$ or $\tilde{\ell}$ are lighter than the chargino, the decay
modes $\chipma\to \, \ell \tilde{\nu}_{\ell}$ or $\chipma\to \, \tilde{\ell}
\nu_{\ell}$ are accessible. The decays to lepton-sneutrino or slepton-neutrino
would yield signals similar to those from slepton pair production and could be
looked for in similar ways. As mentioned earlier, the region in which the
chargino cross section is suppressed corresponds to a light sneutrino and a
gaugino-like chargino. In this situation, the decay mode $\chipma\to \, \ell
\tilde{\nu}_{\ell}$ becomes dominant, see fig.~\ref{fchplot}. 
The efficiency for chargino
detection is therefore improved and can compensate for the lower production
cross section.

The experimental studies of the four LEP experiments that we present
here~\cite{experim} have only considered the three body decays 
in~(\ref{chardec}), so three kinds of events can be observed depending on
whether the charginos decay to leptons or quarks: a pair of acoplanar leptons
that may have different lepton flavour (\ll\ mode); a relatively isolated
lepton with two hadronic jets and missing energy (\jjl\ mode) and hadronic
events with missing energy (\jjjj\ mode). These modes may lead to 
different experimentally observed topologies. For example 
a \jjl\ event where
the lepton is a $\tau$ that decays hadronically may look rather like a four-jet
event than two jets plus a lepton. 

In the Monte Carlo studies carried out by the LEP experiments, the BR of the
\chipma\ have been assumed to be those of the \w, so that the probability of
the above modes are taken to be 10.3\%, 43.5\% and 46.0\%, respectively. 

\begin{table}[ht]
\begin{center}
\begin{tabular}{||l||r|r||} \hline \hline
 & \multicolumn{2}{|c||}{Cross Section (pb)} \\
 \multicolumn{1}{||c||}{Type of event} & 
\multicolumn{1}{c}{}& \multicolumn{1}{c||}{}\\
 \multicolumn{1}{||c||}{} &
\multicolumn{1}{c}{$\sqrt{s}= 175$ GeV} & 
\multicolumn{1}{c||}{$\sqrt{s}= 192$ GeV}  \\ \hline \hline  
$e^+ e^- \rightarrow f \, \bar{f} \, (\gamma)$   & 164.9 &128.1 \\ \hline
$e^+ e^- \rightarrow W^+ W^- \, (\gamma)$        & 13.8  & 17.1 \\ \hline
$e^+ e^- \rightarrow Z^0 Z^0 \, (\gamma)$        & 0.4   &  1.1 \\ \hline
$e^+ e^- \rightarrow W e \, \nu \, (\gamma)$     & 0.5   &  0.7 \\ \hline
$e^+ e^- \rightarrow Z^0 \, e \, e \, (\gamma)$  & 2.7 & 2.9 \\ \hline \hline
\end{tabular}
\ccaption{}{\label{BACKGROUNDS}
Relevant backgrounds to chargino production 
and their cross sections at $\protect \sqrt{s}= 175$ and $192$ GeV.
The cross section for the \gamgam\ process is not given since it
is highly dependent on the initial cuts used.} 
\end{center}
\end{table}
 The relevant backgrounds to this process are given in table~\ref{BACKGROUNDS}.
The most dangerous background is \WW\ pair production due to the
presence of missing energy and visible final states similar to 
those of the signal.
Even though \gamgam\ events are very sensitive to the cuts used
and should
be a manageable background, care should be taken to have them 
well under control through a good knowledge of the apparatus. Indeed, the
extremely high cross section of this kind of events makes it unfeasible
to generate luminosities of simulated events large enough to match
the number of expected real events. The experiments have used
a {\it preselection} in order to discard at the generation level those events
that would be in any case rejected after simulation. 
This procedure 
keeps only the tails of the distributions which may be a dangerous 
background of fake missing energy events. Cross-checks have been performed
by the experiments to make sure that the different generators used to
produce this background according to the various theoretical models
(VDM, QPM and QCD) agree with data at LEP1 energies. The backgrounds
have been generated using standard Monte Carlo programs~\cite{jetset,twogam}.
Chargino events were generated using SUSYGEN~\cite{SUSYGEN}.
For a typical point in SUSY parameter space             
the L3 experiment has checked that using the 
Monte Carlo program DFGT~\cite{SUSYGEN}, which properly takes into account
spin-correlations, gives essentially the same detection efficiency.

%

%
\subsubsection{Mode \jjl}

  The \jjl\ mode has been studied with full simulation of the detector
for several points of the SUSY parameter space by DELPHI, L3 and OPAL 
and with fast simulation by ALEPH.                

 The main features of the \jjl\ mode are the presence of an isolated
lepton in a high multiplicity environment, a low hadronic mass, 
a high missing mass and high missing \pt.
 The initial selection of this mode
is carried out with a cut in multiplicity
and the identification of an isolated electron or muon (OPAL also
includes a simplified $\tau$ identification by looking for 3 or 5 
charged tracks in a $5^\circ$ cone with an invariant mass smaller than
2 \gev).
 
 The large \gamgam\ 
background is rejected demanding a high missing \pt\ 
(cuts range from 5 to 10~\gev), the absence of electromagnetic
energy in the very forward-backward regions, and  
a missing momentum vector not pointing to the 
very forward-backward regions.       
\begin{figure}
\centerline{\epsfig{file=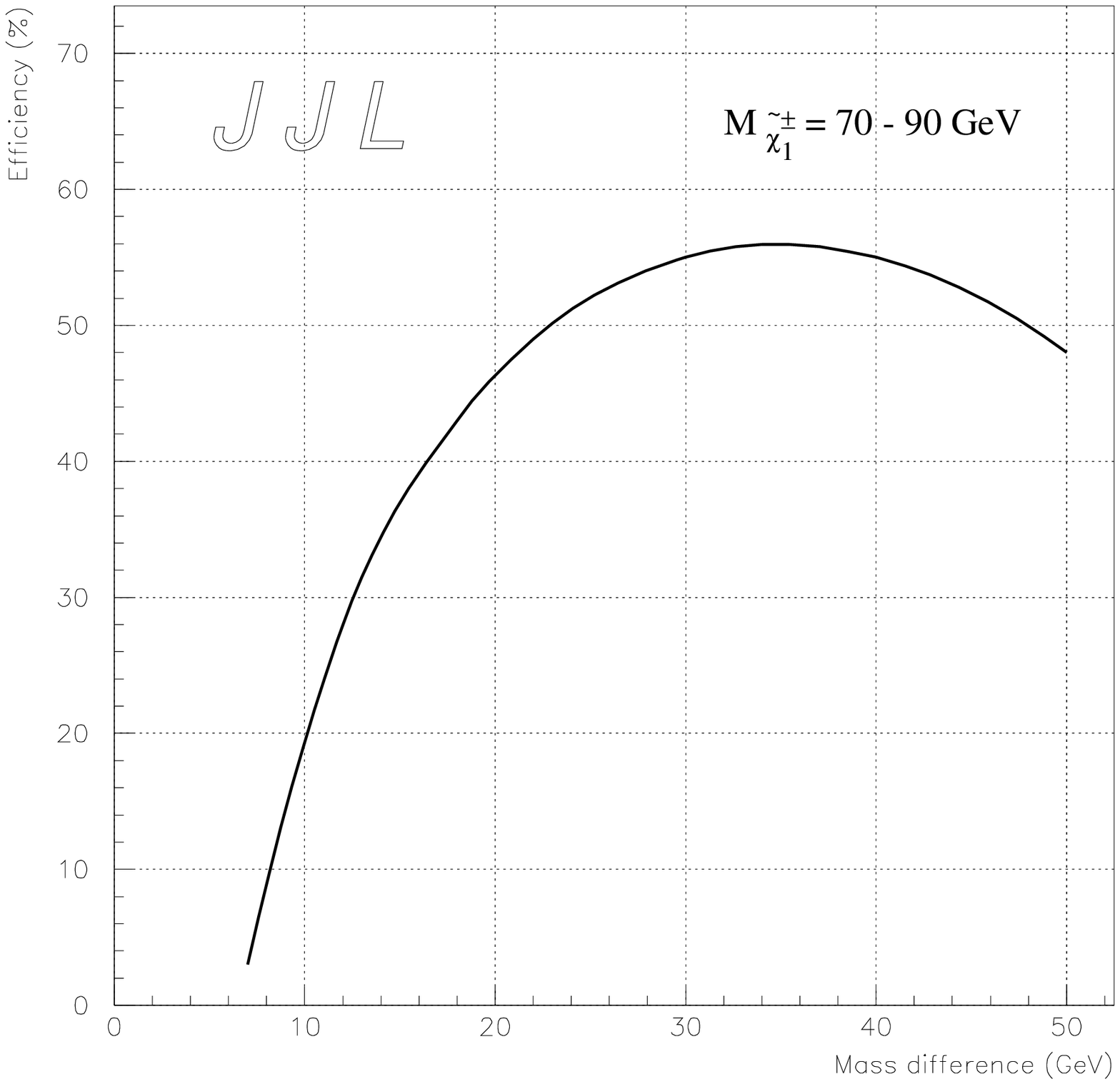,width=.40\textwidth}
            \epsfig{file=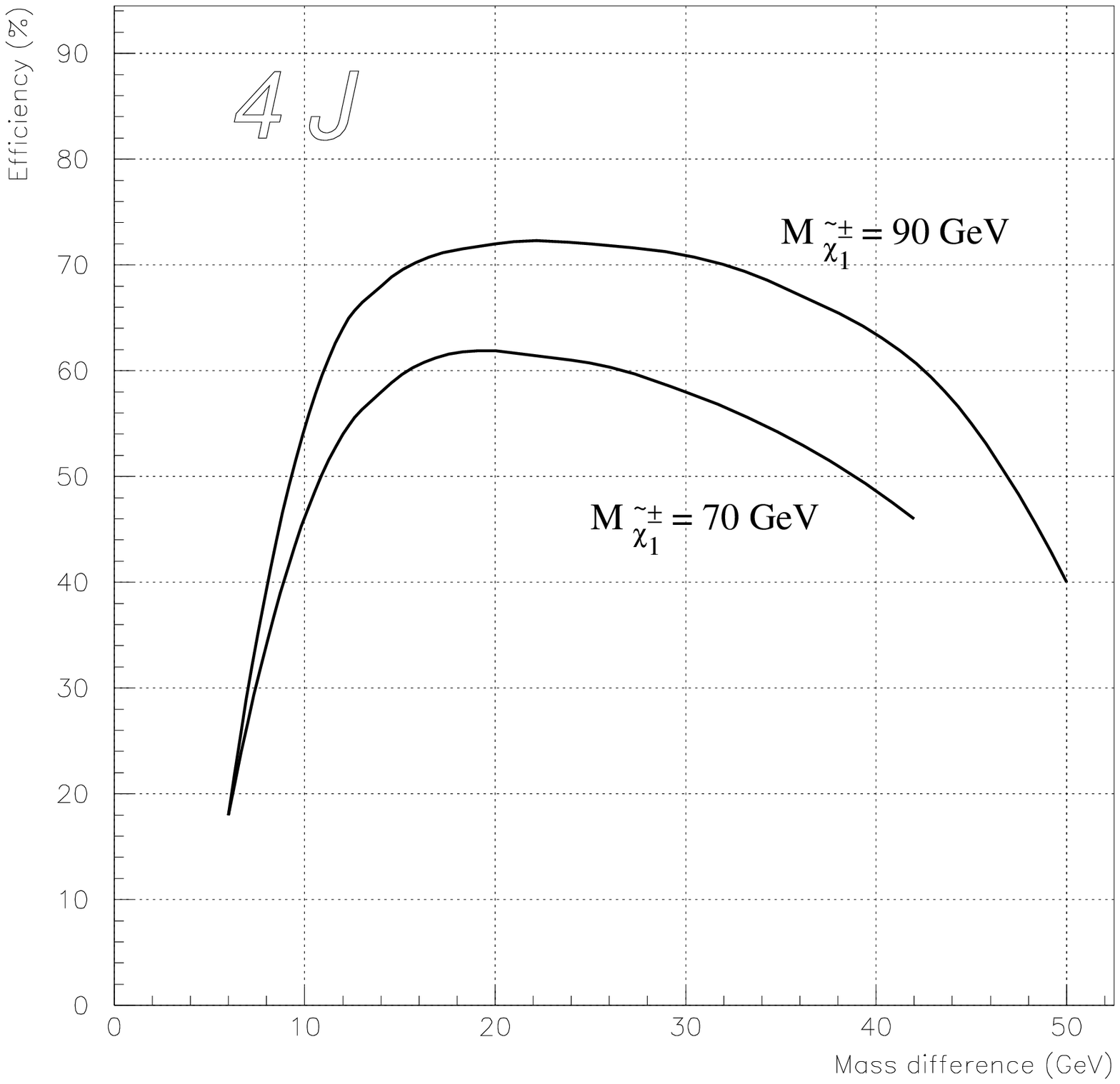,width=.40\textwidth}}
\ccaption{}{ \label{charcross}
Left: selection efficiency as a function of the chargino--neutralino 
mass diffe\-ren\-ce in the \jjl\ mode for an average experiment.
Right: same efficiency for the \jjjj\ mode. The two curves 
correspond to chargino masses of 70 and 90~\gev.}
\end{figure}
 For the non-\gamgam\ background, in particular due to \WW\ production,
a variety of cuts in visible mass, visible energy, missing mass, 
hadronic mass and invariant mass of the lepton and the unseen neutrino
have been used by the four experiments. 
All these selection criteria take advantage of 
the fact that for \WW\ events the missing mass is small (in fact 
with perfect resolution it would vanish since it is the neutrino mass), 
while the hadronic mass 
and the invariant mass of the lepton and unseen neutrino (as reconstructed 
from energy-momentum conservation) should be that of the \w. On the other 
hand, in the chargino events the missing mass is large
and the hadronic mass small due to the presence of the neutralino, while 
there is no particular reason why the ``lepton--neutrino'' invariant mass 
should peak at the \w\ mass. Cuts in acoplanarity, acollinearity and maximum
momentum of the isolated lepton are also included in the selection criteria of
some of the experiments. The very same cuts devised for the \WW\ background
also 
reject the rest of non-\gamgam\ backgrounds, which are less important in terms
of cross section, except for the \qqg\ background. At \rs= 190 GeV all the 
four experiments are able to reduce the background 
to 50--100~\fb. Around 30--50\% of this background,
depending on the experiment, is \WW. The remaining background is 
mostly \qqg\ (20--30\%) with contributions below the 10~\fb\ level  
from the other processes.

Concerning the signal, the efficiency is in general only mildly dependent 
on the SUSY point considered. In fact, it is mostly dependent 
on the chargino and neutralino masses, and much less on their field composition.
It is only when the chargino and neutralino are close in mass 
($\mchipma - \mchioa < 10$~\gev) or the mass of the neutralino 
is very light ($\mchioa <~20$~\gev) that the efficiency starts 
to degrade (fig.~\ref{charcross}, left).
In the first case this is due to the decrease in multiplicity, 
visible energy and missing \pt; in the second, to the increase
in visible energy and the decrease in missing mass, which makes these
events more similar to the \WW\ background. 
For the remaining chargino and neutralino mass values, the efficiency is quite
stable. For the \jjl\ mode, in particular, the purely detector efficiencies
for the electron and muon final states ({\it i.e.} excluding BR) range
from 40 to 60\% depending on the experiment. Taking into account 
the BR to the electron and muon final state (including those through 
the~$\tau$ decay) the overall efficiency in this mode is 
in the 10 to 20\% range, under the assumption that the chargino BR are
equal to those of the \w.

\subsubsection{Mode \jjjj}

 The \jjjj\ mode has been studied by DELPHI, L3 and OPAL 
for several points in the SUSY parameter space with full 
simulation of the detectors.

 In this mode, high multiplicity and the absence
of an isolated lepton are the first requirements. Again, the \gamgam\ 
background is rejected by means of cuts in minimum missing \pt\ and 
minimum acoplanarity. It is also required that only a limited amount 
of the total energy be present in the far forward-backward region and
that the missing momentum vector does not point to this region. 
For the non-\gamgam\
background cuts have been applied on the  minimum missing mass, on the maximum
visible energy and on the maximum hadronic mass. As before, these cuts rely on
the fact that the missing mass is low for the background and high for the
signal, while the hadronic mass is around the center of mass energy for
the background and low for the signal. Even though the background is in this
case slightly higher than in the \jjl\ mode, it is nevertheless still in the 
100--200~\fb\ range.
The experiments agree on the three main sources of background:
\WW, \qqg\ and $W \, e \, \nu$ events, but the percentage of each one
changes from experiment to experiment, reflecting the different
choices of cuts. It is worth noting that for this mode 
the $W \, e \, \nu$ process is  a non-neglible source of background 
amounting to $\sim30\%$ of the total. 
        
The signal efficiency depends only very slightly on the
chargino and neutralino masses unless                
we are close to the limit of a small neutralino mass ($\sim20$ \gev) 
or small chargino--neutralino mass difference. 
However, in this mode the search can be extended down to 
chargino--neutralino mass differences of the 
order of 5 \gev\ (see fig.~\ref{charcross}, right).
This lower mass difference is due to             
the higher multiplicity of the \jjjj\ mode, 
to the absence of the requirement of
an isolated lepton, and to the smaller missing \pt\ needed
to reject the \gamgam\ background.
Another feature of this mode is that it recovers events
missed in the \jjl\ mode. Indeed, some of the \jjl\ events that are lost 
during the \jjl\ selection are kept in the \jjjj\ selection. 
In case of discovery this ``migration'' 
might be a serious problem to compute branching ratios with the present
selection criteria, but during the search stage it is a way of increasing
the overall efficiency.                            
The efficiency to select \jjjj\ events is, as in the preceding mode,
in the 40--50\% range. Additionally, around 15 to 20\% of the \jjl\ events
are classified also as \jjjj. If we take into account also these 
events in our final sample, we obtain an ``efficiency'' of 60 to
70\%.

\subsubsection{Mode \ll }

 The \ll\ mode has been studied with full simulation of the detector by
DELPHI, L3 and OPAL for several points in the SUSY parameter space.

 The \ll\ mode is characterized by two acoplanar leptons, low multiplicity,
low visible energy, large missing mass and missing \pt. The final leptons
might have different leptonic flavour, a feature that distinguishes this
mode from the associated production of sleptons.           
The selection starts
with a cut in multiplicity which is more or less stringent depending on
whether the experiment additionally demands explicit lepton 
identification. Various ways of avoiding leptonic radiative return events
have been used, all of them relying on some sort of restriction on 
the magnitude of the isolated electromagnetic energy in 
the event. Apart from this, cuts in the maximum energy in the far 
forward-backward cone are used to reject the \gamgam\ background.

 \WW\ events are mainly rejected by cuts in maximum visible energy, 
minimum missing mass
or in the maximum momentum of the leptons.
Apart from the acoplanarity, other variables in the 
transverse plane have also been used.  
In particular, cuts have been applied in ``transverse thrust'' 
and in the \pt--acoplanarity plane. 
\begin{figure}
\centerline{\epsfig{file=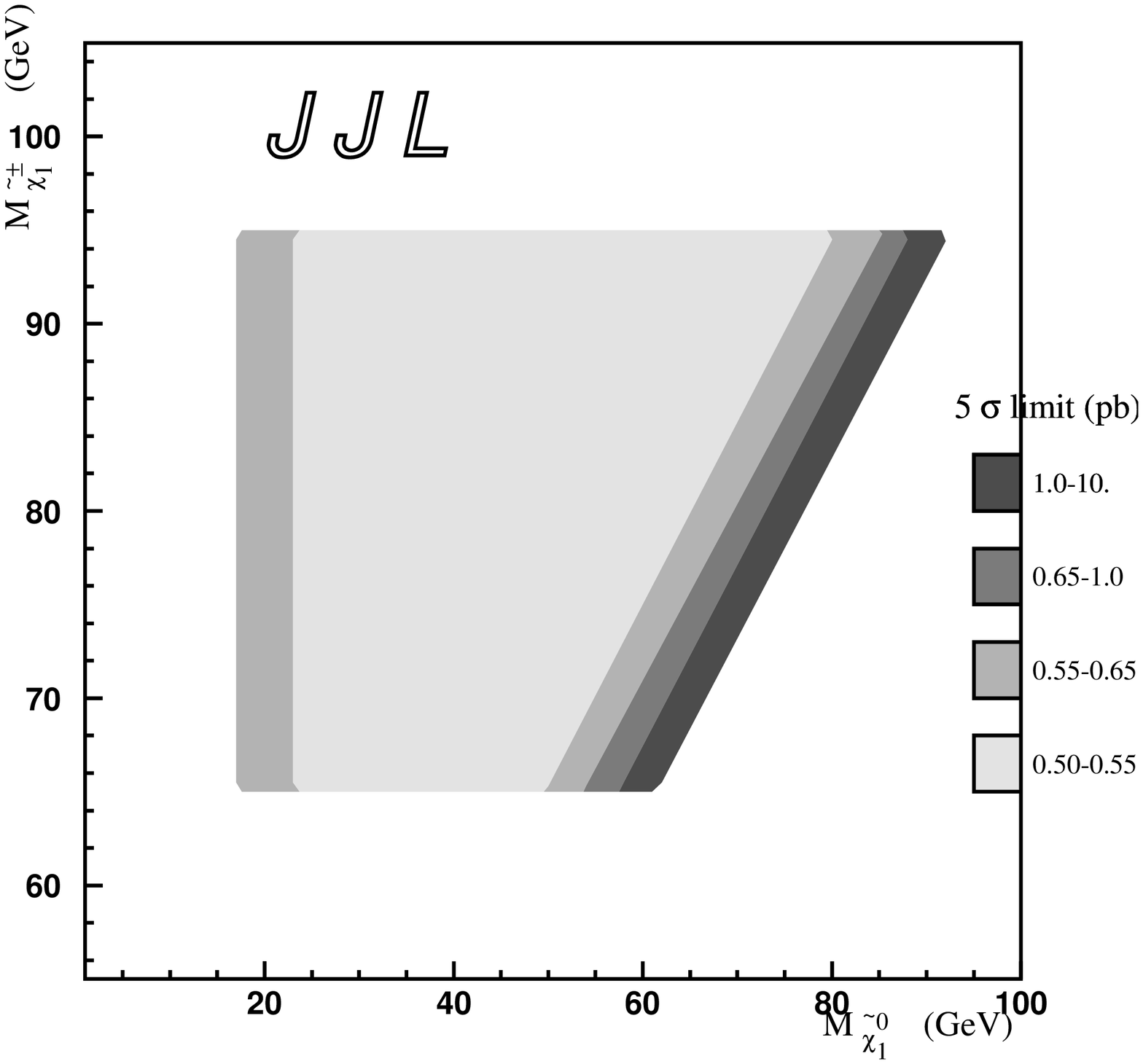,width=.40\textwidth}
            \epsfig{file=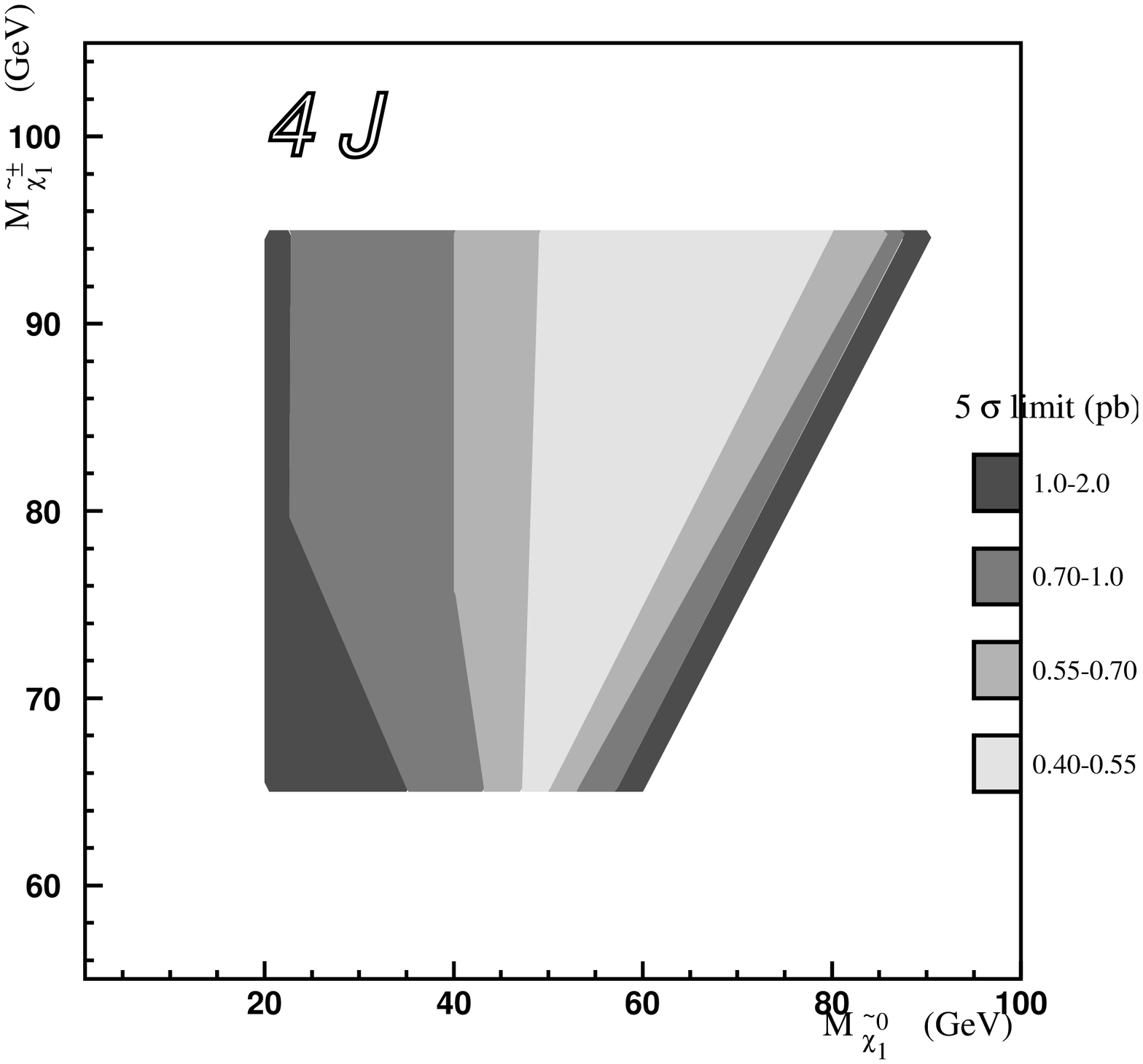,width=.40\textwidth}}
\ccaption{}{ \label{mincross}
Minimum signal cross sections required for 5-$\sigma$ chargino 
discovery in the (\mchipm,\mchioa)
plane, with 150~\ipb\ integrated luminosity. The case of the \jjl\ mode 
(left) and of the  \jjjj\ mode (right) are shown.}                                                         
\end{figure}                                            
 The final background can be reduced to several tens of \fb.
The most serious background in this channel is \WW\ production, 
which represents in all cases more than half the final background. The 
remaining background consists of radiative return and $Z \, e^+ \, e^-$ 
events.

  Concerning the signal, the average detector efficiency is in the 25--35\% 
range. Again, this efficiency is rather stable throughout the SUSY parameter
space, unless the neutralino is very light or very close in mass to the
chargino.  
If we assume that the leptonic BR of the chargino is equal to that of the \w, 
the overall efficiency in this mode is around 3\%. In spite of this 
low efficiency, this mode is particularly important since its enhancement 
might indicate the possible presence of nearby sleptons. 

\subsubsection{Results}
  For \rs= 190 \gev, an integrated luminosity of 150~\ipb, and assuming 
for the chargino the BR of the \w, the LEP experiments are able to 
discover at the $5 \sigma$ confidence level 
a chargino signal in the \jjl\ mode provided that its production cross section 
is above 0.5 to 0.8~\pb, depending on the experiment. An integrated 
luminosity of 500~\ipb\ would allow the exploration of the chargino signal 
in this same mode down to cross sections of 0.2 to 0.4~\pb.
Although the search can be carried out almost to the kinematical limit 
for both luminosities, when the chargino and neutralino are close in mass 
($\mchipma-\mchioa~<~10$~\gev) or the mass of the neutralino 
is very light ($\mchioa < 20$~\gev), the experimental efficiencies 
decrease very quickly. However, it is not excluded that more specialized 
selection criteria can be envisaged to recover, at least partially, 
these regions.

  In the \jjjj\ mode, the minimum reachable cross section at \rs= 190 \gev\ 
to discover the chargino at the $5 \sigma$ confidence level 
is in the 0.4--0.7~\pb\ range, which can be reduced to 0.2--0.4~\pb\
with an integrated luminosity of 500~\ipb. In this mode, 
chargino--neutralino mass differences down to 5 \gev\
can be explored. 
 
In fig.~\ref{mincross} we show the minimum cross-section at the $5\sigma$
confidence level for an integrated luminosity of 150~\ipb\ for an average
experiment in the chargino-neutralino mass plane, for the \jjl\ 
and the \jjjj\ mode.                                            
                             
  In the \ll\ channel, the corresponding minimum cross section
at \rs= 190 \gev\ is in the range 4--5~\pb\ with 150~\ipb\ and
2--3~\pb\ with 500~\ipb. The minimum chargino-neutralino mass
difference is 10~\gev.

 Combining the three modes and assuming an integrated luminosity of
150~\ipb, the chargino search can go down to a minimum cross section
of 0.3--0.5~\pb\ depending on the experiment.
This conclusion is reached under the assumption that the
chargino BR are the same as those of the \w. A variety of enhancements
and suppressions of the leptonic and hadronic BR of the \chipma\  
can take place depending on the relevant SUSY parameters, as was pointed 
out above. For these cases the above results can be properly rescaled.
                                                         
\subsection{Scalar Lepton Searches}
\label{susysleptons}
\begin{figure}
\centerline{
\epsfig{figure=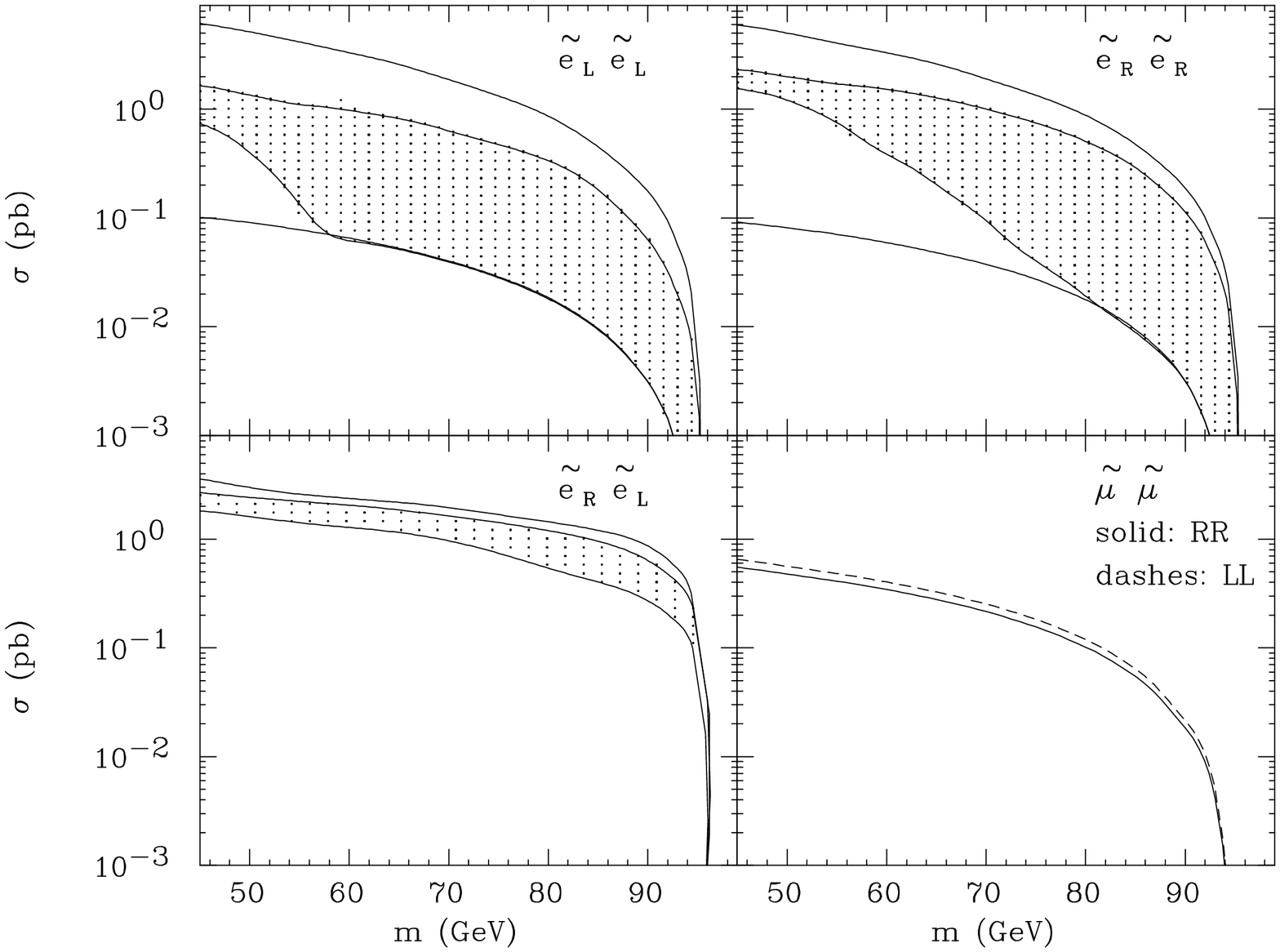,width=0.5\textwidth,angle=90}}                                  
\ccaption{}{\label{slepxs}                    
Cross sections for the production of various slepton pairs at $190~\gev$,
as a function of                                                         
the slepton mass. In the case of $\tilde e_L \tilde e_R$ production we assume
$m_{\tilde e_L}=m_{\tilde e_R}$. ISR corrections are included throughout.
For the selectron processes, the solid lines represent minimum and maximum
rates obtained by varying $M_2$, $\mu$ and \tanb\ in the range
allowed by the LEP1 constraints. The shaded areas have the additional
requirement $\mchipm>95$~GeV.
Notice that the minimum cross section for $\sele_L\sele_R$ is off scale when
the $\mchipm>95$~GeV requirement is not applied.}             
\end{figure}

Each SM charged lepton has two scalar partners, which will be called
right and left sleptons. As mixings among different slepton states are
assumed to be proportional to the corresponding Yukawa couplings, 
$\sele_{L,R}$ and $\smu_{L,R}$ are approximately mass eigenstates. On
the other hand, a
non-trivial mixing between the two $\stau$ states can be expected, especially
for large $\tan \beta$.
Smuons can be pair produced at LEP2 via $Z$ and $\gamma$ 
$s$-channel exchange.                       
Their production cross section, corrected for ISR, is shown in
fig.~\ref{slepxs}, as a function of $\msmu$. Identical results hold for the
$\stau$, in the limit of vanishing left-right mixing. In the case 
of the selectron,
neutralino-exchange in the $t$-channel can also
contribute to the production. Now the cross section is not uniquely
determined by $\msele$, but it depends also on the neutralino parameters $M_2$,
$\mu$, and \tanb. As an effect of the $t$-channel exchange, the
associated        
production of $\sele_L$ and $\sele_R$ is possible, even if the selectron
mixing angle vanishes. Figure~\ref{slepxs} shows the range of
$\sele$-production cross sections          
in the three possible channels ($\sele_L \sele_L$, $\sele_R \sele_R$,
$\sele_L \sele_R$). The minimum and maximum cross sections are obtained 
by varying the neutralino parameters within
the range allowed by LEP1 constraints on charginos and neutralinos from
the visible and invisible \z\ widths, consistently with the requirement
that the neutralino is the LSP ($\mchioa <\msele_{L,R}$).
We also show the minimum and maximum
cross sections obtained
with the further requirement $\mchipm > 95$ GeV, expected to hold if
the chargino search at LEP2 turns out to be unsuccessful. Because of
the interference between $s$-channel and $t$-channel contributions, the
$\sele$-production cross sections vary by more than an order of magnitude.
Notice in particular that they could be
significantly smaller than the \smu\ cross section. Knowledge of
(or constraints on) the parameters $M_2$, $\mu$, and $\tan \beta$ from
chargino and neutralino searches at LEP2 can be of great help to
sharpen the predictions on the expected $\sele$ cross section.

Both left and right sleptons can decay into the corresponding charged
lepton and a neutralino:
\begin{equation}
\tilde{\ell}^\pm \rightarrow \ell^\pm + \chio_i, 
~~~~~  i=1,2,3,4.                            
\end{equation}
Left sleptons can also decay into the corresponding neutrino and a chargino:
\begin{equation}
\tilde{\ell}^\pm \rightarrow \nu_\ell + \chipm_i, 
~~~~~ i=1,2.                                  
\end{equation}
If $m_{\tilde \nu_\ell}<m_{\tilde\ell}<\mchio$ then
$\tilde \ell$ will have only three-body decays.
In our analysis we will not consider this possibility.

\begin{figure}
\centerline{
\epsfig{figure=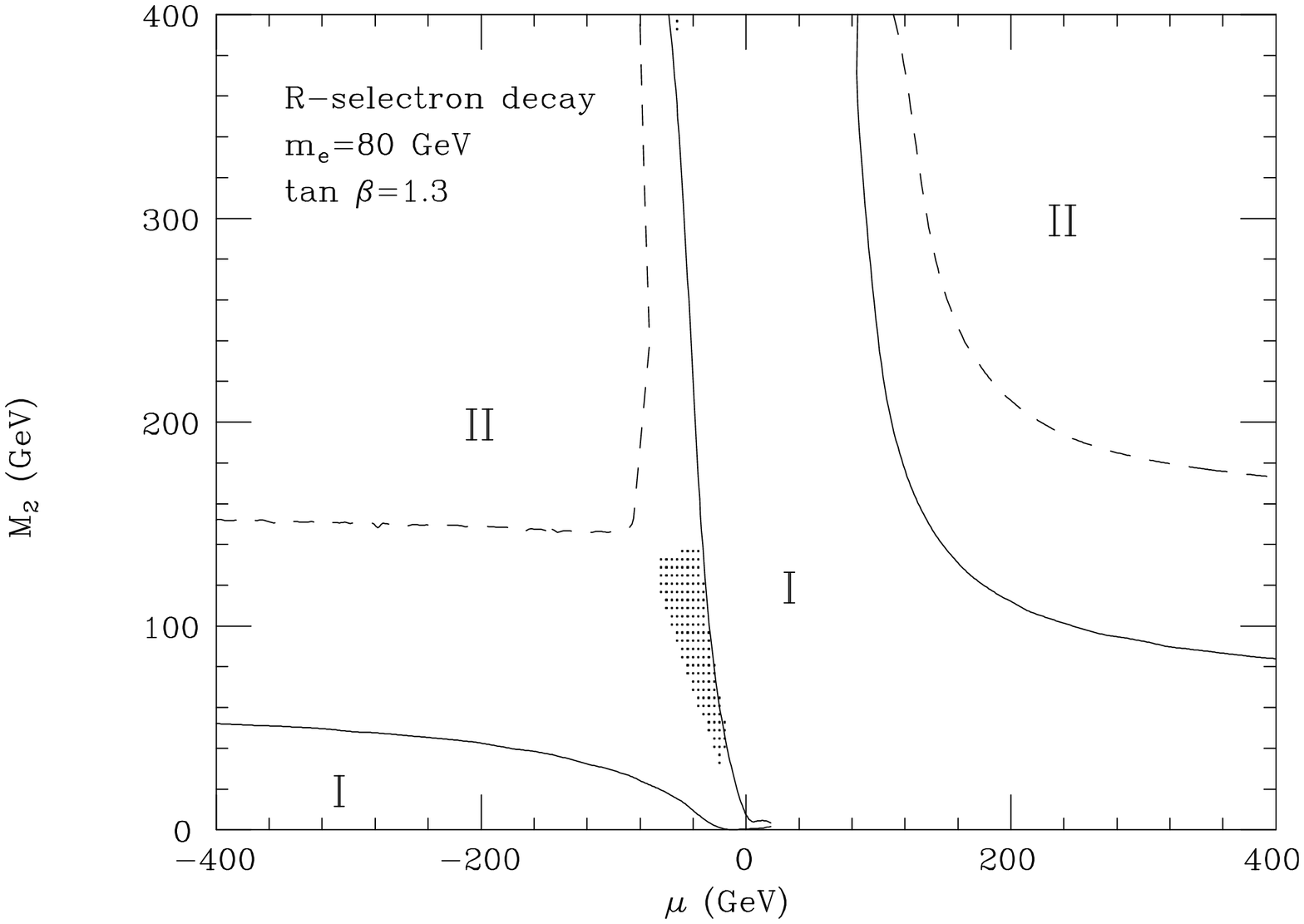,width=0.4\textwidth,angle=90.}
\epsfig{figure=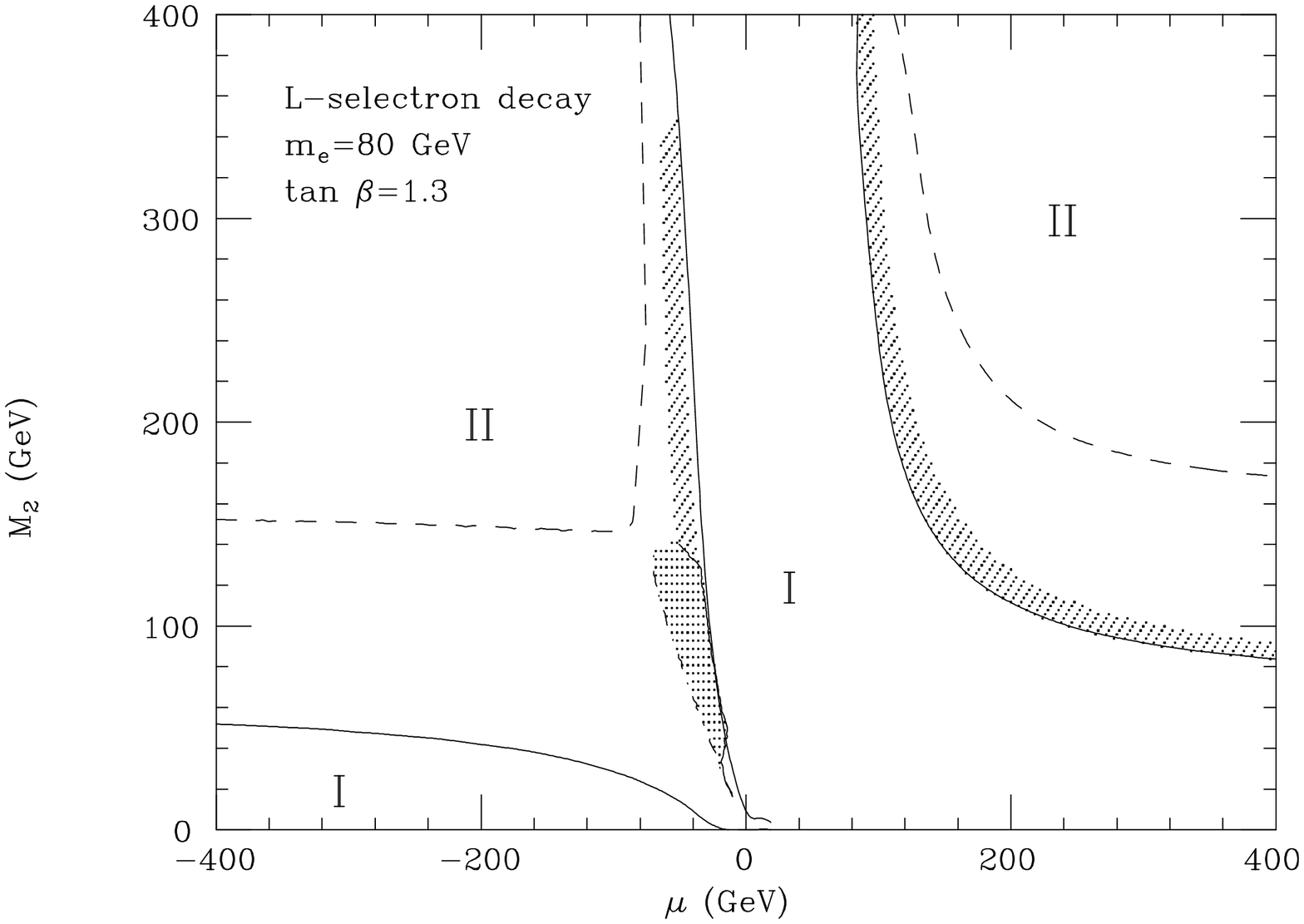,width=0.4\textwidth,angle=90.}
}                                             
\ccaption{}{\label{slepbr} 
Dominant slepton decay modes in the $(M_2,\mu)$ plane, for
$\mslep=80$~GeV and \tanb=1.3. The regions labeled I are excluded by LEP1
data. The regions II do not satisfy $\mslep>\mchioa$. 
The large unmarked regions correspond to $\slep\to \ell \chioa$ being the
dominant decay mode. In the dotted regions the dominant decay is
$\slep\to \ell \chiob$, while in the hatched area the dominant decay is
$\slep\to \nu_\ell \chipma$.}                                                      
\end{figure}                

In the case of the right slepton, the decay mode 
$\tilde{\ell}_R^\pm \rightarrow \ell^\pm + \chioa$ is always
dominant, aside from a small region of parameters shown in 
fig.~\ref{slepbr}{\em a} where                   
$\tilde{\ell}_R^\pm \rightarrow \ell^\pm + \chiob$ is the main
decay process. This region rapidly disappears as we increase \tanb.
For left sleptons, the decay mode
$\tilde{\ell}_L^\pm \rightarrow \nu_\ell + \chipma$ can also
become the dominant process in the region of parameters where the chargino
is rather light, as illustrated in fig.~\ref{slepbr}{\em b}.
Contrary to the case of
$\tilde{\ell}_R$, the regions of the dominant $\tilde{\ell}_L$ decay modes 
do not significantly depend on $\tan \beta$. 

The slepton decays into $\chipm$ and $\chio_i$ ($i>1$) give rise to cascade
processes, which may have very characteristic signatures \cite{leptwo}.  In
case of slepton discovery, these decay modes can give important information
about the  values of the relevant supersymmetric parameters.

The signature which was used in the studies that will now be presented is
an acoplanar pair of charged leptons of
the same flavour, accompanied by a large missing momentum.
%
%
In order to investigate the detectability of selectrons at LEP2,
a study was performed for the L3 experiment~\cite{r-selL3}, 
relying  on a detailed
description of the detector in form of a fast simulation, with detector
response and acceptance checked against data and against a full GEANT
description.
For the generation of selectron pairs, the BABY slepton generator
described in ref.~\cite{r-BABY} was used, taking into account initial state
radiation. In this generator, cascade decays are not included. Its use is
nevertheless justified for the studied R-selectrons, which
almost entirely decay to an electron and the lightest neutralino.
The following background reactions have been considered:
\begin{displaymath}      
e^+e^- \rightarrow l\bar{l}, W^+W^-, ZZ,
\end{displaymath}
together with the photon interactions:
\begin{displaymath}
\gamma\gamma \rightarrow l\bar{l}~{\rm and}
 ~e\gamma \rightarrow \nu W^\mp, e^{\mp}Z. 
\end{displaymath}
Looking for selectrons, only two electrons are expected in the detector, and
some energy has to be missing, due to the neutralinos escaping detection.
To suppress the background sources, while keeping a high signal efficiency,
two sets of selections were used.
For large values of the difference $\Delta$m
between $\msele$  and $\mchioa$, an
acoplanarity angle smaller than $130^o$ and a total transverse
momentum larger than 15~\gev are required (selection I).
For the parameter space region where $\Delta$m is small, the acoplanarity
angle was required to be below $160^o$, the total transverse
momentum larger than 5~\gev and the missing energy in the event larger
than 150~\gev\ (selection II).

\begin{figure}
\centerline{
\epsfig{figure=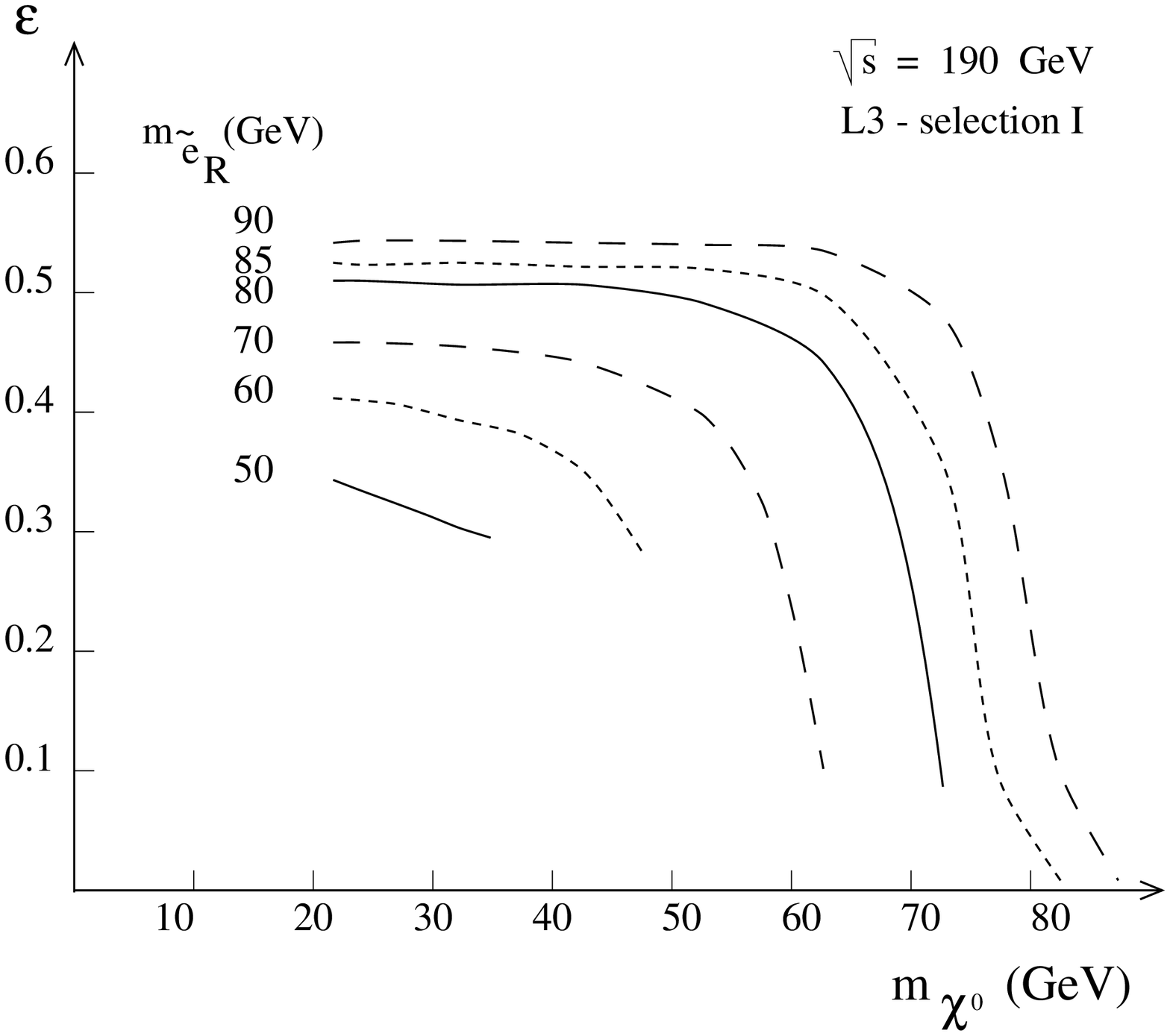,height=6.cm,angle=0.}
\epsfig{figure=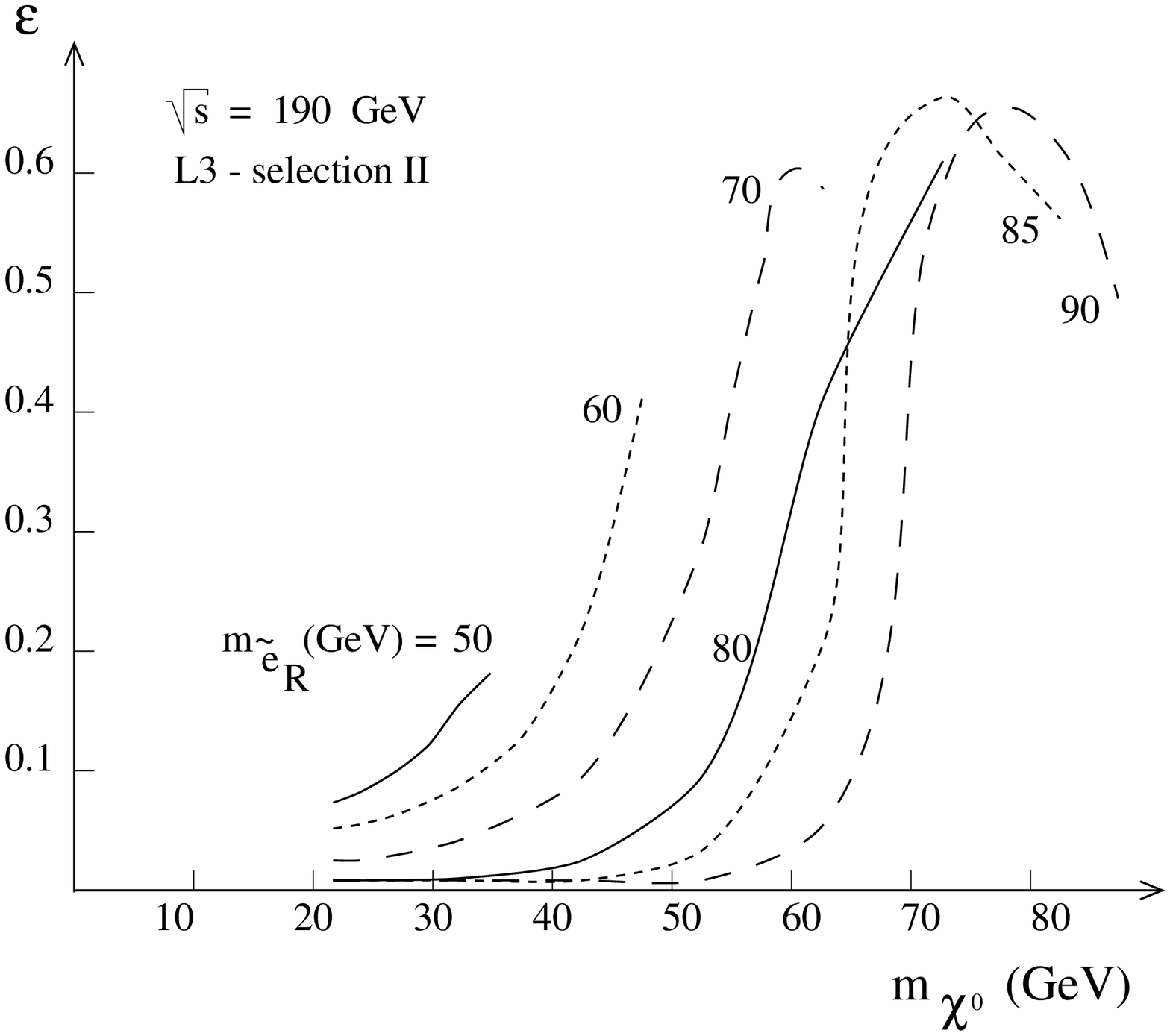,height=6.cm,angle=0.}
}
\ccaption{}{\label{effi}
Efficiency of the selection cuts I and II on selectron 
pairs of various masses, as a function of $\mchioa$.
}
\end{figure}

The remaining background cross sections are 82 fb for selection I and 7.2 fb
for selection II.
The selection efficiencies for the signal are shown in fig.~\ref{effi}, 
for different                                                         
selectron masses, as a function of the neutralino mass. For $\msele \geq
60~\gev $, they are                                            
larger than 40\% for $\Delta$m $>15$~\gev, and of the
order of 60\% for $5<$$\Delta$m$<10$~\gev, where the tight \missE\  cut
is most efficient.

\begin{figure}
\centerline{
\epsfig{figure=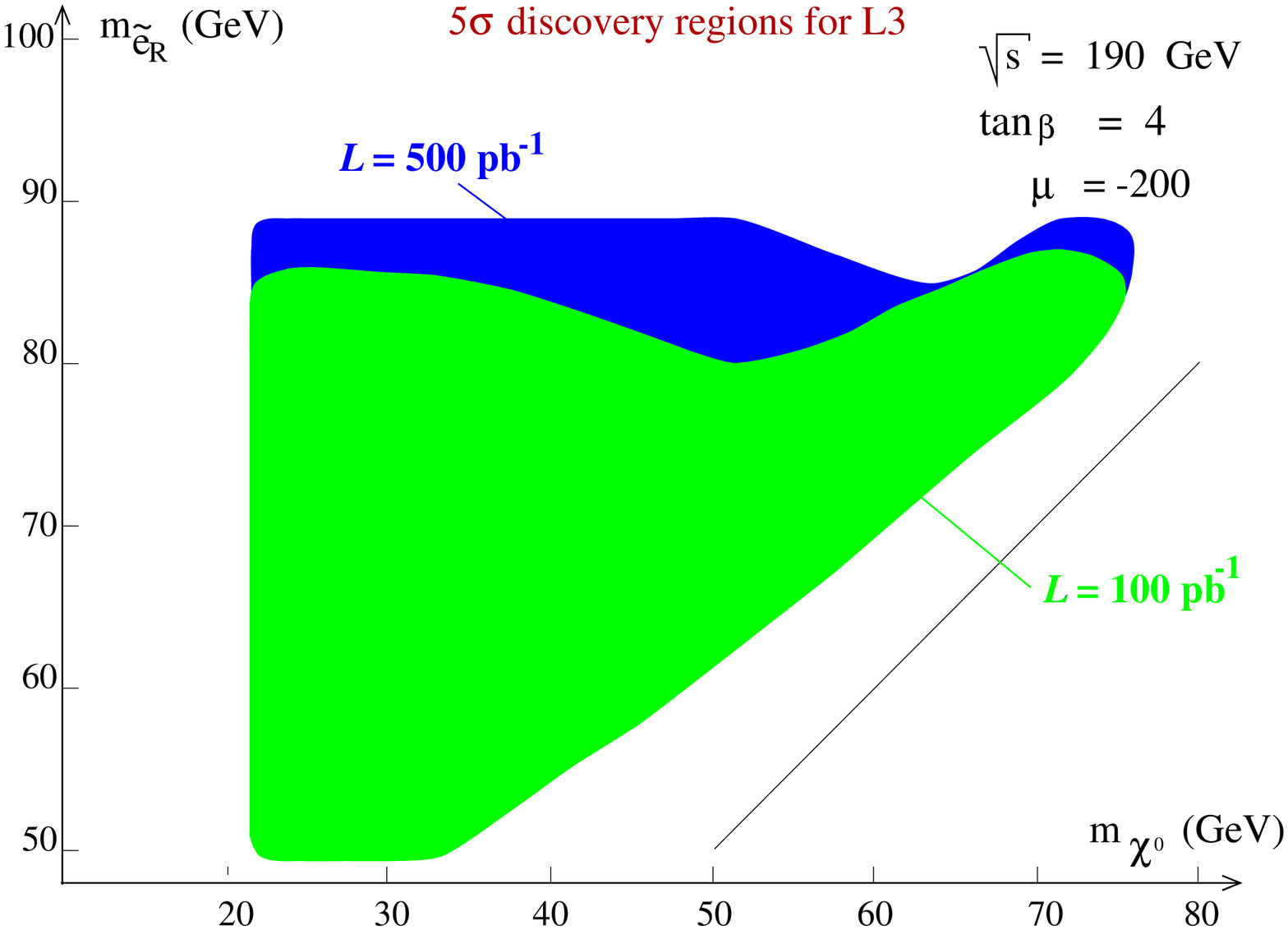,height=6.cm,angle=0.}
\epsfig{figure=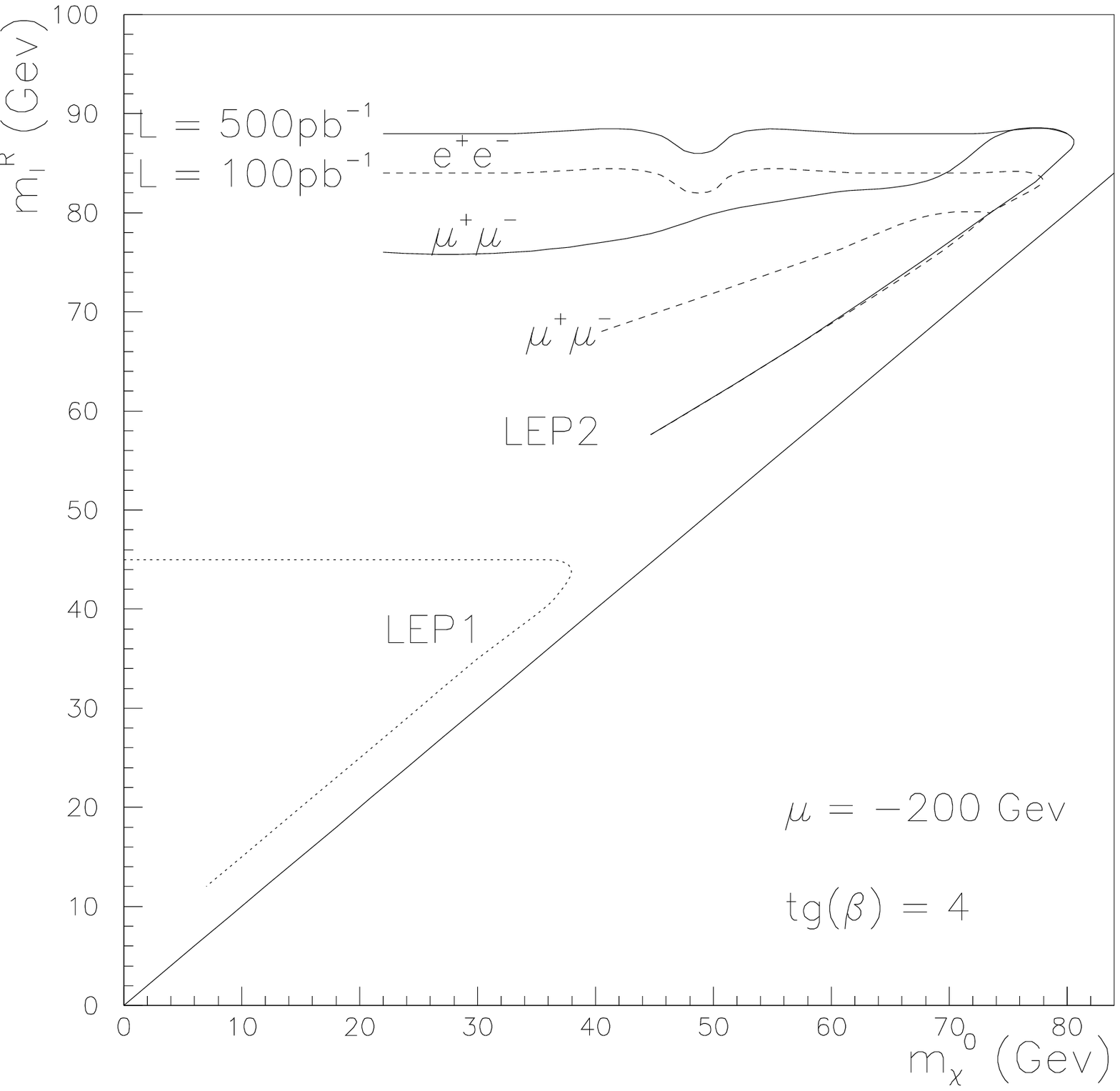,height=6.cm,angle=0}
}
\ccaption{}{\label{limits}
Limits of detectabily of sleptons with $5\sigma$,
at 190~\gev, $\tanb = 4$ and $\mu = -200~\gev$, for integrated
luminosities of $100 pb^{-1}$ and $500 pb^{-1}$, in the plane
$\mslep-\mchioa$, (a) for $\tilde{e}_R^+\tilde{e}_R^-$ 
in the L3 detector at LEP II and (b) for right 
selectrons and smuons, in an ideal  LEP detector.
}
\end{figure}

Thus, the minimum signal cross sections needed for a $5\sigma$ statistical
significance are, 143 fb (64 fb) in
the region where selection I is applied, and 42 fb (19 fb) where selection II
is used, for an integrated luminosity of $100\; pb^{-1} (500\; pb^{-1})$.
For $\mu=-200$, $\tan \beta =4$ at $\sqrt{s}=190$~\gev, the resulting
discovery region in the (\msele, \mchioa) plane is shown in
fig.~\ref{limits}a.
The slight dip along the $\mchioa$ axis for high values of $\msele$ is due
to the transition from one set of cuts to the other one.
The lack of sensitivity observed in the region $\msele-\mchioa<15$ GeV at low
\msele\ is mostly due to the particular choice of SUSY parameters and to the
mass relations built into the BABY Monte-Carlo, which artificially
restrict the possible values of the neutralino mass.
The precise value of $\msele-\mchioa$ at which the detection efficiency will
drop below the 10\% level depends critically on the analysis cuts and on the
accuracy of the detector simulation. Fig.~\ref{effi} suggests that the L3
analysis could be sensitive to mass differences down to values of the
order of $5-10$ GeV.
%
%

A similar  exploratory
study has been performed for selectron, smuon 
and stau pair production. In this preliminary study, 
the detector acceptance and performance have been crudely simulated by 
various cuts and by smearing the kinematical variables to take into
account the measurement errors, see ref.~\cite{IIHE}. The production of
sleptons has been simulated using the generator SUSYGEN~\cite{SUSYGEN}, 
taking into account ISR and cascade decays. The                      
background processes have been simulated using PYTHIA~\cite{jetset}.
In addition to the background processes generated in the preceeding
analysis, the following reactions have been considered:
\begin{displaymath}
e^+e^- \rightarrow f\bar{f}, \nu\bar{\nu}Z~{\rm and}
~\gamma\gamma \rightarrow f\bar{f}.
\end{displaymath}
The selection cuts used for selectrons and smuons are very similar 
to those used in the L3 analysis. However three different sets are used 
instead of two, in order to try to suppress the dip observed 
in the contour of fig.~\ref{limits}a. Selection~3 is
identical to selection II of L3 analysis. In selection~2, 
the missing energy cut is moved from $150~\gev$ to     
 $130~\gev$. In selection~1 the acoplanarity angle has to be
smaller than $160^\circ$, the total transverse momentum
larger than 15~\gev and the 
difference of longitudinal momentum between the negative and the
positive lepton, has to
be smaller than 40~\gev. 
As the $\tau$ can decay either leptonically
or hadronically, the final state searched for is either 2 jets or
a jet and a lepton. Two lepton final states, representing 13\% of
the total, would hardly be distinguished from smuon or selectron
production. The total number of charged particles is limited
to four, the acoplanarity angle must lie between $5^\circ$ and
$150^\circ$, the total transverse momentum between 15 and 
29~\gev and the missing energy must be larger than 150~\gev.

After applying the above selections, the cross section corresponding to
the total remaining background, $\sigma_b$, is reduced to the values
shown in table~\ref{backgr}.
The difference in background remaining for selectron production, compared
with the L3 study, is due to the more realistic detector 
efficiencies used there.

%
%
\begin{table}[hbtp]
\begin{center}
\begin{tabular}{|c|l|l||l|}
\hline                   
Final & $\epem$     & $\mu^+\mu^-$         & $\tau^+\tau^-$ \\
state &             &                      &                \\
\hline
selection 1 & 115 fb & 104 fb &       \\
selection 2 &  23 fb &  21 fb & 55 fb \\
selection 3 &  11 fb &  10 fb &       \\
\hline
\end{tabular}
\end{center}
\caption{Cross sections of remaining background, 
for \epem, $\mu^+\mu^-$ and $\tau^+\tau^-$ final
states.}
\label{backgr}
\end{table}

The $5\sigma$ detectability ranges obtained 
after applying the above cuts on simulated signal events
are shown in fig.~\ref{limits}b, assuming $\tanb = 4$ and $\mu = -200~\gev$.
The results for selectrons and smuons are shown           
in the $\mslep-\mchioa$ plane  for 
$\cal{L} =$ $100~pb^{-1}$      
and $\cal{L}$ $= 500~pb^{-1}$. 
The agreement between selectron limits
in both analyses brings some confidence in the cruder analysis of
fig.~\ref{limits}b, which is the only one existing for smuons.
With the above selection staus are not
observed with $5\sigma$ at these luminosities,
for this set of MSSM parameters. It is only possible in the most
favourable case: $\mu < -50$ or $\mu > 300~\gev$ and $\tanb \simeq 1$, at 
$\cal{L} =$ $500 pb^{-1}$, for
$m_{\stau}$ 
$< 60~\gev$~\cite{Favart}.

A further study involving a realistic simulation of the detector response has
been performed at $\rs = 190~\gev$ for stau pairs in the OPAL
detector~\cite{Koji}.
It relies on a full detector simulation, with both background and signal 
generated with PYTHIA~\cite{jetset}; the stau mixing is neglected.
A different set of cuts has been used than for the preliminary study above.
They essentially differ by looser cuts on the transverse momentum
and on the acoplanarity of the two jets in the final state: 
$\pt> 10$ GeV and $\theta> -3\pt({\rm GeV}) +60^o$. This is
compensated by a cut on detected gammas that must have less than $10~\gev$.
The missing energy cut is replaced by a cut on the two jet invariant mass
that must be smaller than $50~\gev$.

\begin{figure}
\centerline{
\epsfig{figure=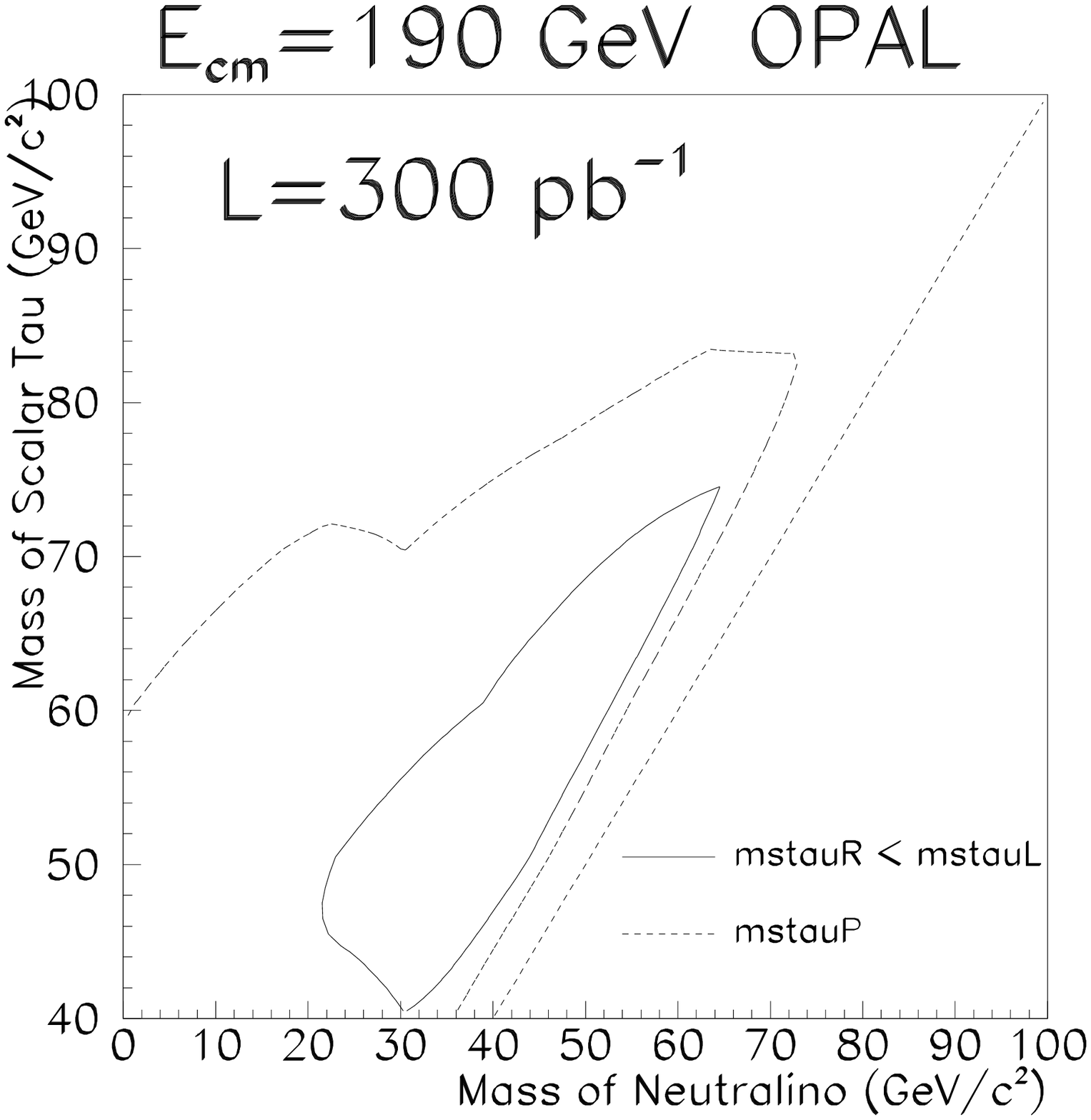,width=0.4\textwidth,angle=0}}                                          
\ccaption{}{\label{Koji-limit}                      
OPAL's $5\sigma$ limits of detectability for stau pairs in the 
$\mslep-\mchioa$ plane.                      
$\sqrt{s}=190$~GeV, for an integrated        
luminosity of $300~pb^{-1}$.}
\end{figure}

After these cuts, only $2~\fb$ of background remain, allowing for
a much lower value of the observed signal cross section,
of the order of $12.5~\fb$,
for $\cal{L} =$ $300 pb^{-1}$, leading to a domain of
detectability of stau pairs at $5~\sigma$, shown in 
figure~\ref{Koji-limit}. The two contours correspond respectively to 
the case where the mass of the $\stau_R$ is much smaller than
the mass of the $\stau_L$ and to the case where the
two masses are degenerate.
The above studies show that with 500 pb$^{-1}$ selectron pairs could be 
detected at LEP2 with                         
$5~\sigma$ up to masses about $5~\gev$ below the kinematical limit,
with $\Delta$m $> 10~\gev$. The reach could be significantly reduced if
the SUSY parameters were such as to induce large destructive interference
among the production diagrams.                  
The domain of detectability for smuons is less model dependent, but 
limited to about 20~GeV below the kinematical limit.
In the case of staus the reach is still more constrained,
because of the lower efficiency in the detection of the  
final state taus.
\clearpage
\subsection{Stop and Sbottom}
\label{susystop}
\subsubsection{Phenomenological Aspects}
The SUSY partners $\st_L$ and $\st_R$ of the top quark
are expected to be mixed due to the large top Yukawa coupling.
Therefore, the lighter mass eigenstate $\stopone$ will most likely be the
lightest squark and may even be the lightest visible SUSY 
particle.
If $\tan\b\,$ is large $(\tan\b > 10)$ also the sbottom $\sb_1$ can
be rather light~\cite{wwien}.
Thus, it may well be that $\stopone$ or/and $\sb_1$ 
are within the reach of LEP2.

The mass matrix for the stops in the ($\st_L,\:\st_R$) basis is given 
by~\cite{susyrev}:
\beq
  {\cal M}^2_{\st} = \left( \baa{ll} 
     M_Q^2 + m_t^2 +  m_Z^2 \cos 2\b\,(T_{3t} - Q_t \sin^2\tW ) \:\:
        & m_t\, (A_t - \mu \cot\b )  \\ 
     m_t\, (A_t - \mu \cot\b ) 
        & M_U^2 + m_t^2 + m_Z^2 \cos 2\b\, Q_t \sin^2\tW  
  \eaa \right)
\eeq
where $T_{3t}$ and $Q_t$ are the third isospin-component and electric charge
of the top quark, respectively.
For the $\sb$ system analogous formulae hold with $M_U^2$ replaced 
by $M_D^2$ and with the off-diagonal element replaced by 
$m_b\, (A_b - \mu \tan\b )$.
The mass eigenstates are
$\stopone = \cth_{\st}\;\st_L + \sth_{\st}\;\st_R$,  
$\st_2 = -\sth_{\st}\;\st_L + \cth_{\st}\;\st_R$, and analogously
for $\sb_{1,2}$. Thus, the experimental determination of the mass 
eigenvalues and the mixing angles provides information on the
soft SUSY-breaking parameters $M_Q$, $M_U$, $M_D$, $A_t$, and $A_b$. 
                   
\noi
\begin{minipage}[b]{60mm}
\hspace{4mm} 
The reactions $\eeto\stst$ and $\eeto\sbsb$ proceed via s-channel
$\gamma$ and $Z^0$ exchange. The $Z^0$ couplings to
$\sq_1\q{\sq_1}$ are proportional $T_{3q} \cthq_{\sq} - Q_q \sthq_W$. 
In fig.~\ref{fig:sqfig1} (a) and (b) we show the total cross sections 
for these processes at  $\sqrt{s} = 175$ GeV and $\sqrt{s} = 192.5$ GeV 
as a function of $\cth_{\st}$ and $\cth_{\sb}$ for several mass 
values of $\stopone$ and $\sb_1$.
Here we have included QCD radiative corrections and initial 
state radiation (ISR)~\cite{drees,zerwas}. 
At $\sqrt{s} = 192.5$ GeV, for masses of 70~GeV the cross sections 
reach values of 0.7~pb and 0.55~pb for stop and sbottom production, 
respectively.                                   
There is a pronounced dependence on the mixing angle 
for $\cth_{\st,\sb} \gsim 0.6$.
Radiative corrections are impor-
\end{minipage}
\hspace{2mm}
\begin{minipage}[b]{106mm} {\setlength{\unitlength}{1mm}   
\begin{picture}(106,85)
  \put(0,2){\mbox{\epsfig{figure=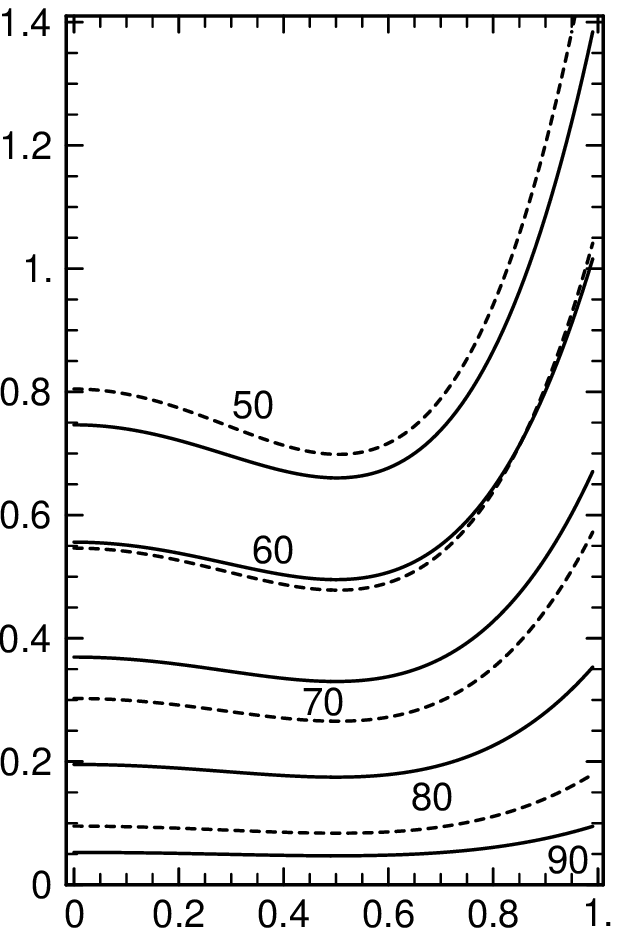,width=5.15cm}}}
  \put(55,2){\mbox{\epsfig{figure=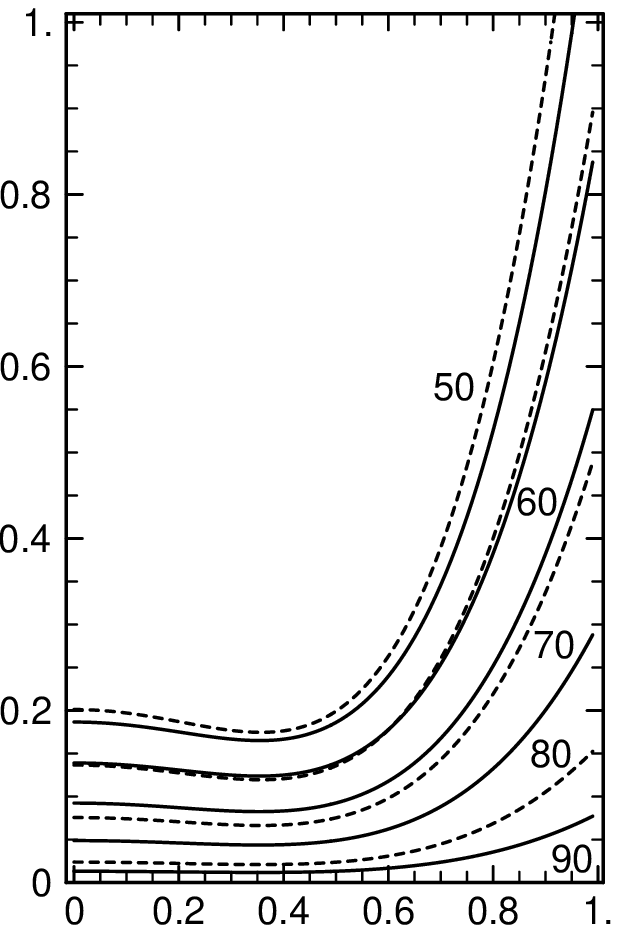,width=5.15cm}}}
  \put(13,71){\makebox(0,0)[bl]{{\small (a)}}}
  \put(68,71){\makebox(0,0)[bl]{{\small (b)}}}
  \put(53,0){\makebox(0,0)[br]{{\small $\cth_{\st}$}}}
  \put(106,0){\makebox(0,0)[br]{{\small $\cth_{\sb}$}}}
  \put(1,84){\makebox(0,0)[bl]{{\small $\sigma$~[pb]}}}
  \put(55,84){\makebox(0,0)[bl]{{\small $\sigma$~[pb]}}}
\end{picture}}                                 
\refstepcounter{figure}   
\label{fig:sqfig1}
\begin{small} 
  {Figure~\arabic{figure}:}~\it Total cross section in pb 
  at $\sqrt{s} = 175$ GeV (dashed lines) and $\sqrt{s} = 192.5$ GeV 
  (solid lines) as a function of the mixing angle for squark masses of 
  50, 60, 70, 80, and 90 GeV, for (a) $\eeto\stst$ and (b) $\eeto\sbsb$.
\end{small} 

\end{minipage}

\noi
\begin{minipage}[b]{98mm}
tant as can be seen in fig.~\ref{fig:sqfig2}, where 
we show the QCD and ISR corrected cross section together 
with the Born approximation for $\eeto\stst$. \\


\hspace{4mm}
If $\,\stopone$ is the lightest charged SUSY particle, it will decay with
100\% branching ratio according to $\,\stopone\to c\,\chioa$.
If $\,\mstopone > m_{\ch_1} + m_b$, then the decay $\,\stopone\to b\,\chipa$
has practically 100\% branching ratio in the mass range of LEP2.
In fig.~\ref{fig:sqfig3} we show the domains of the $\,\stopone$ decay modes 
in the ($M_2,\:\mu$)~plane for $\mstopone = 80~\gev$ and $\tan\b = 2$.
There is a small strip where the decay $\stopone\to c\,\nt_2$ is also possible.
The signature from the $\,c\,\chioa$ decay is two jets and missing energy    
$(E\llap/ )$, whereas for $\,\stopone\to b\,\chipa$ one has two $b$ jets 
accompanied by two leptons + $E\llap/$, or lepton + jet + $E\llap/$,
or two jets + $E\llap/$. 
In the case of $\,\stopone\to b\,\chipa$ it will obviously be useful if the
$\chipa$ has already been observed and its main properties are 
\end{minipage}                     
\hspace{2mm}
\begin{minipage}[b]{68mm} {\setlength{\unitlength}{1mm}   
\begin{picture}(68,40)
  \put(0,6){\mbox{\epsfig{figure=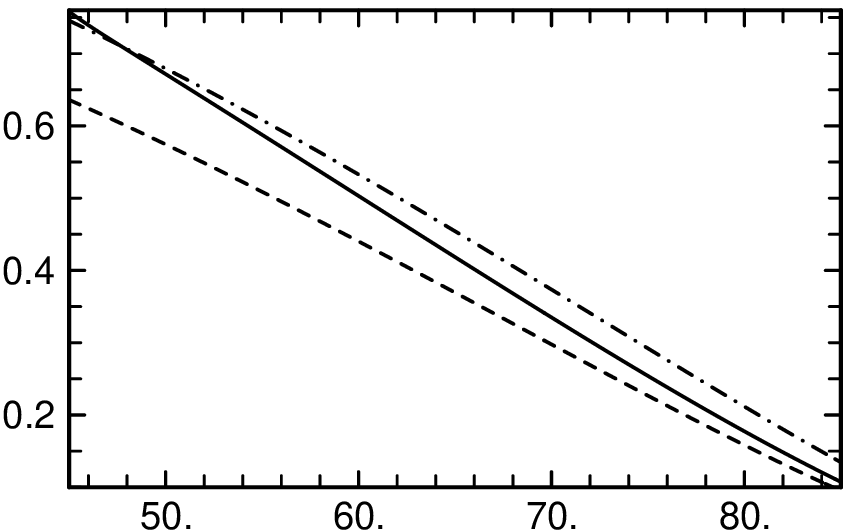,width=6.8cm}}}
  \put(63,34){\makebox(0,0)[br]{{\footnotesize $\cth_{\st}=0.4$}}}
  \put(63,39){\makebox(0,0)[br]{{\footnotesize $\sqrt{s}=192.5$~GeV}}}
  \put(68,0.5){\makebox(0,0)[br]{{\small $\mstopone$}{\footnotesize ~[GeV]}}}
  \put(0,50){\makebox(0,0)[bl]{{\small $\sigma\,(\stst)~[pb]$}}}
\end{picture}}                                
\refstepcounter{figure}   
\label{fig:sqfig2}
\begin{small} 
  {Figure~\arabic{figure}:}~\it Born approximation (- - -), 
  QCD corrected (-.-), and QCD+ISR corrected (---) cross section for
  $\eeto\stst$ as a function of $\mstopone$ for $\sqrt{s}=192.5$~GeV 
  and $\cth_{\st}=0.4$. 
\end{small}
\end{minipage}
known.

The lifetime of $\stopone$ is expected to be larger than the
hadronization scale \cite{hikasa}; thus $\stopone$ hadronizes first into 
a colourless $(\stopone \q q)$ bound state before decaying.
This affects the spectrum and multiplicity of final state
hadrons, as discussed in ref.~\cite{SUSYGEN}.
The decay width for $\stopone \to b\,\chipa$ becomes 
larger than 0.2 GeV for $\mstopone - \mchipma \gsim 25~\gev$
and $\cth_{\st} > 0.9$.                                 

\noi
\hspace{5mm}\begin{minipage}[t]{77.5mm}   
{\setlength{\unitlength}{1mm}  
\begin{picture}(80,76)                        
\put(6,4){\mbox{\psfig{figure=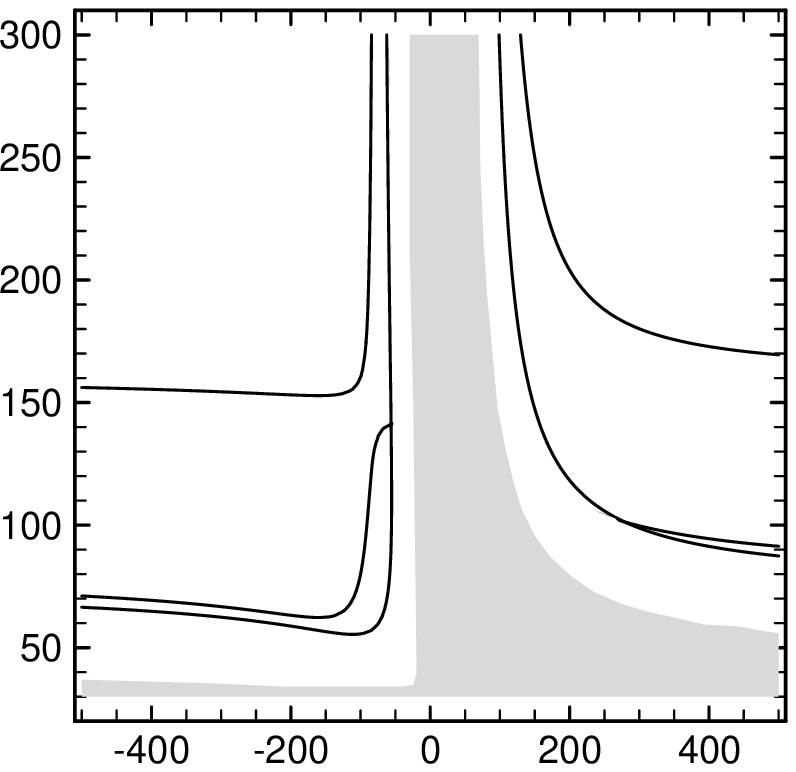,height=6.6cm}}}
\put(72.5,1){\makebox(0,0)[br]{{\small $\mu$~[GeV]}}}
\put(6,70){\makebox(0,0)[bl]{{\small $M_2$~[GeV]}}}
\multiput(16,58)(36,0){2}{
  \makebox(0,0)[bl]{{\footnotesize $\mstopone < \mchioa$}}}
\put(16,30){                                      
  \makebox(0,0)[bl]{{\footnotesize $\stopone \to c\, \chioa$}}}
\put(54,34){
  \makebox(0,0)[bl]{{\footnotesize $\stopone \to c\, \chioa$}}}
\put(16,22){
  \makebox(0,0)[bl]{{\footnotesize $\stopone \to c\, \nt_{1,2}$}}}
\put(28,21){\vector(2,-1){8}}
\put(53,28.5){
  \makebox(0,0)[bl]{{\footnotesize $\stopone \to c\, \nt_{1,2}$}}}
\put(65,27.5){\vector(1,-1){4.5}}
\put(18,12){
  \makebox(0,0)[bl]{{\footnotesize $\stopone \to b\, \chipa$}}}
\put(55,18.5){
  \makebox(0,0)[bl]{{\footnotesize $\stopone \to b\, \chipa$}}}
\put(43.5,12){\makebox(0,0)[bl]{{\footnotesize excluded}}}
\end{picture}}
\refstepcounter{figure}   
\label{fig:sqfig3}
\begin{small} 
  {Figure~\arabic{figure}:}~\it Parameter domains in the $(M_2,\:\mu)$ plane 
  for the various $\stopone$ decay modes, for $\mstopone=80$ GeV and 
  $\tan\b=2$. The grey area is excluded by LEP1.                 
\end{small} 
\end{minipage}
\hspace{5mm}
\begin{minipage}[t]{77.5mm} {\setlength{\unitlength}{1mm}   
\begin{picture}(80,76)
\put(6,4){\mbox{\psfig{figure=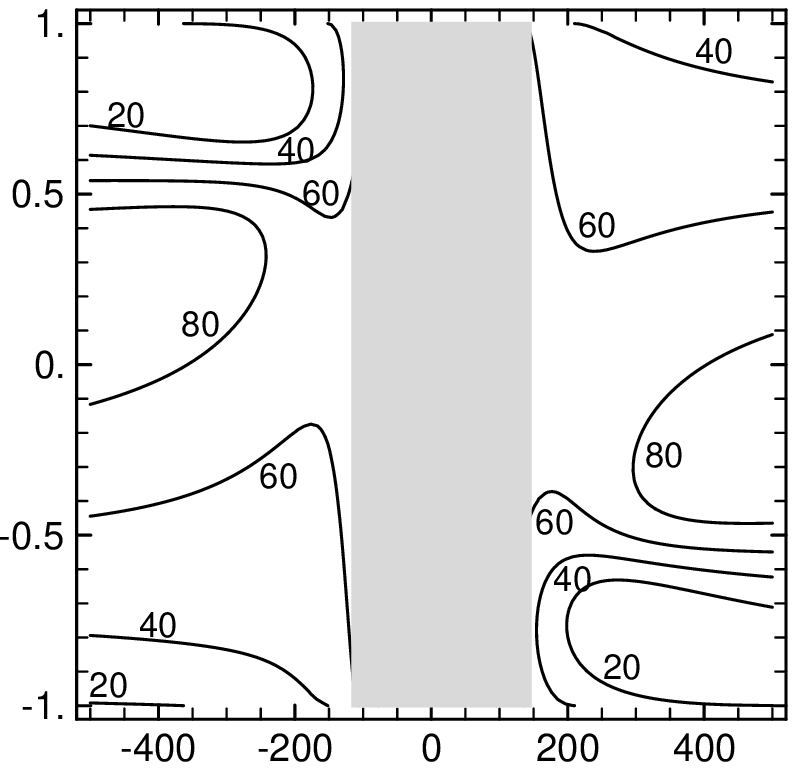,height=6.6cm}}}
\put(72.5,1){\makebox(0,0)[br]{{\small $\mu$~[GeV]}}}
\put(6,70){\makebox(0,0)[bl]{{\small $\cth_{\sb}$}}}
\put(37,12){\makebox(0,0)[bl]{{\footnotesize excluded}}}
\end{picture}}
\refstepcounter{figure}   
\label{fig:sqfig4}
\begin{small}   
  {Figure~\arabic{figure}:}~\it Contour lines for the branching ratio  
  (in \%) of $\sb_1\to b\,\chioa$, for $m_{\sb_1}=80$ GeV, $\tan\b=30$, 
  and $M_2=60$ GeV. The grey area is excluded by LEP1.
\end{small} 
\end{minipage}
\hspace{5mm}

The main decay modes of the $\sb_1$ are $\sb_1\to b\,\chioa$ and 
$\sb_1\to b\,\nt_2$, the second decay being possible in the parameter 
region approximately given by $M_2 < m_{\sb_1}-m_b$
or $|\mu| < m_{\sb_1}-m_b$. In fig.~\ref{fig:sqfig4} we show the 
branching ratio for $\sb_1\to b\,\chioa$ as a function of $\cth_{\sb}$ 
and $\mu$ for $m_{\sb_1} = 80$~GeV, $\tan\b = 30$, and $M_2 = 60$ GeV. 
As can be seen, the decay $\sb_1\to b\,\nt_2$ plays an important role 
for $|\cth_{\sb}| > 0.5$. Moreover, there is a certain dependence on the sign
of $\cth_{\sb}$.

\begin{figure}
\centerline{
\epsfig{figure=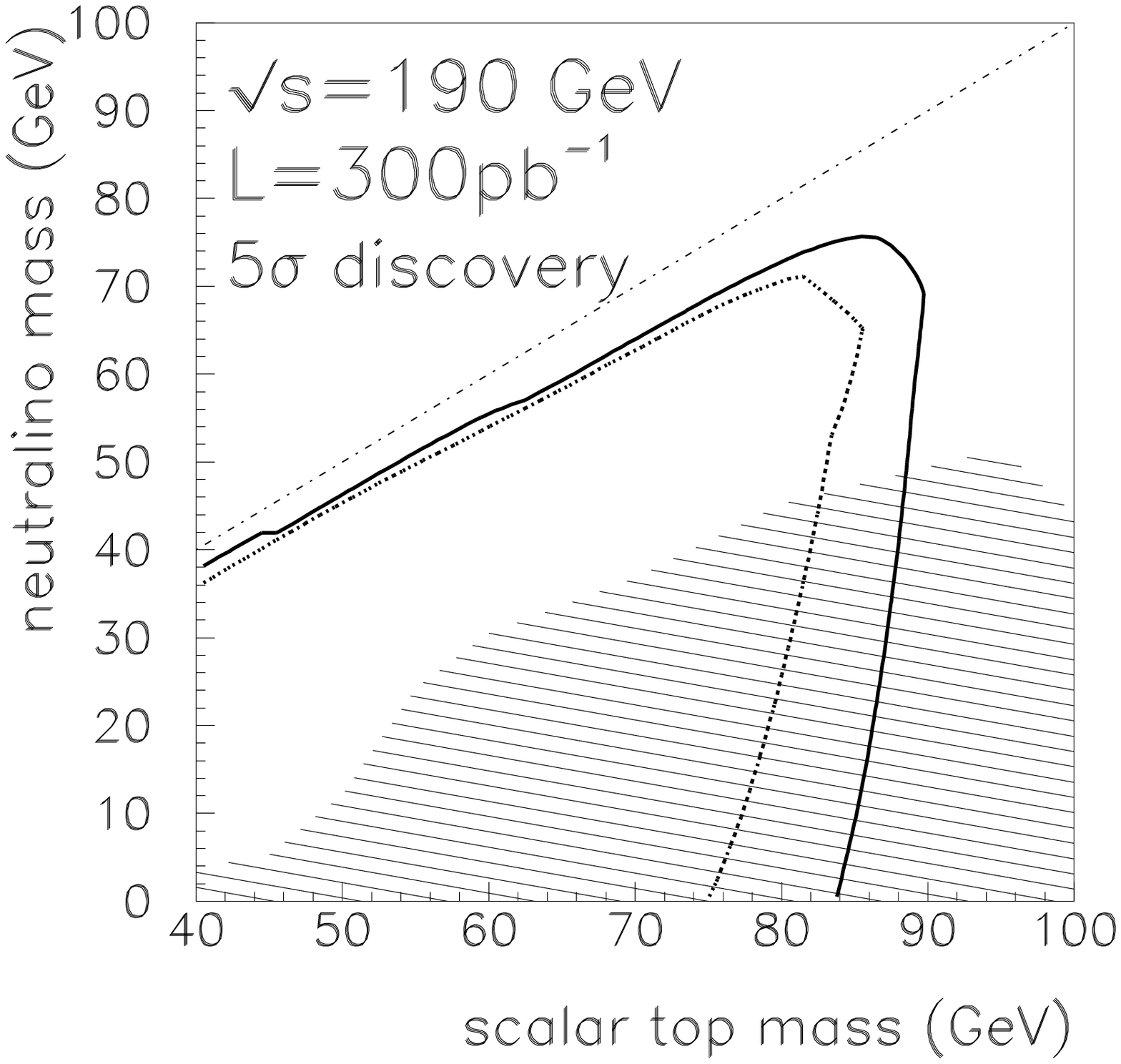,width=0.4\textwidth,angle=0}
\epsfig{figure=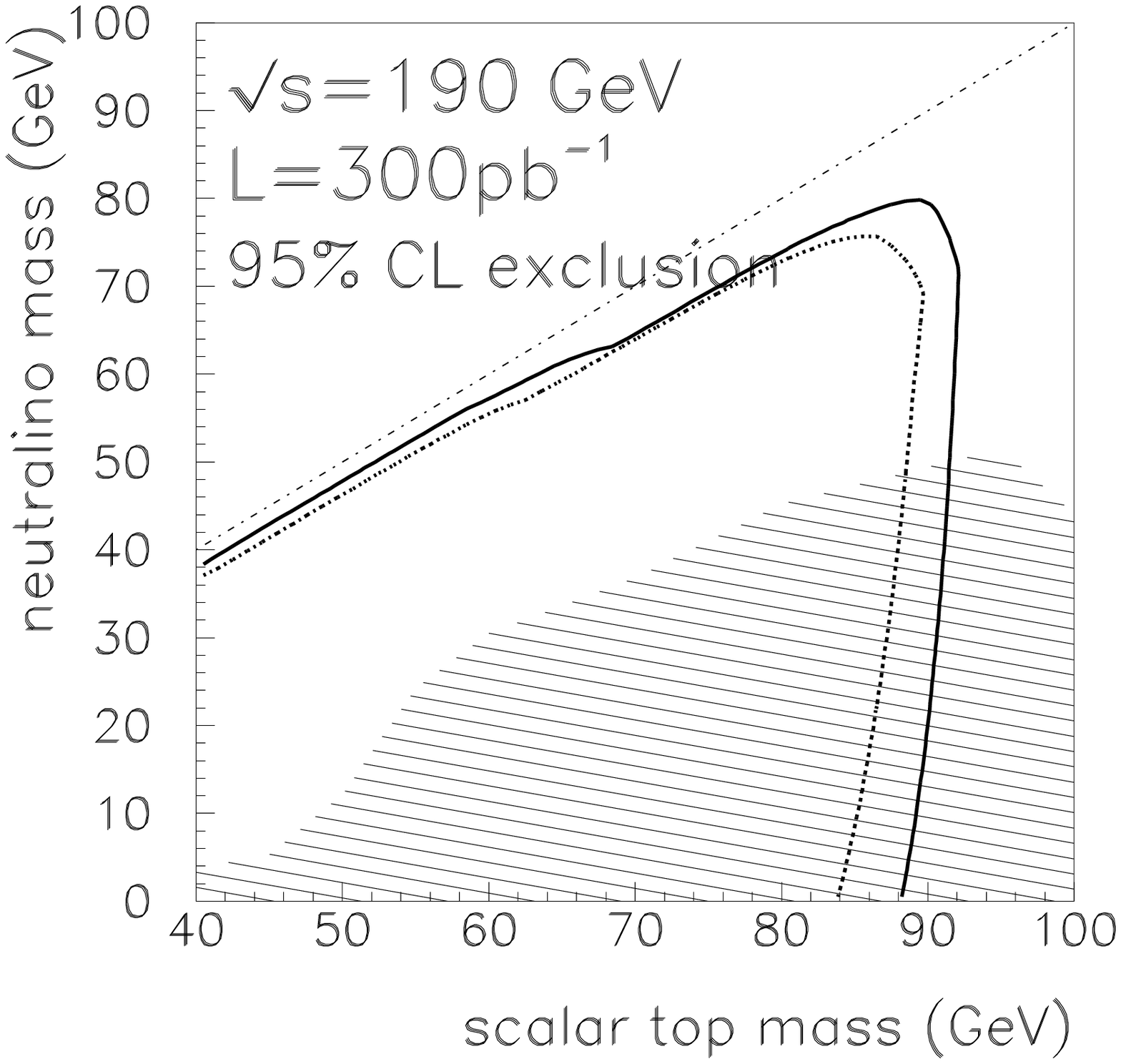,width=0.4\textwidth,angle=0}
}                      
\vspace{-2cm}
\ccaption{}{\label{stopfig1}
5 $\sigma$ discovery reach (left) and
95\%~CL limits (right) for the \stop\ search at 190 GeV.
The solid (dashed) lines correspond to maximal (minimal) coupling to the $Z$.
The shaded area corresponds to the                    
current 95\%~CL Tevatron limits \cite{D0stop}.}
\end{figure}

\subsubsection{Search Strategy for Stop}
\begin{table}         
\begin{center}
\begin{tabular}{|l||c||c|c|c|c|}\hline\hline
Luminosity   & coupling    & 5$\sigma$ & 95\% C.L. & 5$\sigma$ & 95\% C.L.\\
\hline\hline
$\Delta m$   &  & 20 GeV    & 20 GeV    & 10 GeV     &  10 GeV \\
\hline\hline
75~pb$^{-1}$ &  full & 83 GeV    & 87 GeV    & 79 GeV     & 83 GeV   \\
             &  zero & 73 GeV    & 82 GeV    & 67 GeV     & 77 GeV   \\
\hline
150~pb$^{-1}$&  full & 87 GeV    & 90 GeV    & 83 GeV     & 87 GeV   \\
             &  zero & 81 GeV    & 86 GeV    & 77 GeV     & 82 GeV    \\
\hline
500~pb$^{-1}$&  full & 91 GeV    & 91 GeV    & 88 GeV     & 91 GeV    \\
             &  zero & 88 GeV    & 90 GeV    & 84 GeV     & 88 GeV  \\
\hline\hline
\end{tabular}
\end{center}
\vspace*{-0.5cm}
\ccaption{}{\label{tab:stlimit} The maximum $\stopone$ mass
                for 5$\sigma$ discovery and 95\% C.L. exclusion 
                at $\sqrt{s} = 190$~GeV as a function
                of the integrated luminosity.
                The numbers are for one typical LEP experiment for
                the two cases of  the full and zero coupling of $\stopone$
                to the Z boson.}
\end{table}
A search for pair production of $\stopone$ with $\stopone$ decaying into
$c \chioa$ was made at LEP1 \cite{opalstop,alephstop}.
The background situation at LEP2 is, however, more severe.
DELPHI, L3 and OPAL have studied the search strategy for
$\stst \ra c \chioa \bar{c} \chioa$ events.
To determine the selection criteria, an integrated luminosity of
300-500 pb$^{-1}$ was assumed at $\sqrt{s} = 190$~GeV.
The $\stst$ event generators which are used for the three
experiments 
are described in ref.~\cite{SUSYGEN}.
The analyses were performed with realistic detector simulation.
L3 \cite{Lthree} and OPAL \cite{OPAL} have applied cuts
to the event shape and kinematical variables.
DELPHI \cite{DELPHI} has used a statistical method
to set the cuts.

It is important to have a sensitivity for small values of the \stop\-\chioa\
mass difference ($\Delta m$) down to values of the order of 5 GeV,
as it is difficult for the Tevatron experiments \cite{D0stop} to                
fully cover this region because of small missing \pt.
The visible energy of $\stst$ events is small for small $\Delta m$,
as the $\chioa$'s carry a large fraction of
energy, and the energy of particles form fragmentation in small. 
L3 and OPAL accept events with typical $E_{\mathrm vis} > 0.1 \sqrt{s}$.
In this case the main background is due to two photon processes.
These background events can be reduced by a veto of the scattered
electrons in the forward detectors (luminosity monitors).
Typically, LEP detectors cover the forward region down to
25-50~mrad.
The maximum $\pt$ observed for two photon processes is
determined by the forward coverage.
Therefore the total transverse momentum of the events
can be required to be larger than
$\approx \theta_{\mathrm min} \sqrt{s}$, where $\theta_{\mathrm min}$
is the minimum polar angle of the coverage.
The balance between the two photon background rejection and the
signal efficiency determines the $E_{\mathrm vis}$ or $\pt$ cuts.

For large $\Delta m$ ($\gsim 20$ GeV), the main background sources are
$W^+ W^- \ra \ell \nu q \bar{q}'$ and  
$Z Z \ra \nu \bar{\nu} q \bar{q}$ events.
These events can be reduced by requiring
the visible invariant mass of the events to be larger than
$\approx 50$~GeV.
A cut against isolated leptons can be used to reject
events with $W \ra \ell \nu$ decays.
The lower value of the visible energy cut can be tightened to
eliminate the background from the two photon processes.

Since $\stst$ events are expected to have two narrow jets, the number of
reconstructed jets is required to be two.
The jet resolution parameter are optimized to have
a good rejection of background and a good signal efficiency.
This requirement reduces four jet events from $W^+ W^-$ or
$Z Z$ background as well as QCD multi-jet events.
L3 used the JADE jet algorithm.
OPAL used the Lund jet algorithm with the jet resolution parameter
$d_{\mathrm join} = (2.5 + 2 E_{\mathrm vis}/\sqrt{s})$~GeV.
The $E_{\mathrm vis}$ dependence of the $d_{\mathrm join}$
parameter is needed for a good jet reconstruction over a wide
range of $\mstopone$.
Since the jet resolution parameter used for OPAL is tight,
multi-jet events are effectively reduced and the other
cuts can be loosened.
Likewise, the L3 two-jet requirement is very efficient in removing multi-jet
backgrounds.
Another possibility to eliminate $q \bar{q} (\gamma)$ events
or four jet events from $W^+ W^-$ or $Z Z$ events is setting a cut
on the maximum value of $E_{\mathrm vis}$.
To reduce the two jet $q \bar{q}$ background, the acoplanarity angle
($\phi_{\mathrm acop}$) cut is very powerful.
The acoplanarity angle is defined to be 180$^\circ$ minus the
opening angle between two jet momenta projected on to the
plane pependicular to the beam axis.
The events which have back-to-back topology in the plane
perpendicular to the beam direction are reduced in this way.
Therefore, $Z \gamma \ra q \bar{q} \gamma$ events where the
$\gamma$ escapes into the beam pipe are also reduced.
L3 has applied the cut $\phi_{\mathrm acop} > 34^\circ$, OPAL
has taken $\phi_{\mathrm acop} > 15^\circ$ for
$E_{\mathrm vis} > 0.2 \sqrt{s}$ and linearly increased the
$\phi_{\mathrm acop}$ cut value to $40^\circ$ as $E_{\mathrm vis}$
decreased to $0.1 \sqrt{s}$.

If the two jets are emitted close to the beam axis, the acoplanarity angle
measurement becomes poor.                           
Therefore a cut $| \cos \theta_T |\lsim$ 0.7 is
usually applied to the polar angle $\theta_T$    
of event thrust axis.
Events with a jet in the forward direction may also
cause a large acoplanarity angle.
The azimuthal angle of the forward jet cannot be measured with
good precision because the particles belonging to the jet
spread over a large azimuthal angle range.
Events are rejected if one of the two jets points close to the
beam axis.
To further reduce $q \bar{q} \gamma$ events where the $\gamma$ escapes
into the beam pipe, L3 required the longitudinal momentum balance
$| \Sigma p_z |/E_{\mathrm vis} < 0.5$, where $\Sigma$ runs over
all the particles.

The detection efficiency of $\stst$ events is about 25\% for
$\mstopone = 70$~GeV and $\Delta m = 20$~GeV, and large background rejection is
maintained. For $\cal L$=500 pb$^{-1}$ only 2--3 background events are
expected.
The efficiency increases gradually to 40\% at $\mstopone = 85$~GeV
with the same $\Delta m$.
For $\Delta m = 10$~GeV, the efficiency is about 20\% at
$\mstopone = 70$~GeV, as the visible energy decreases.
It goes down to a few \%  for $\Delta m = 5$~GeV.

DELPHI has chosen a very different approach. A statistical method is used to
extract the signal from background. After a loose pre-selection of events, one
applies a cut discriminating analysis {\it \`{a} la} Fisher with up to 31
kinematical and event shape variables: $V_i$ $(i=1,2,...,31$). The
discrimination function $F$ is calculated as a linear combination of the
variables: $F = a_0 + \sum^{31}_{i=1} a_i V_i$, where the coefficients $a_i$
are optimized to have the best separation between signal and background. 
The discovery potential is evaluated from the number of expected signal to
background events after having applied a cut on $F$, which is optimized in
order to maximize the discovery potential. The analysis is repeated for
different masses of $\stopone$ and $\chioa$ as well as for different $\sqrt{s}$
(175~GeV and 190~GeV) focusing on small $\Delta m$. 

The five sigma discovery and 95\%~CL exclusion regions in the
$\mstopone$-$\mchioa$ plane, which can be reached by a typical LEP analysis at
$\sqrt{s} = 190$~GeV, are shown in fig.~\ref{stopfig1}. The shaded area
corresponds to the region currently excluded at the 95\%~CL by searches at the
Tevatron collider~\cite{D0stop}.
The minimum luminosity needed for five sigma discovery
and for 95\% C.L. exclusion in the 
cases of $\Delta m= 10$ and 20~GeV are given in
table~\ref{tab:stlimit}.                                    

No effort has been made so far to study $\stopone$ identification. For this
purpose it is important to identify the charm quark in the $\stopone \ra c \chioa$
decay. After applying appropriate selection cuts, the enhancement of the soft
$\pi^\pm$ from  $D^*$ decay, the reconstructed $D$ or $D^*$ mass peak and the
leptons from charm semileptonic decays can be studied to identify the charm
quark. A direct mass reconstruction is not 
straightforward due to the large missing mass arising form the two neutralinos.                                                               
In the case of $\stopone \ra b \chipa$ the visible energy is larger than in
$\stopone \ra c \chioa$, and b-tagging can be used to enhance the
signal. If one of the $\chipma$'s decays into $\chioa~\ell^+ \nu$, an isolated
lepton is expected in the event. These leptons are softer than those
from $W^+ W^-$ or $Z Z$ events. If the \chipma\ is lighter than \stopone, it
is most likely that it will be discovered first. The information of the 
\chipma\ characteristics can be used to identify $\stopone$ by the decay into
$b \chipa$.  
                             
DELPHI has studied the decay $\stopone \ra b \chipa$ using the same method as
described above. Additionally, the vertex information has been used to enhance
events with b-quark. The experimental reach for $\mstopone$ is 85~GeV at
$\sqrt{s} =190$~GeV, with a luminosity of 300~pb$^{-1}$. Moreover, DELPHI has
studied $\sb_1$ search when $\sb_1$ decays into $b \chioa$. The discovery
potential for $\sb_1$ is similar to that of the $\stopone \ra c \chioa$.
\subsection{Neutralinos}
\begin{figure}
\centerline{ 
\epsfig{figure=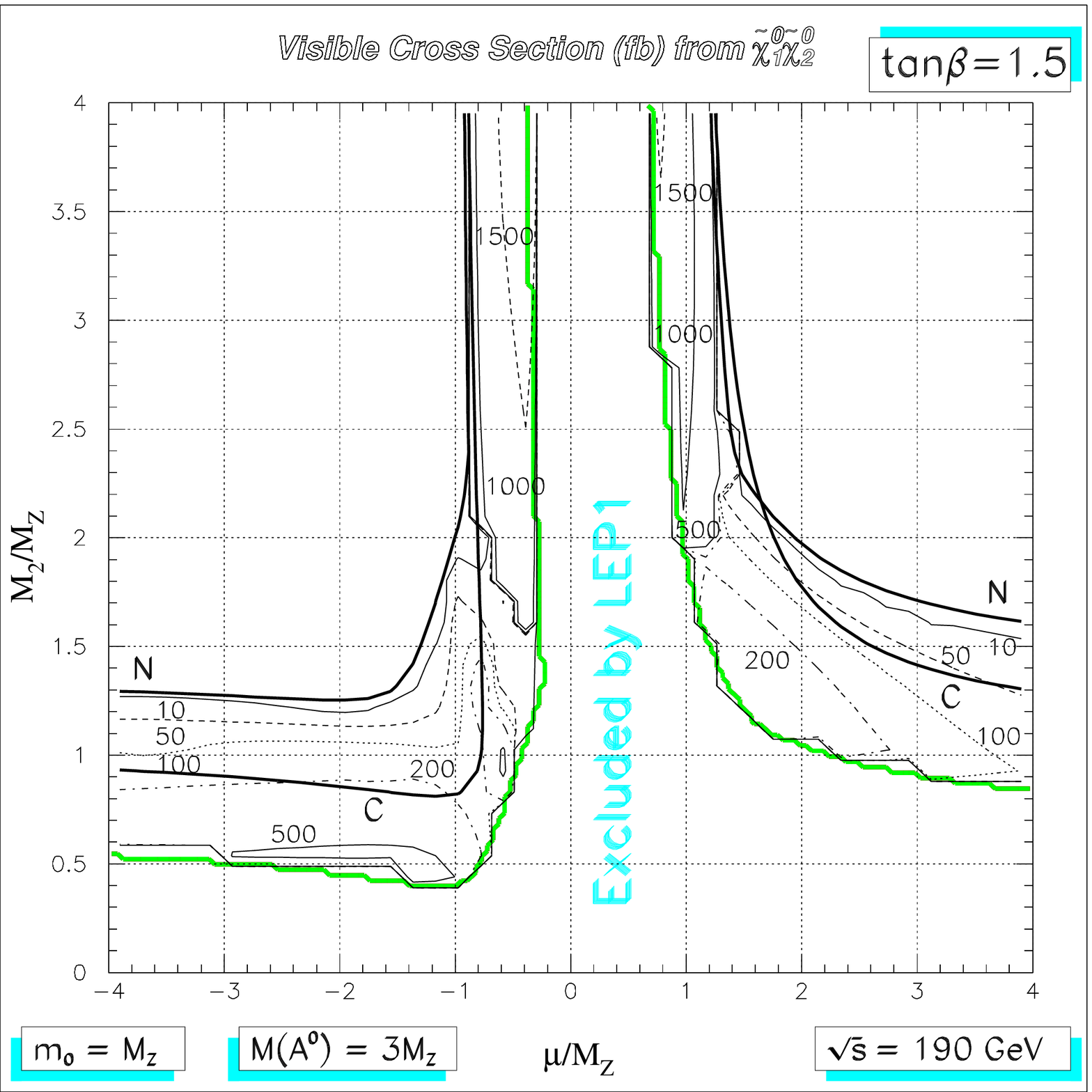,height=8.0cm,
        width=10.0cm,angle=0}     
}
\ccaption{}{\label{neut-1}
Contour lines for the cross section (fb) of $\nno \rar$ visible, in the $(\mu,
M_2)$ plane, for $\tgb = 1.5$, $m_0 = M_Z$, $M_{A^0} = 3M_Z$. The central
empty region is excluded by LEP1 data. The bold lines represent the
kinematical limits for $\nno$ (`N') and $\cc$ (`C') at LEP2 ($\protect\sqrt{s} =
190$ GeV).} 
\end{figure}

\begin{figure}
\centerline{ 
\epsfig{figure=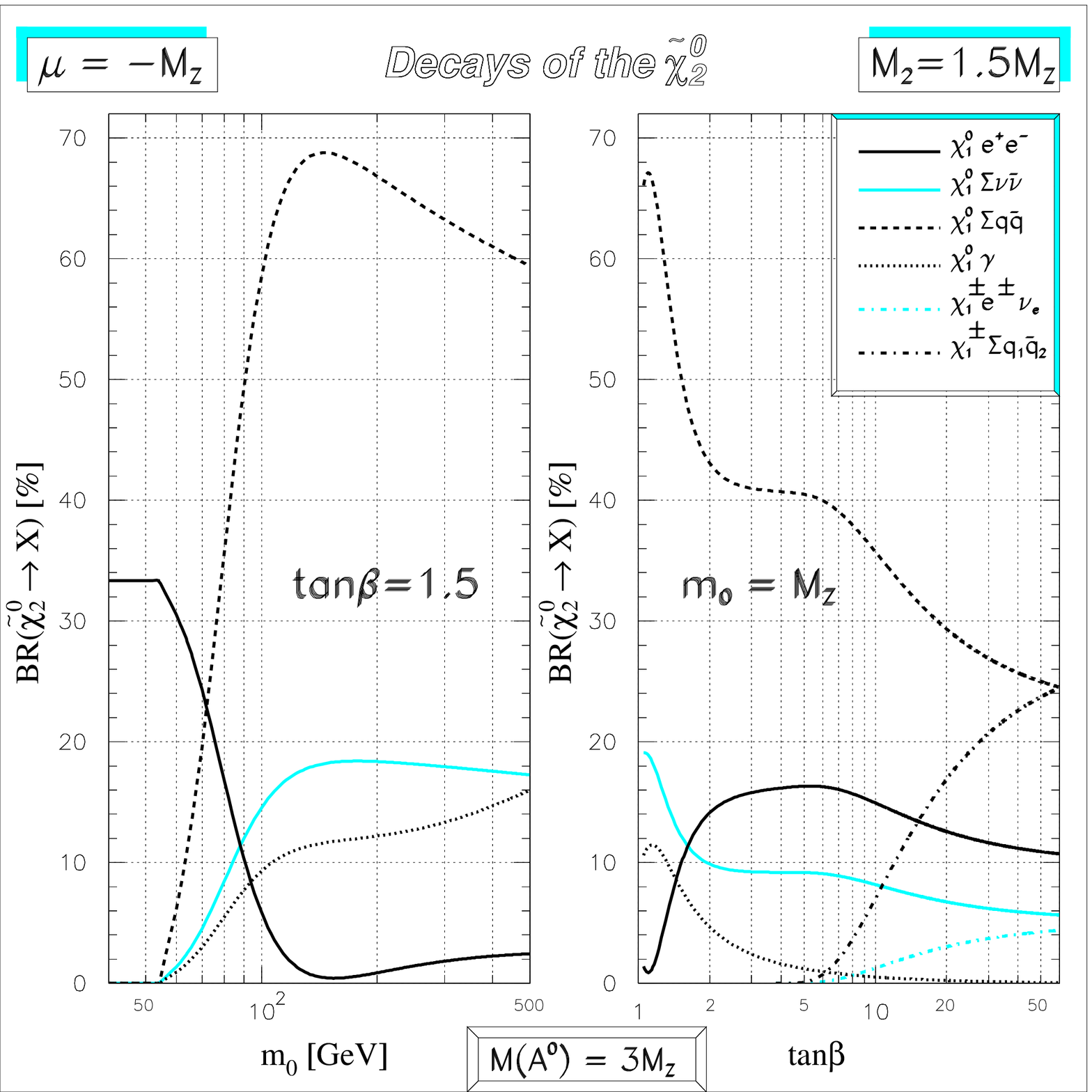,height=8.0cm,
        width=10.0cm,angle=0}     
}
\ccaption{}{\label{neut-2}
        B.R.'s of all $\n{2}$ decay channels for $\mu = \mz$,  
        $M_2 = 1.5 \mz$, $m_{A^0} = 3\mz$ as functions of $m_0$ 
        (with $\tgb = 1.5$) and of $\tgb$ (with $m_0 = \mz$).} 
\end{figure}
\label{susyneutralinos}
The most promising neutralino production process at LEP2 is the production of a
lightest plus a next-to-lightest neutralino, $\nno$. In fact, most of the times
the $\n{2}$ decays in some visible final state (plus a $\n{1}$) while the
lightest neutralino goes undetected, producing a large imbalance in the final
state momentum and missing mass. 

A thorough study of the production rates at LEP2 for the process $\nno$, which
takes into account all the possible signatures that can derive from the
different $\n{2}$ decays, has been performed as a function of the SUSY
parameters $\mu$, $M_2$, $\tgb$, and the common scalar mass $m_0$
\cite{ambr-mele1}. The $\n{2}$ can decay into a $\n{1}$ plus: $\ell^+\ell^-$,
$\nu_\ell\bar{\nu}_\ell$, $q\bar{q}$, $\ell^+\ell^{\prime
-}\nu_\ell\bar{\nu}_{\ell^\prime}$, $\ell^\pm \nu_\ell q\bar{q}^{\prime}$,
$q_1\bar{q}^{\prime}_1 q_2\bar{q}^{\prime}_2$ (the last three arising from
cascade decays through a light chargino) or a photon. Possible decays into a
light Higgs boson have also been included. A typical scenario is shown in
fig.~\ref{neut-1}, where the  rates for all visible $\n{2}$ final
states are included. At LEP2, one finds that:
                                                 
a) In regions where $\n{1}$ and $\n{2}$  are mostly higgsino-like 
({\it i.e.} for $|\mu| \ltap M_Z$ and $M_2 \gtap 1.5 M_Z$)
production rates are large (more than 1 pb) and comparable to
the chargino rates.
                   
b) In regions where $\n{1}$ and $\n{2}$  have non-negligible
gaugino components (\ie for $|\mu| \gtap M_Z$), the kinematical reach 
(marked by `N' in fig.~\ref{neut-1}) of the reaction $\nno$ in the 
($\mu,M_2$) plane is larger than the one relative to the light 
chargino-pair production (marked by `C' in fig.~\ref{neut-1}).

c) For $|\mu| \gtap M_Z$, the neutralino production rates are much 
lower than the chargino rates (mainly due to the absence of the 
$\gamma$-exchange channel for neutral particles) and 
critically dependent on selectron masses.
For $\tilde e$ close to the LEP2 reach ({\it i.e.} $m_0 \simeq M_Z$ GeV),
cross sections 
up to about 0.5 pb can be reached outside the regions excluded by LEP1, and 
up to 0.2 pb in the area not covered by chargino searches at LEP2.

Hence, neutralino production provides a remarkable  additional 
tool to explore the SUSY parameter space in the regions 
covered by light-chargino pair production and beyond.
Comparative chargino/neutralino studies in more constrained
SUSY models, which assume radiative breaking of the EW 
symmetry, are
reported in sect.~\ref{susybaer}.
                      

A detailed analysis of the $\n{2}$ partial decay widths and branching 
ratios is necessary to establish the actual fraction of detectable events
from $\nno$. The results of this analysis \cite{ambr-mele2}, are critically 
dependent on the physical  $\n{2}$ compositions and on the assumed 
s-fermion spectrum of the theory. An example is given in fig.~\ref{neut-2},
where the $m_0$ and $\tgb$ dependence of the $\n{2}$ branching ratios
for all possible decays is shown for a typical scenario of interest for 
LEP2 searches, {\it i.e.}, $\mu = -\mz$ and $M_2 = 1.5 \mz$.

\begin{figure}
\begin{center}    
\mbox{\epsfig{file=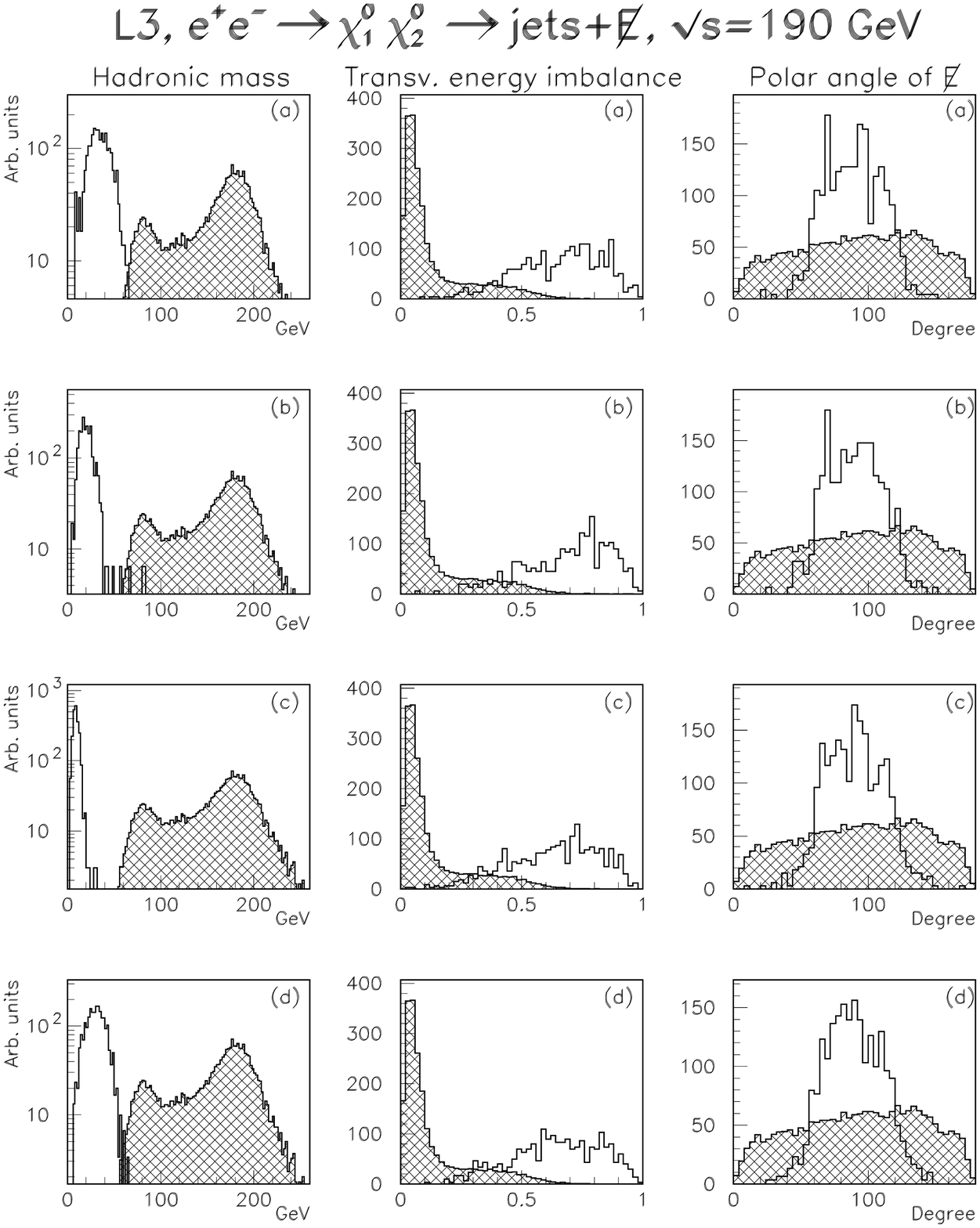,height=16cm}}
\ccaption{}{       \label{neuton}         
Distributions of the most discriminating variables for the
signal from $e^+e^-\rightarrow\chioa\chiob\rightarrow\ jets +
missing\  momentum$ for cases (a), (b), (c) and (d)
corresponding respectively to                     
$(\mchioa,\mchiob)$= (49.7, 107), (51.5, 85.2), (73.7, 89.8) and
(56, 108.2) GeV.
The distributions for signal (empty histogram) and
the sum of the backgrounds from standard physics processes
(hatched histogram) are normalized to the same integral.}
\end {center}
\end{figure}
As already discussed in this report, a critical parameter for the detection of
the neutralino and chargino production is the mass splitting between the
decaying particle and the lightest neutralino. In the parameter region where
$M_2$ is much larger than both $\mu$ and \mw, the lightest chargino and two
lightest neutralinos are mainly higgsino-like and are nearly mass degenerate.
However the mass difference $\mchiob -\mchioa$ is about twice as large as
$\mchipm -\mchioa$ \cite{last}. This is approximately true even after radiative
corrections are included \cite{pomar}. Since, in the case of higgsino-like
neutralinos, the cross section for the process $e^+e^-\to \chioa\chiob$ is
large (of the order of a pb), neutralino searches are very useful in studying a
parameter region where chargino identification can be problematic \cite{last}.
Notice also that the higgsino-like region ($M_2\gg\mu ,\mw$) is favoured by the 
present $R_b$ measurements at LEP1, as discussed in sect.~\ref{susyrb}. 

Another neutralino channel that could be of interest for LEP2 is the radiative
single-photon process $\nng$, where a hard, large-angle photon accompanies the
invisible $\n{1}$ pair. After applying typical experimental cuts on the photon,
optimized for background suppression, ({\it i.e.}, $p^T_\gamma/E_{beam} > 0.065$,
$|\cos\theta_{\gamma}| < 0.95$, $E_\gamma/E_{beam} < 0.5$), one finds rates up
to 40 fb in the gaugino regions, for moderate $\tilde e$ masses~\cite{pavia}. 
Unfortunately, the corresponding rate of the
main background process, $\epem\rar\gamma\nu_\ell\bar{\nu}_\ell$, is about 900
fb and makes the radiative neutralino production very hard to isolate. 

\subsubsection{Search Strategy for Neutralinos}
We have investigated the possibility of detecting a signal
from the $\chioa\chiob$ associate production
with the L3 experiment at $\sqrt{s}=190$ GeV~\cite{L3neutr}.
We have studied detection efficiencies and background rejection for the
four points A, B, C, D
of the MSSM, discussed in ref.~\cite{ambr-mele1},
corresponding
to the parameter region not covered by chargino searches, for which
($\mchioa,\mchiob$)= (49.7, 107), (51.5, 85.2), (73.7, 89.8) and
(56, 108.2) GeV.

The signal of $\chioa\chiob$ associate production, with
$\chiob\rightarrow\chioa+q\bar{q}\ or\ \ell^+\ell^-$ decay,
is a pair of acoplanar jets or
leptons and missing momentum from the two undetected $\chioa$'s.
Since the branching ratio of the $\chiob$ in the hadronic or leptonic
mode depends on the values of the parameters, we have studied separately
the detection efficiencies and background rejections for the two modes.
                        
We consider as background the known physics processes
which can produce events with
acoplanar jets or leptons + missing momentum.
They are (in parenthesis the total cross section)
$e^+e^-\rightarrow\ q\bar{q}\gamma$ (94~pb), $WW$ (18~pb),
$We\nu$ (0.92~pb), $ZZ$ (1.1~pb), $Zee$ (3.2~pb) and
$ee\rightarrow ee\gamma\gamma\rightarrow\ f\bar{f}ee$ (3000~pb).
The $ee\rightarrow ee\gamma\gamma\rightarrow\ f\bar{f}ee$ process,
hereafter called the $\gamma\gamma$ process,
despite the very large cross section, gives a
negligible contribution to the total background, after the cuts which we
describe in the following.

The signal events were generated with SUSYGEN \cite{SUSYGEN},
while the backgrounds were generated with PYTHIA.
The events were then passed through the standard simulation and
reconstruction of the L3 detector.

In the hadronic events, we require at least 4 tracks and 10 calorimetric
clusters. Hadronic jets are reconstructed from the calorimetric clusters
with the JADE algorithm, using $y_{cut}$=0.02.
In order to reduce the background from $WW$ ($ZZ$) with an isolated
high momentum lepton in the final state,        
we require that there are no isolated
electrons or muons with momentum higher than 20 GeV.
In addition we require that there is no energy deposited in the very forward
detectors.  This cut reduces mainly $\gamma\gamma$ and $Zee$ events,
where one or both of the electrons are emitted in the very forward
direction, and they can either hit the forward detectors or go undetected
in the beam pipe. To reduce events with particles escaping in the beam pipe
(from $qq(\gamma)$, $Zee$, $\gamma\gamma$),                                 
we require that the longitudinal imbalance
be less than 40\% of the total visible energy in the event.
Exploiting the fact that the acoplanarity
(defined as the complementary of the angular difference
in the transverse plane among the two highest energy jets) is
typically small for the background and large for the signal,
we require the acoplanarity to be larger than 40$^o$.

After the above cuts the most discriminating quantities are: ({\it i}) the
hadronic mass, which is usually large for most of the backgrounds and small for
the signal; ({\it ii}) the transverse imbalance, usually smaller in the
background than in the signal; ({\it iii}) the polar angle of the missing
momentum.
The distributions of the most discriminating variables for
the neutralino signals analyzed
and the sum of the backgrounds are shown in fig.~\ref{neuton}.
                                                      
The requirement that the hadronic mass be between 5 and 50 GeV,
the transverse imbalance above 40\% of the total visible
energy, and the polar angle of
the missing momentum point between $20^{o}$ and $160^{o}$,
reduces the detected background cross
section to 16 fb while the signal detection efficiencies
is around 40\% , almost independently of $\mchiob$ and
$\mchioa$. It should be noted that, in the four cases analyzed,
the $\mchiob-\mchioa$ mass difference
is always above 15 GeV, so detection efficiency is not problematic
as it is in the low mass difference region (see sect.~\ref{susycharginos}).

Similar cuts to those applied in the hadronic channel are
used in the leptonic channel, where two isolated electrons or muons
are required. Cuts on the total energy, the longitudinal and transverse
imbalance, the angle of the missing momentum, the acoplanarity and the
mass of the lepton pairs allow to reduce the detected background cross sections
to 25 fb and 20 fb for acoplanar electron and muon events, respectively.       
The detection efficiency is 40\% and 30\% for electrons and muons,
respectively, almost independently of the neutralino masses and for mass
differences above 15 GeV.                                       

We conclude that 300 (100) pb$^{-1}$ allow to detect a
$\chiob\chioa$ signal in      
regions of the parameter space were the signal production cross
section is above $\sim$ 100 (200) fb, for
a $\mchiob-\mchioa> 15$ GeV.
                        
\subsubsection{Neutralinos in the NMSSM}
\label{susyfranke}                
The Next-to-Minimal SUSY Standard Model (NMSSM) is a
simple extension of the MSSM, obtained by adding a singlet superfield 
and by allowing only cubic superpotential couplings~\cite{nmssm}.           

It contains five neutralinos with mass
eigenstates and mixings determined by 
the singlet vacuum expectation value $x$ and the
couplings $\lambda$ and $k$ in the superpotential,
in addition to the MSSM parameters $M_1$, $M_2$, and
$\tan\beta$. Neutralino and Higgs sectors 
are strongly correlated in the NMSSM.
In scenarios with light singlet-like neutralinos there often
exist also light Higgs bosons with masses below the MSSM bounds. 
Contrary to the MSSM, the experimental data do not
exclude very light or even massless neutralinos and Higgs bosons
\cite{franke1}.

\begin{figure}
\begin{center}
\epsfig{file=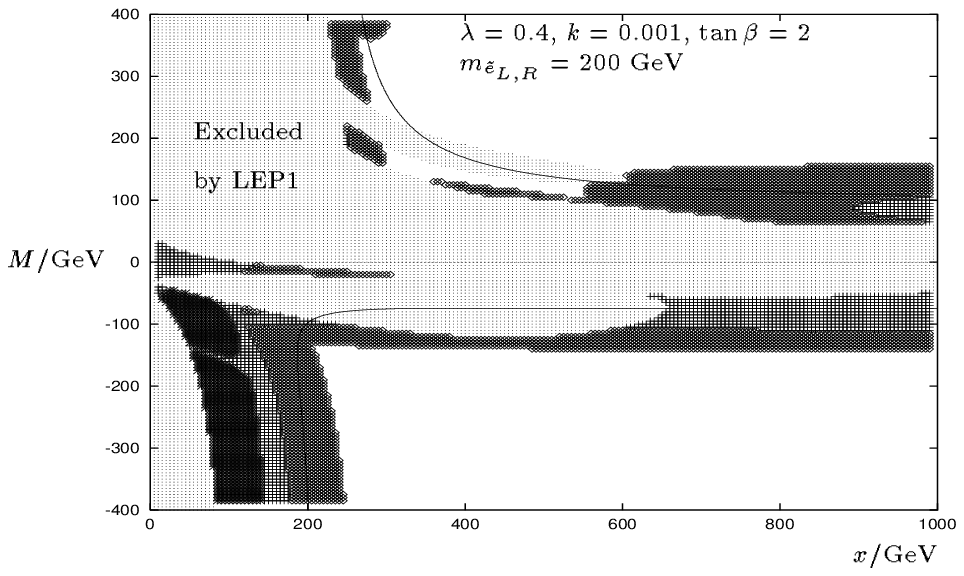,width=0.8\textwidth}
\end{center}    
\ccaption{}{\label{fignmssm}
Accessible parameter space at LEP2 ($\sqrt{s}=190$ GeV).
The different shadings denote regions where the \chiob\-production
cross section is larger than 500, 200, and 100 fb (from dark to light).
The contour line for $\mchipm = 95$ GeV is also shown.}
\end{figure}

Fundamental differences between the NMSSM and the MSSM scenarios 
may occur when the  neutralinos accessible at LEP2
have significant singlet components. Since the singlet superfield with zero
hypercharge does not couple to (s)fermions and gauge bosons, neutralino pair
production 
just differs because of the neutralino mixing.

For a typical NMSSM scenario,
fig.~\ref{fignmssm} shows the areas of the parameter space not 
excluded by LEP1, 
where at least one production channel with a visible                           
neutralino reaches a cross section above 100, 200 or 500 fb. 
In most of this region the lightest neutralino is singlet-like.
A NMSSM scenario                        
could be tested even for parameter regions beyond the kinematical limit for
chargino production, \eg in the large $x$ domain. Here the lightest
neutralino (assumed to be the LSP) is very light ($\approx 10$ GeV) while the
visible next-to-lightest neutralino is produced at LEP2 and may be
distinguished from the MSSM by its decay modes, as discussed below. On the
other hand, there exists a region in fig.~\ref{fignmssm},
where a chargino and a neutralino in a corresponding MSSM scenario with $\mu =
\lambda x$ can be found, but for which in the NMSSM 
all light neutralinos have singlet components
large enough to suppress their production at LEP2. In general, however, 
LEP2 considerably extends the parameter space which can be probed.
In particular, small $x$
values ($x \lsim 80$ GeV) can be excluded independently of the concrete NMSSM
scenario if no neutralino is found. On the other hand, 
one cannot expect a stronger lower bound on
the neutralino mass.

\begin{figure}
\centerline{
\epsfig{file=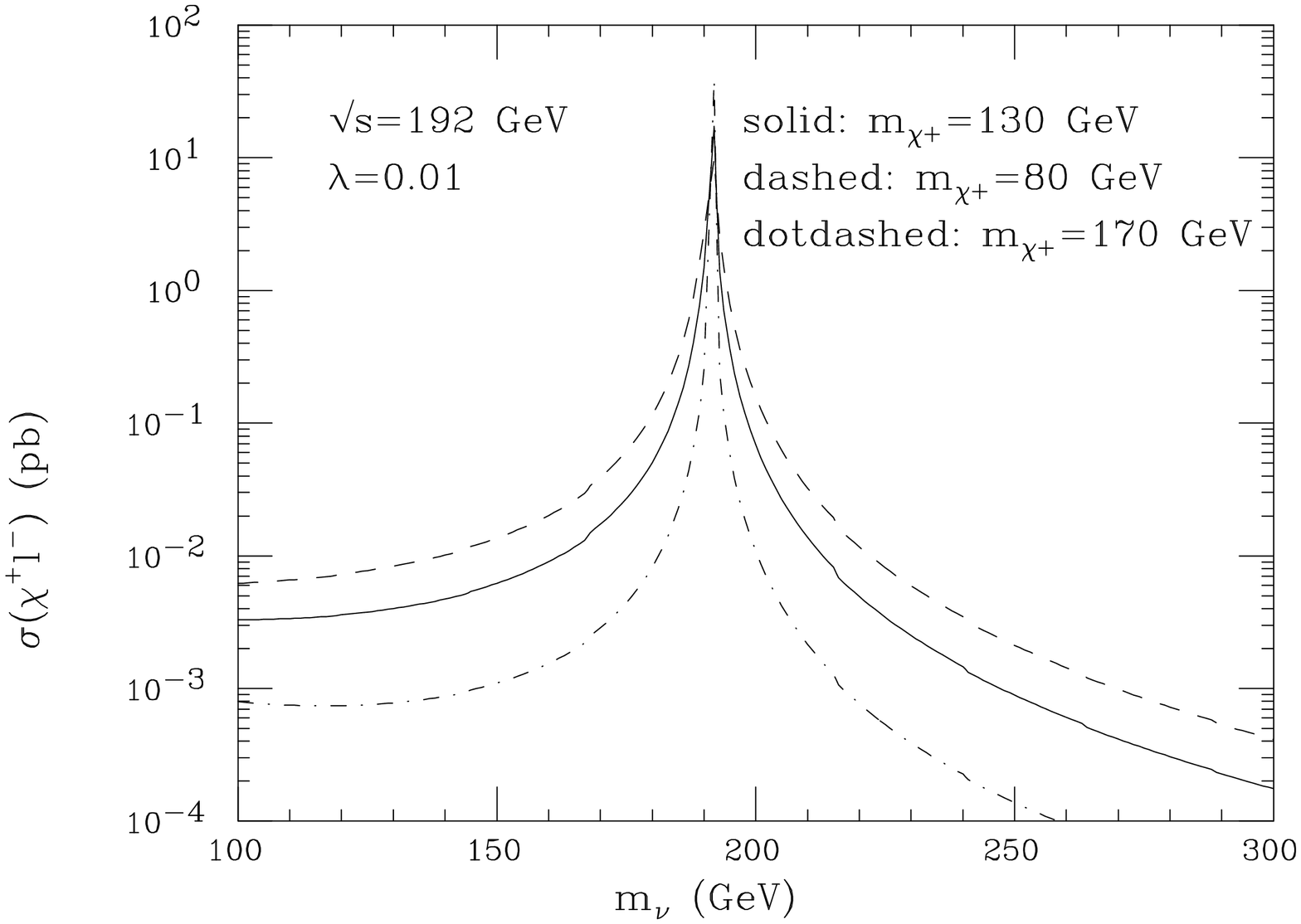,width=.40\textwidth,angle=90}
\epsfig{file=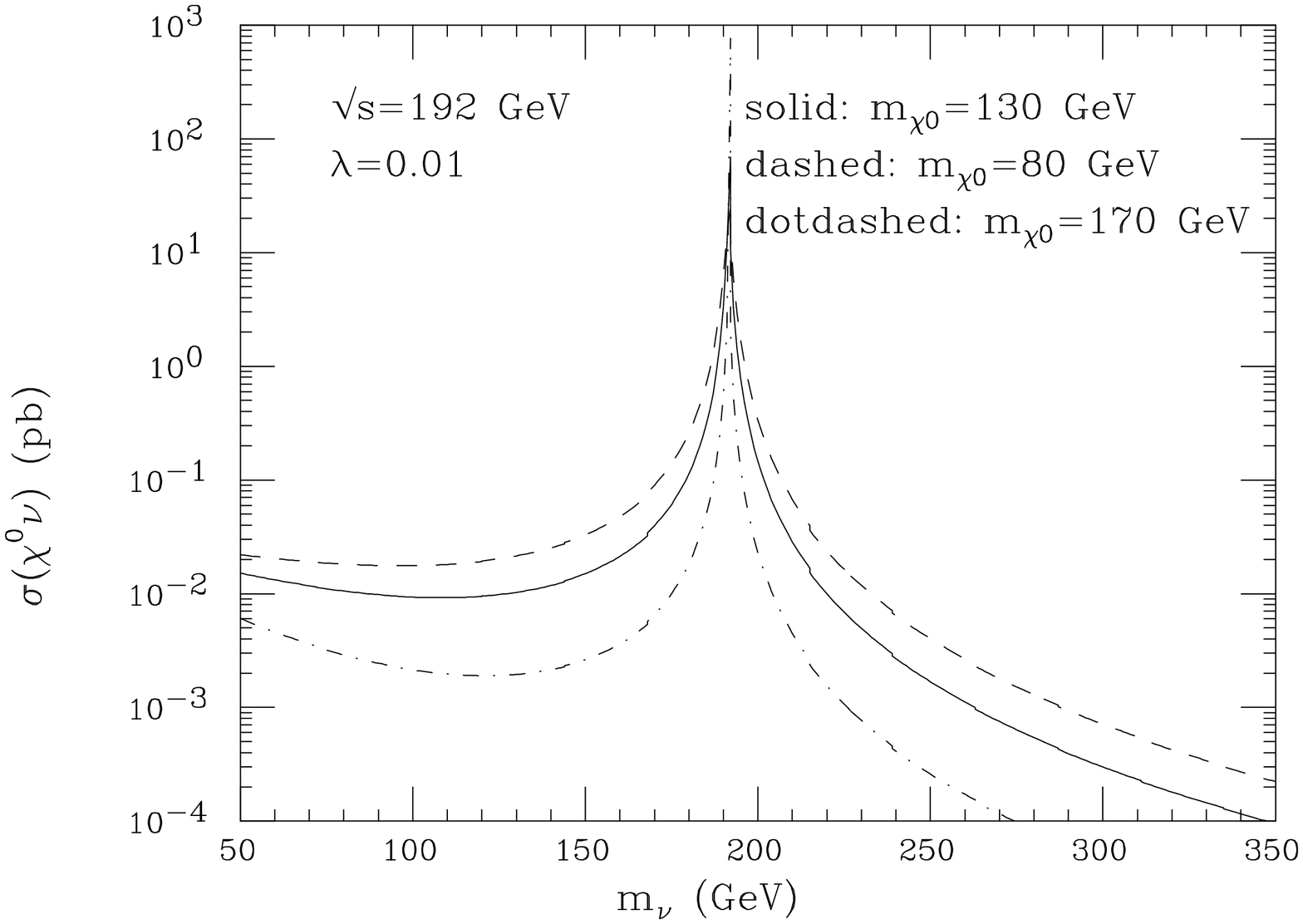,width=.40\textwidth,angle=90}}
\ccaption{}{\label{f:rp} 
Production cross section for $(\nu\chioa)$ (a) and 
$(\ell^\mp\chipm)$ (b) fianl states produced via R-parity violating $s$-channel
\snu\ exchange.}                                 
\end{figure}

In order to estimate the prospects for identifying a NMSSM neutralino and
distinguishing between NMSSM and MSSM, an analysis of the dominant decay
channels of the produced neutralinos is essential. Compared to the MSSM, the
decay of the next-to-lightest neutralino into a Higgs boson plus the LSP and
the loop decay into a photon and the LSP is enhanced in typical NMSSM scenarios
\cite{franke3}. This happens because the singlet neutralino does not couple
to the gauge sector, but only to the Higgs sector for $\lambda \ne 0$.
Generally, the decay into a Higgs boson becomes dominant if
kinematically allowed. Then, depending on the parameters of the Higgs sector,
the Higgs may decay dominantly into $b$-quark pairs, but also into two
invisible LSPs. If the next-to-lightest neutralino cannot decay into a Higgs
boson, three-body decays with a fermion pair and the loop decay with a
photon in the final state become comparable. In typical NMSSM scenarios as in
fig.~\ref{fignmssm} the latter dominates for large $x$-values and vice versa. 
A detailed simulation study of these possibilities has not yet been performed.

In conclusion, it may be possible to distinguish a NMSSM neutralino from the
minimal model at LEP2 if the LSP is mainly a singlet and the next-to-lightest
neutralino is pair produced at the available center-of-mass energy. 
However unfavourable neutralino mixings may also prevent the discovery of
a NMSSM neutralino at LEP2.
\subsection{R-Parity Violation}
\label{rparitysection}
The SUSY extension of the SM can contain the following renormalizable
terms in the superpotential \cite{susyrev}\ beyond those present in the 
MSSM:                          
\be
\lambda_{ijk} L_iL_j{\bar E_k}+\lambda_{ijk}'L_iQ_j{\bar D_k}
+\lambda_{ijk}''{\bar U_i}{\bar D_j}{\bar D_k}.
\label{eq:rpops}                               
\ee
Here $Q_i$, $\bar U_i$, $\bar D_i$, $L_i$ and $\bar E_i$
($i=1,2,3$) denote the three generation quark and lepton superfields. 
These terms are allowed by gauge symmetry but forbidden by R-parity
conservation, with 
$R=(-1)^{2S+3B+L}$ ($S$, $B$ and $L$ representing spin, baryon and
lepton number).
The interactions in eq.~(\ref{eq:rpops}) lead in general to unacceptable rates
for proton decay. However, if only a subset of these terms were allowed, {\it
e.g.} by imposing some discrete symmetries, the proton can be stable but the
superymmetric experimental signatures would be dramatically different. 
Examples of such discrete symmetries are 
baryon parity, lepton parity \cite{ibanross} or family
symmetries \cite{ali1}. In this section we study the phenomenology of 
including $R$-parity violating interactions. In this analysis it is 
assumed that
(i) for each process considered, no more than one of the 45 coupling 
constants in eq.~(\ref{eq:rpops}) 
is non-vanishing and (ii) the LSP
is the lightest neutralino.  
The phenomenology of SUSY searches then changes in two important 
aspects. \newline
(1) The LSP decays in the detector if any of the couplings in 
eq.~(\ref{eq:rpops}) satisfies \cite{LSPdecay}
\be                                            
\lambda> 8\times 10^{-6} \sqrt{\gamma_L} \left(m_{\tilde f}/{150\,{\rm GeV}}
\right)^2\left({45\,{\rm GeV}}/{M_{LSP}}\right)^{5/2},
\label{eq:LSP}                                                                   
\ee
where $\gamma_L$ is the Lorentz boost and $m_{\tilde f}$ is the relevant
sfermion mass.
\newline
(2) At LEP200 single resonant sneutrino production can occur via the 
operators $L_1L_{2,3}{\bar E}_1$.                            
\be
e^+e^-\rightarrow {\tilde \nu}^*_{2,3}\rightarrow (\nu \chi^0_1),
(e^\pm \chi^\mp_1)
\label{eq:snuprod}
\ee
The resulting two sets of signals are discussed below.

\noindent{\bf MSSM-Production followed by LSP Decay.}
The MSSM production mechanisms are unaltered by the operators of
eq.~(\ref{eq:rpops}). However, the final state now includes the decay of
the LSP, provided eq.~(\ref{eq:LSP}) is satisfied. Thus the pair production of
the two lightest neutralinos leads to a visible signal. For LEP2 and LEP140
this has been discussed in detail in ref.~\cite{tatavernon}. Furthermore, the
full LSP decay as described in ref.~\cite{morawitz}\ is included in SUSYGEN
\cite{SUSYGEN}. It should be clear that if the LEP2 energy is above {\it any}
SUSY particle pair threshold, and no LSP decay is detected, then {\it all} the
R-parity violating couplings in eq.~(\ref{eq:rpops}) can be excluded down to
the value given by eq.~(\ref{eq:LSP}). 

\noindent {\bf Resonant Sneutrino Production.}
The operator $L_1L_{2,3}{\bar E}_1$ offers the unique possibility of resonant
sneutrino (\snu) production. The BR(${\tilde{\nu}}_{2,3}\to e^+e^-$) 
is strongly constrained by
present bounds \cite{bargerhan}\ on the coupling constants of
eq.~(\ref{eq:rpops}). Thus the sneutrino typically decays via a gaugino to an
R-parity odd final state: ${\tilde{\nu}}_{2,3}\rightarrow \nu_{2,3}\chioa$ or
$\ell_{2,3}^-\chipa$. The decay widths are given in ref.~\cite{bargerhan}.
Explicitly the amplitude squared for the chargino production (\ref{eq:snuprod})
is given by: 
\begin{eqnarray}
|{\cal M}|^2(e^+e^-\rightarrow \chi^-_1\ell_{2,3}^+)&=& 
g^2\lambda^2|V_{11}|^2
\left(\frac{s(M^2_{\chi^-}-s)}{|R(s)|^2}+\frac{t(M^2_{\chi^-}-t)}
{|D(t)|^2}-{\cal R}\frac{I(s,t,u)}{R(s)D(t)}\right)
\end{eqnarray}
where $R(s)=s-m^2_{\tilde\nu}+i\Gamma_{\tilde\nu}m_{\tilde\nu}$,
$D(t)=t-m^2_{\tilde\nu}$, and $I(s,t,u)=s(M_{\chi^-}-s)-u(M_{\chi^-}-u)+
t(M_{\chi^-}-t)$. Here, $|V_{11}|^2<1$ is the gaugino fraction in the final
state chargino, and $t=(p(e^-)-p(\chi))^2$. The cross sections are plotted in
fig.~\ref{f:rp} as functions of \msnu\ for $\sqrt{s}=190$ GeV
and $\lambda=0.01$. 


The \chioa\ will decay to $e^+e^-\nu$ or $e^\pm e^\mp_{2,3}\nu$.
The signal consists of two charged leptons typically in the same
hemisphere and a large amount of missing $\pt$. 
The chargino can either decay directly via an R-parity violating coupling
or it can decay to \chioa. The amplitudes squared for the R-parity
violating ($L_1L_2{\bar E}_1$) decay $\chi^-(k)\rightarrow e^-(p_1)\mu^-(p_2)
e^+(p_3)$ is given by ($s=(k+p_1)^2,\,t=(k-p_2)^2$)
\be
|{\cal M}|^2_{\not R_p}=g^2\lambda^2|V_{11}|^2\left(\frac{s(s-M^2_{\chi})}
{|R(s)|^2}+\frac{t(t-M^2_{\chi})}{|D(t)|^2}
+{\cal R}e\frac{I(s,t,u)}{R(s)D(t)}\right) ~.
\ee
The branching fraction depends strongly on the size of $\lambda$ and on the
kinematical suppression due to the \chioa\ mass. In both cases the signal
contains four charged leptons which should be clearly visible. Thus the
chargino production is preferable to the LSP production. Since the cross
sections are comparable, it is the main signal for R-parity violation. 
\subsection{Multi-mode Search for Minimal Supergravity at LEP2}
\label{susybaer}                                       
We summarize in this section prospects for detecting SUSY at LEP2 assuming the
paradigm case of the minimal supergravity (SUGRA) model. We assume gauge
coupling unification and radiative EW symmetry breaking, along with
universal soft SUSY-breaking terms at a GUT scale $M_X$ taken to be equal to
the scale at which the $U(1)$ and $SU(2)$ gauge couplings unify. The model
parameters are thus $m_0$, the universal scalar mass, $m_{1/2}$, the universal
gaugino mass, $A_0$, the universal trilinear term, $\tan\beta$, the ratio of
Higgs field VEVs, and the sign of the SUSY conserving Higgsino mixing term
$\mu$. Along with \mt, these parameters suffice to determine the weak-scale
sparticle masses and mixings, which in turn allow all sparticle and Higgs boson
production cross sections and decay modes to be calculated. 
This allows to delineate the regions of parameter space accessible to
searches at LEP2 and to distinguish the different SUSY signals from
one another. It also allows to compare the potential of LEP2 with that of the
Tevatron Main Injector $p\bar{p}$ collider, which is expected to begin
collecting data around 1999.
           
\begin{figure}
\centerline{  
\epsfig{figure=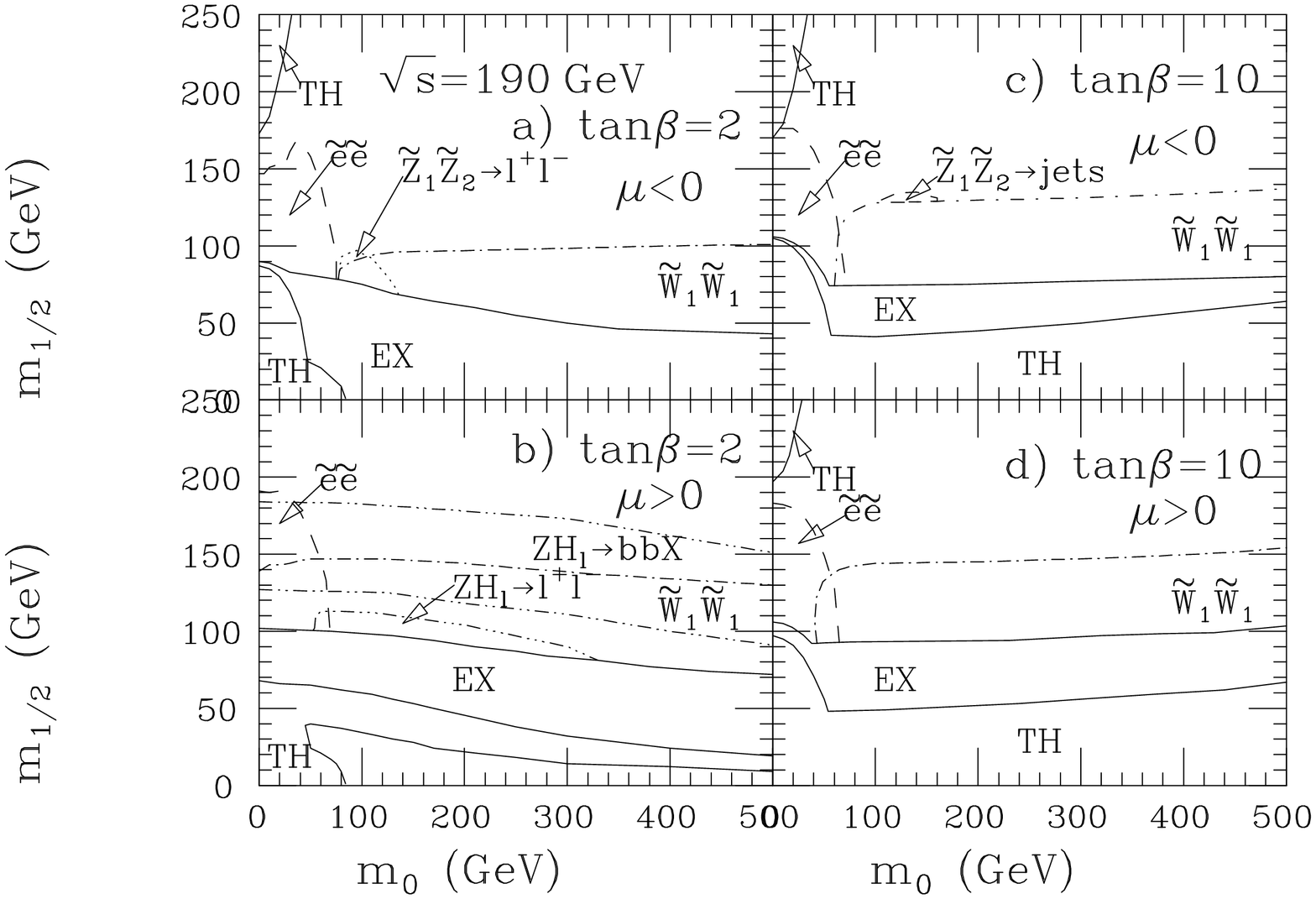,width=0.5\textwidth,angle=90}
}                                             
\ccaption{}{\label{fig3}
\it Regions of the minimal supergravity parameter space explorable at LEP2
with $\sqrt{s}=190$ GeV.}
\end{figure}
\begin{figure}
\centerline{
\epsfig{figure=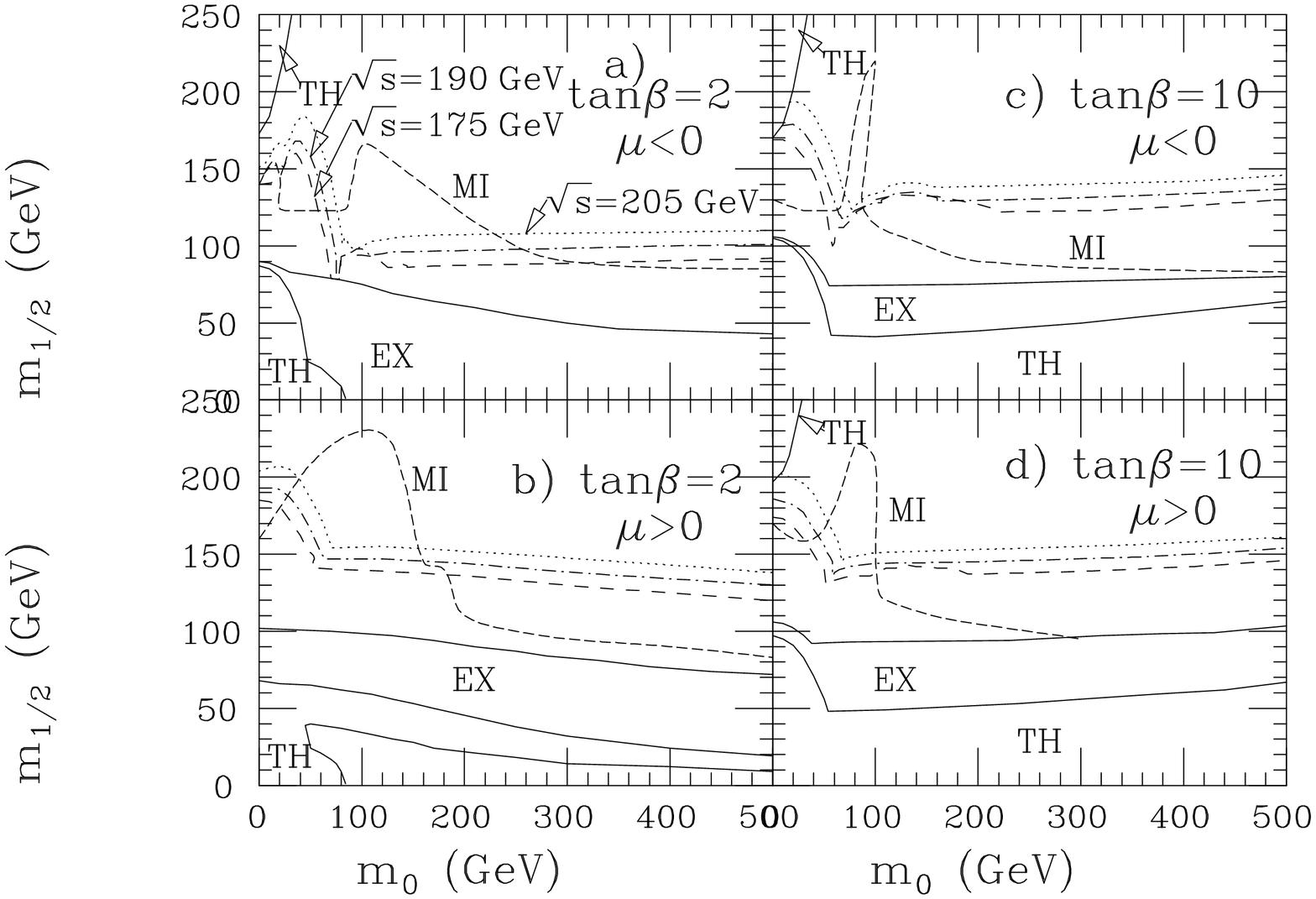,width=0.5\textwidth,angle=90}
}                                 
\ccaption{}{\label{fig7}
\it Cumulative reach of various LEP2 options (and Tevatron MI) for
SUSY particles (excluding Higgs bosons).}
\end{figure}

The cross sections
and decay modes have been embedded into the event generator
ISAJET \cite{isajet}, \cite{SUSYGEN}. 
All allowed SUSY and Higgs
boson production mechanisms for the sampled points in SUGRA parameter space
have then been generated,
and compared against SM backgrounds such as
$\tau\bar{\tau}$, $WW$ and $ZZ$ production. The latter two were calculated with
complete spin correlations using the HELAS \cite{helas} package. 
Further details and results may be found in ref. \cite{leptwo}, along with
references to related studies.               

In fig.~\ref{fig3}, we show regions of the 
$m_0$ vs. $m_{1/2}$ plane explorable at LEP2 given $\sqrt{s}=190$ GeV, 
and $\int {\cal L}dt=500$ pb$^{-1}$. In all frames, we
take $A_0=0$. In {\it a}), we take $\tan\beta =2$, $\mu <0$, while
in {\it b}) we take $\tan\beta =2$ with $\mu >0$.
In {\it c}), we take $\tan\beta =10$, $\mu <0$ and in {\it d}) we take
$\tan\beta =10$, $\mu >0$.
The regions denoted
by TH are excluded by the theoretical constraints built into the model, such as
the requirement of radiative EW symmetry breaking.
The region labelled EX
is excluded by experimental constraints. We have implemented various cuts
designed to select signal from SM background, and to separate SUSY processes
from one another. In the 
lower left regions, $\te\bar{\te}\rightarrow e^+ e^- +\esl$ 
signals ought to be probed, whereas in the 
lower $m_{1/2}$ regions, chargino signals should be detectable via mixed
leptonic/hadronic signatures. In addition, some small regions are noted
where $\chioa\chiob$ can be seen via dilepton or jet signatures.
If in fact a $\chioa\chiob$ signal is seen, then it may be best to run
LEP2 around $\sqrt{s}=150$ GeV, to eliminate backgrounds from $WW$ and
possibly $Zh$ production; then the tiny $\chioa\chiob$ signal may be 
more easily examined in a relatively background free environment.

Higgs boson masses are correlated with the rest of the sparticle mass
spectrum, and $Zh$ production can be seen in significant 
regions of the $\tan\beta =2$ frames.
$h$ is somewhat heavier in the $\tan\beta =10$ frames, and so energies in
excess of $\sqrt{s}=190$ GeV will be needed in this case to see it.

In fig.~\ref{fig7}, we show the cumulative reach of various LEP2 
upgrade options for SUSY particles (excluding Higgs bosons), 
for
$\sqrt{s}=175$ GeV and $\int {\cal L}dt=500$ pb$^{-1}$ (dashed),
$\sqrt{s}=190$ GeV and $\int {\cal L}dt=300$ pb$^{-1}$ (dot-dashed), and
$\sqrt{s}=205$ GeV and $\int {\cal L}dt=300$ pb$^{-1}$ (dotted). Also shown
for comparison is the combined reach of Tevatron Main Injector era
experiments ($\sqrt{s}=2$ TeV and  $\int {\cal L}dt=1000$ pb$^{-1}$)
(dashed curve labelled by MI). We see regions of parameter space
accessible to LEP2 that the MI cannot explore, and regions of parameter
space open to MI that LEP2 cannot explore (except possibly via Higgs searches).
These plots emphasize the complementarity between the two colliders.

In addition, in ref. \cite{leptwo} one can find a study of the regions of
parameter space explorable via $3\ell +$jets and clean $4\ell$ signatures
($\ell =e$ or $\mu$). Such events arise from $\te_L\te_R$, $\chiob\chiob$,
$\tnu_e\bar{\tnu_e}$ production with various cascade decays. These reactions
generally occur within subsets of the regions of parameter space already
delineated in fig.~\ref{fig3}, and so give no additional reach for
SUSY, in the constrained SUGRA model under consideration. The detection
of such events is nonetheless important since it could serve to test the
details of the underlying model. Also ref. \cite{leptwo} contains the
corresponding regions of parameter space accessible at LEP2 via search for the
lightest SUSY Higgs boson $h$. These regions can extend deep into parameter
space; however, over a wide range of parameters, the SUSY $h$ would be very
difficult to distinguish from a SM Higgs boson. 
\section{New Fermions}
\label{newfermions}
\subsection{Introduction}
\label{intr:newferm}
In this chapter we concentrate on the LEP2 potential for discovering 
new fermions (elementary or composite)
other than those predicted by the MSSM. New quarks and 
leptons may arise in a class of theories with quark-lepton 
unification~\cite{exotic,guts} or in those 
that endeavour to answer questions about fermion masses and family 
replication~\cite{4th family,exotic}. It is plausible that the 
symmetry structure of the theory may protect their masses to be 
comparable to the Fermi scale, and indeed certain theoretical 
arguments~\cite{follia} 
do indicate this to be true in a wide class of models. 
 Excited fermions, 
on the other hand, are natural corollaries of models where the 
SM fermions themselves 
are composite particles rather than being elementary 
ones~\cite{hagiwara}. The mass gap is determined by the compositeness scale, 
and, for a not too large value of the latter,
may be bridged by the LEP2 energy.
In either case, then, looking for new fermions  provides
a probe of possible new physics beyond the SM.

\subsection{New Elementary Fermions} \label{elem ferm}
These may  be subdivided into two categories :
\begin{itemize}
\item{sequential},
 \ie with gauge quantum numbers identical 
  to the SM fermions~\cite{4th family}, or,  
\item{exotic}~\cite{exotic,dpfef} \ie all those 
  fermions that have no analog in the SM. Popular examples are provided by
  \begin{itemize}
    \item{mirror fermions} which have chiral properties 
      exactly opposite to those of the SM fermions;
    \item{vector fermions}, whose left- and right-handed 
        components have identical gauge quantum numbers; and
    \item{singlet neutrinos}, which have no coupling with the SM gauge 
        sector except through mixing, and consequently are the hardest to 
        detect.
  \end{itemize}
\end{itemize}
\begin{figure}
\centerline{\psfig{figure=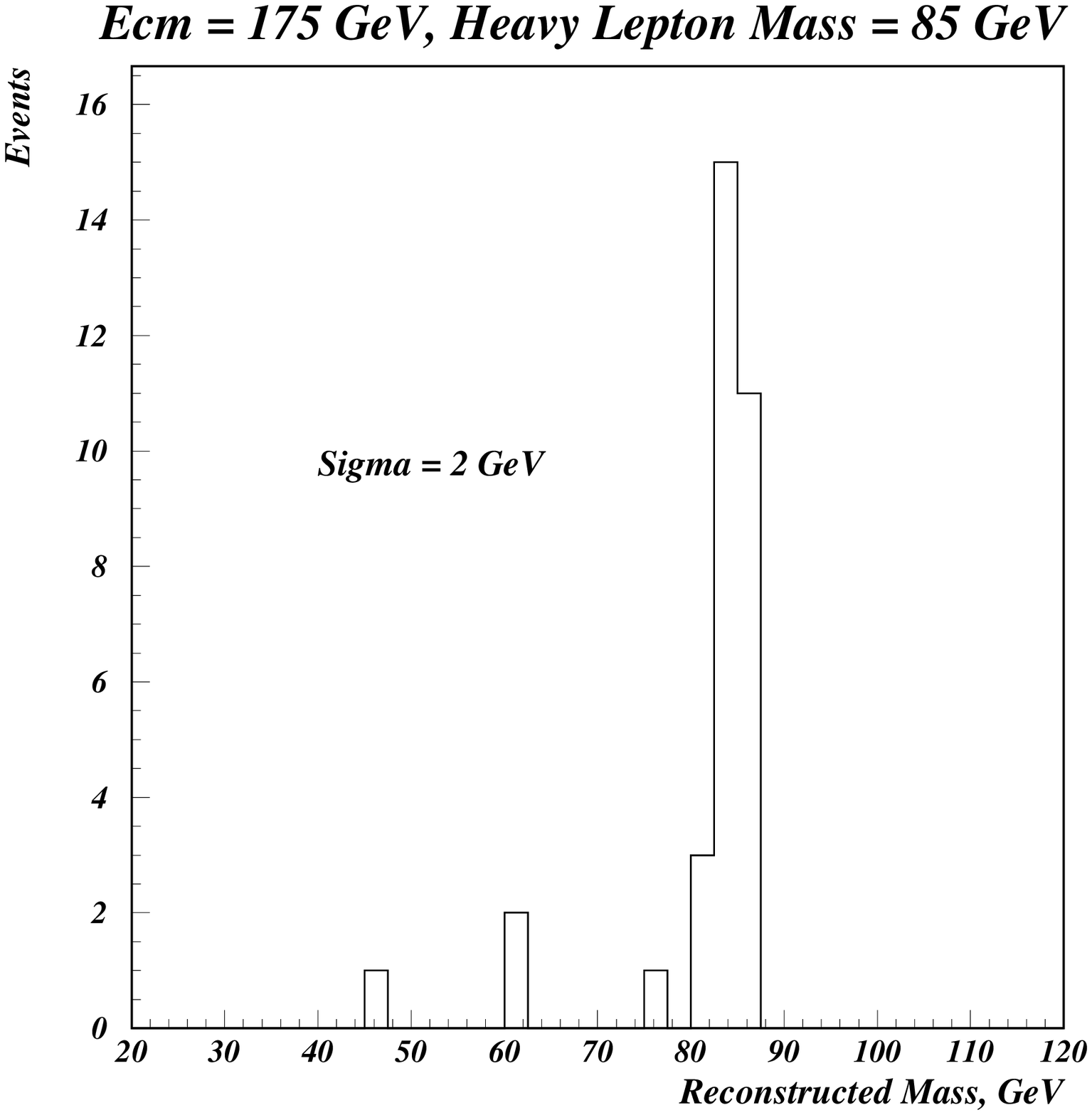,height=7.5cm,width=0.4\textwidth}
            \psfig{figure=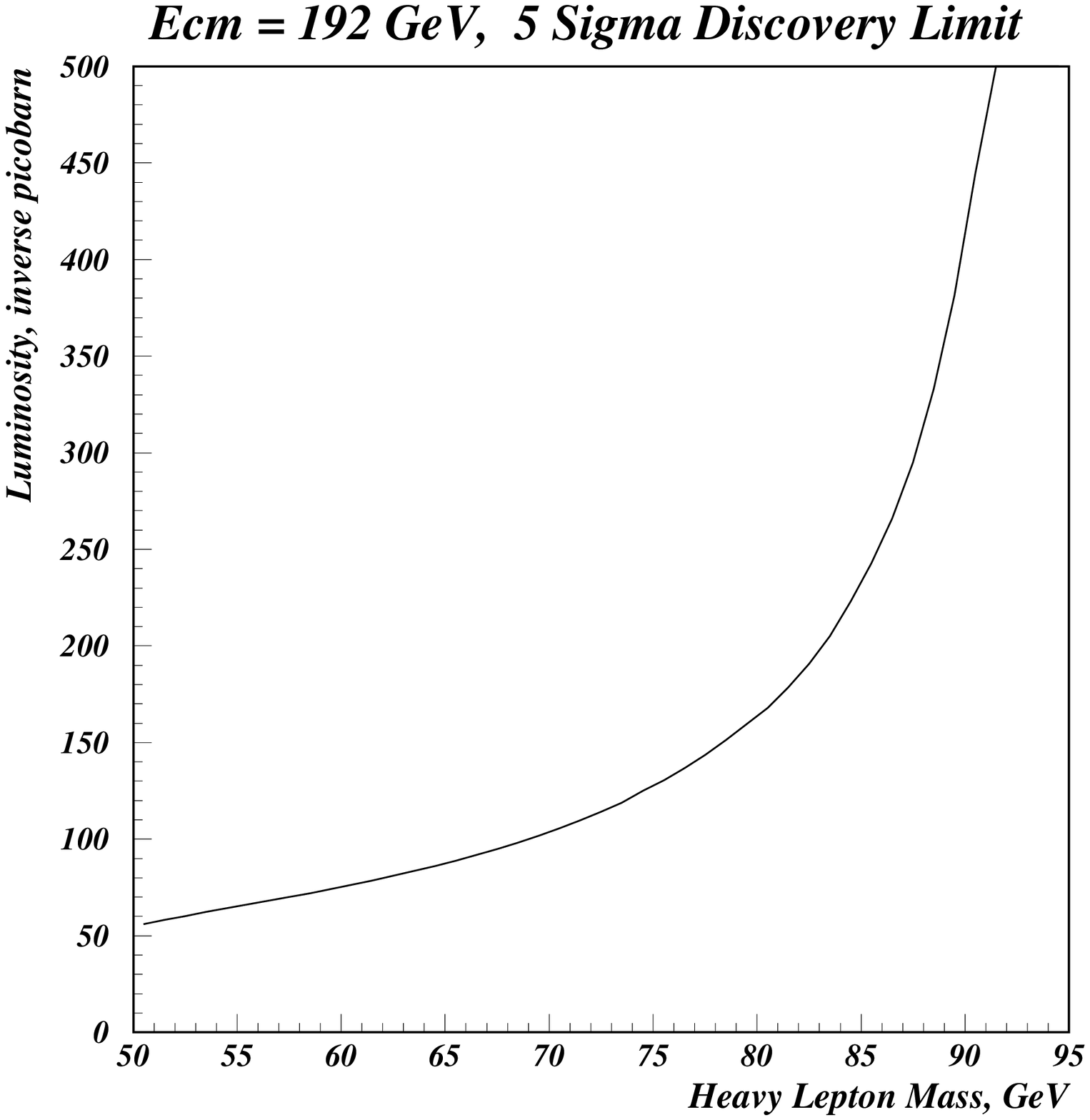,height=7.5cm,width=0.4\textwidth}}
\ccaption{}{\label{L3shev}
Left panel: mass reconstruction for $N\overline{N}$ for the process of
eq.~\protect\ref{NN:sigb}. Right panel: the
luminosity needed for heavy neutrino  $5\sigma$~discovery, as
a function of $N$-mass, for $\sqrt{s}=192~\gev$. The limits for 
$\sqrt{s}=175~\gev$~are similar,
but the mass reach is less.}
\end{figure}
Since the SM gauge interactions of these new fermions are determined 
(apart from mixing effects) by their quantum numbers, 
some of the strongest 
bounds can be inferred from the absence of
unexpected decay channels of the $\Z$ at  LEP.
Thus, for all such fermions other than singlet neutrinos, we have
$    m_f \gsim$ 45 GeV.
At the Tevatron, though, stronger bounds ($\sim 85$ GeV)
can be imposed on new quarks, 
provided they decay within the detectors \cite{pdg,dp_bm}. 
These bounds are likely to improve with the new data. 
As lepton production at the Tevatron proceeds mainly through Drell-Yan-like 
processes or through gauge boson fusion, the corresponding bounds are 
expected to be somewhat weaker. However, such an analysis has not yet been 
reported in the literature.  LEP data, 
along with other low-energy measurements can also be used to place strong 
bounds on the possible mixings that the new fermions may have with the 
SM particles~\cite{nardi}. 
Certain indirect bounds can also be placed from precision 
EW measurements at LEP as well as from some other low 
energy observables~\cite{LEP-precision}. 
For example,  the LEP bounds on the oblique parameter 
$S$ (equivalently $\epsilon_3$) restricts the number of additional chiral 
generations to be at most one. 

\subsubsection{Pair Production} \label{elem:pp}
Since the introduction of sequential fermions would 
still have all flavour changing neutral currents (FCNC) to be vanishing 
at the tree level\footnote{While this is not strictly true for the exotics, 
constraints \protect\cite{nardi} on FCNC imply that pair production 
dominates over single production for much of the range $m_f < \sqrt{s}$.},
at LEP2 such particles may only be 
pair-produced\footnote{The only exception is the process 
$e^+ e^- \rightarrow N \nu$ for a heavy neutrino $N$ which may proceed 
through a $t$-channel $W$-exchange diagram. We shall discuss this case 
later.} through $s$-channel diagram(s) mediated by $\gamma/Z$. Keeping 
in mind the expected improvement in the Tevatron 
bounds, it is thus not very interesting to look for heavy
quarks at LEP2. We may thus safely concentrate on lepton pairs only.
Total cross sections are in the range 1--4 pb until quite close
to the kinematic limit. 

The only tree-level decay mode for a  fourth-generation lepton proceeds via
the charged current interaction.
We then make the assumption that the dominant 
decay is to one of the SM leptons and not to the heavy isospin partner. If the 
latter were the case, the partner itself should be sought. It should be
noted that even if the mixing angle ($\zeta$) with the 
light leptons is small, the heavy lepton would still decay within the 
detector provided  $\zeta \gsim 10^{-6}$. 

\begin{figure}
\vskip 4.2in\relax\noindent\hskip -0.7in\relax{\includegraphics{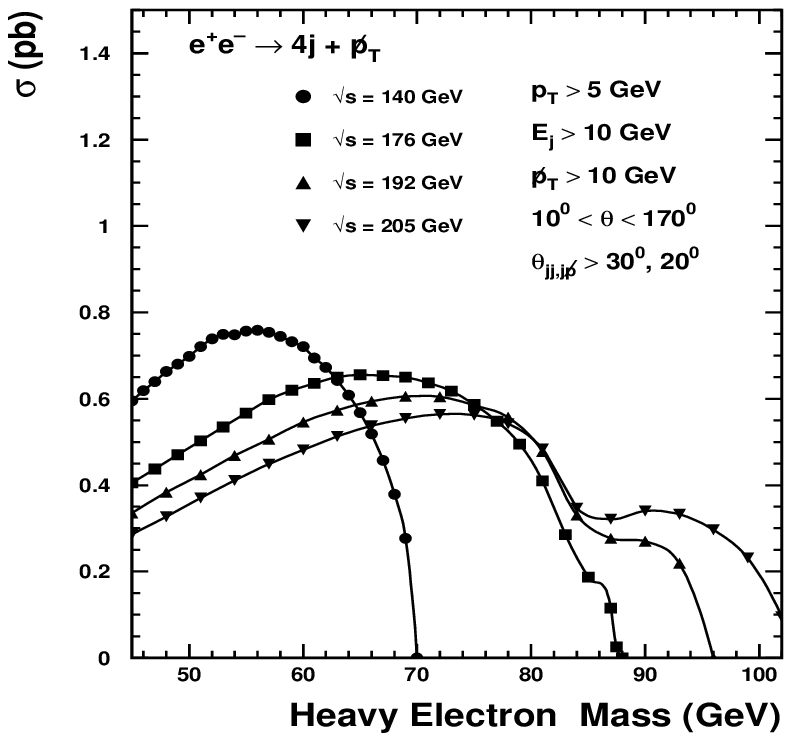}}
          \relax\noindent\hskip 3.5in\relax{\includegraphics{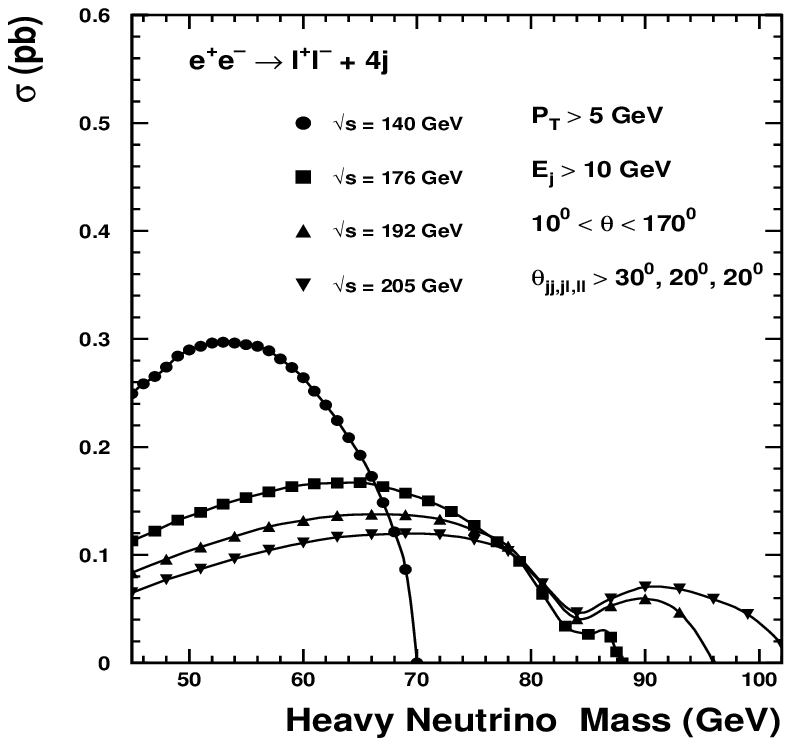}}
\vspace{-18ex}                                              
\ccaption{}{\label{fig:pair_effective}
The effective cross section (at various c.m. energies)
 for the processes of
eqs.\protect\ref{EE:sig} and \protect\ref{NN:siga},
after imposing the cuts of
eqs.\protect\ref{EE-cuts} and \protect\ref{NN-cuts} respectively.}
\end{figure} 
     
While a heavy neutrino can be either a Dirac or Majorana particle, 
in this note we shall concentrate on the former alternative. For,
pair-produced Majorana neutrinos can be easily distinguished by the tell-tale 
signature of like-sign dileptons with hadronic activity 
but without any missing momentum. The  backgrounds 
derive from two sources: ($i$) cascade decays of a heavy 
quark pair (say $b \bar{b} \to 
c e^- \bar\nu \bar{c} q_i \bar{q_j} \to e^- e^- \nu \nu + \; {\rm jets}$)
or ($ii$) effects like $B$-$\overline{B}$ mixing. They
could be eliminated by a combination of isolation cuts and imposition 
of an upper bound on missing momentum.
For a pair of heavy Dirac neutrinos the possible signals are
\ba           
e^+ e^- &\rightarrow& N \overline{N} 
        \rightarrow l_i^+ l_i^- W^{(\ast )} W^{(\ast )} 
        \rightarrow l_i^+ l_i^- + 4 \:{\mbox{\rm jets}}
        \label{NN:siga}
\\
e^+ e^- &\rightarrow& N \overline{N} 
        \rightarrow l_i^+ l_i^- W^{(\ast )} W^{(\ast )} 
        \rightarrow l_i^+ l_i^- l_a + 2 \:{\mbox{\rm jets}} + {\not p}_T
        \label{NN:sigb}
\\
e^+ e^- &\rightarrow& N \overline{N} 
        \rightarrow l_i^+ l_i^- W^{(\ast )} W^{(\ast )} 
        \rightarrow l_i^+ l_i^- l_a^- l_b^+ + {\not p}
        \label{NN:sigc}
\ea
where ${\not p}_T$ denotes missing 
transverse momentum and the generation index $i$ is determined by the dominant
mixing. In this study we concentrate on $l_i = e/\mu$ as the efficiency for 
$\tau$--detection  is low. 
The purely leptonic channel contains two neutrinos, so mass
reconstruction is not possible. Both the other channels have been studied,
and a typical  mass reconstruction~\cite{shev}  is shown in
fig.~\ref{L3shev}. Derived from the $3 l + 2 j$ channel,
the events are selected by requiring 
enough hadronic activity to form two jets plus three isolated leptons
($<5~\gev$~ hadronic energy 
within a cone of $30^{\circ}$ around the track) above $3~\gev$. 
The lepton (one of the two like-sign ones) associated to the 
hadronically decaying $\W$ is identified by 
requiring it to have $E_l \sim E_{beam}-E_{had}$. 
The hadronic energy is then improved by
fitting to this relation, and the neutrino identified with the improved
${\not p}$. Background from $\z\z$ is further reduced by rejecting events in
which the same-flavour 
lepton-pair invariant mass is within $5~\gev$ of the $\z$-mass.
Backgrounds from all SM processes are estimated as $4.8\pm2.0$ events for
500 pb$^{-1}$ and signal efficiencies are around 25\% 
(inclusive of branching ratio, cut efficiency and geometrical acceptance)
over a wide range of
$N$-masses. The expected $5\sigma$ discovery limits at $192~\gev$ are shown in
fig.~\ref{L3shev}. 
While the  cross sections are larger for $E \bar{E}$
production, detection is more difficult. Since the distinctive signature is 
\be
e^+ e^- \rightarrow E \overline{E} 
        \rightarrow \nu_i  \bar{\nu}_i W^{(\ast )} W^{(\ast )} 
        \rightarrow 4 \:{\mbox{\rm jets}} + {\not p}_T,
        \label{EE:sig}
\ee
direct mass reconstruction is impossible and a
detailed fitting procedure would be needed to determine the mass. In
fig.~\ref{fig:pair_effective} we show the effective 
cross section~\cite{gb_dc}, after the cuts
\be
 \pt(j)> 5 \gev, \quad  E(j) > 10 \gev, \quad          
{\not p}_T > 10 \gev, \quad 10^\circ < \theta_j < 170^\circ, \quad
\theta_{jj} > 30^\circ, \quad \theta_{j \not p} > 20^\circ. 
  \label{EE-cuts}
\ee
With these cuts the SM background (estimated with MadGraph and 
verified with GRACE) falls to well below 1~fb.
With an integrated 
luminosity of $300 {\mbox{\rm pb}}^{-1}$, it should be possible to 
probe up to $m_E \approx \sqrt{s}/2$.
With similar cuts,
\be
\pt(j,l) > 5 \gev, \quad     E(j) > 10 \gev,  \quad
10^\circ < \theta_j, \theta_l < 170^\circ, \quad
\theta_{jj} > 30^\circ, \quad    \theta_{ll},  \theta_{lj} > 20^\circ,
  \label{NN-cuts}
\ee
the $N\overline{N}$ channel yields the effective cross section
also shown in fig.~\ref{fig:pair_effective}.
\begin{figure}
\centerline{\epsfig{figure=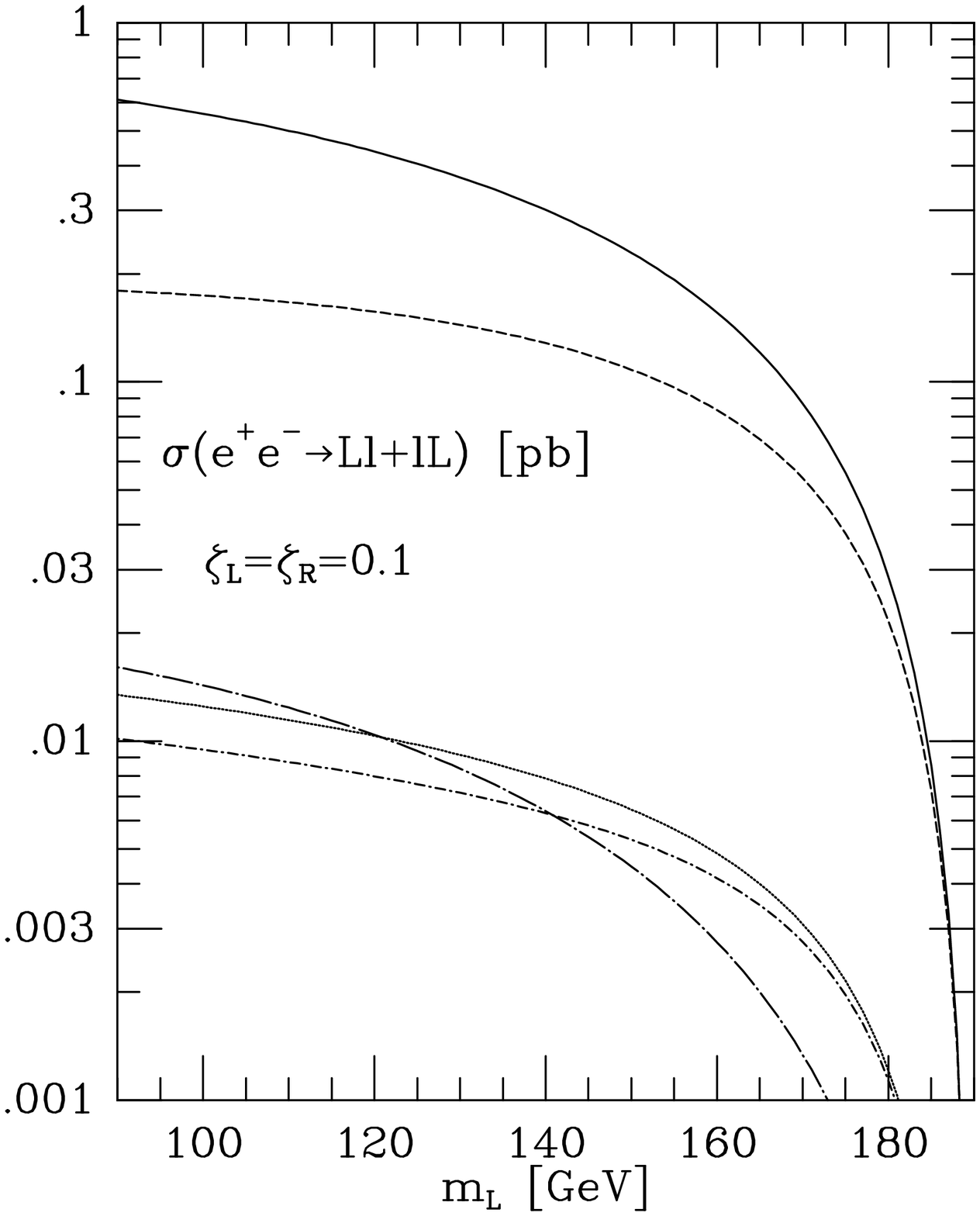,width=0.4\textwidth}
            \epsfig{figure=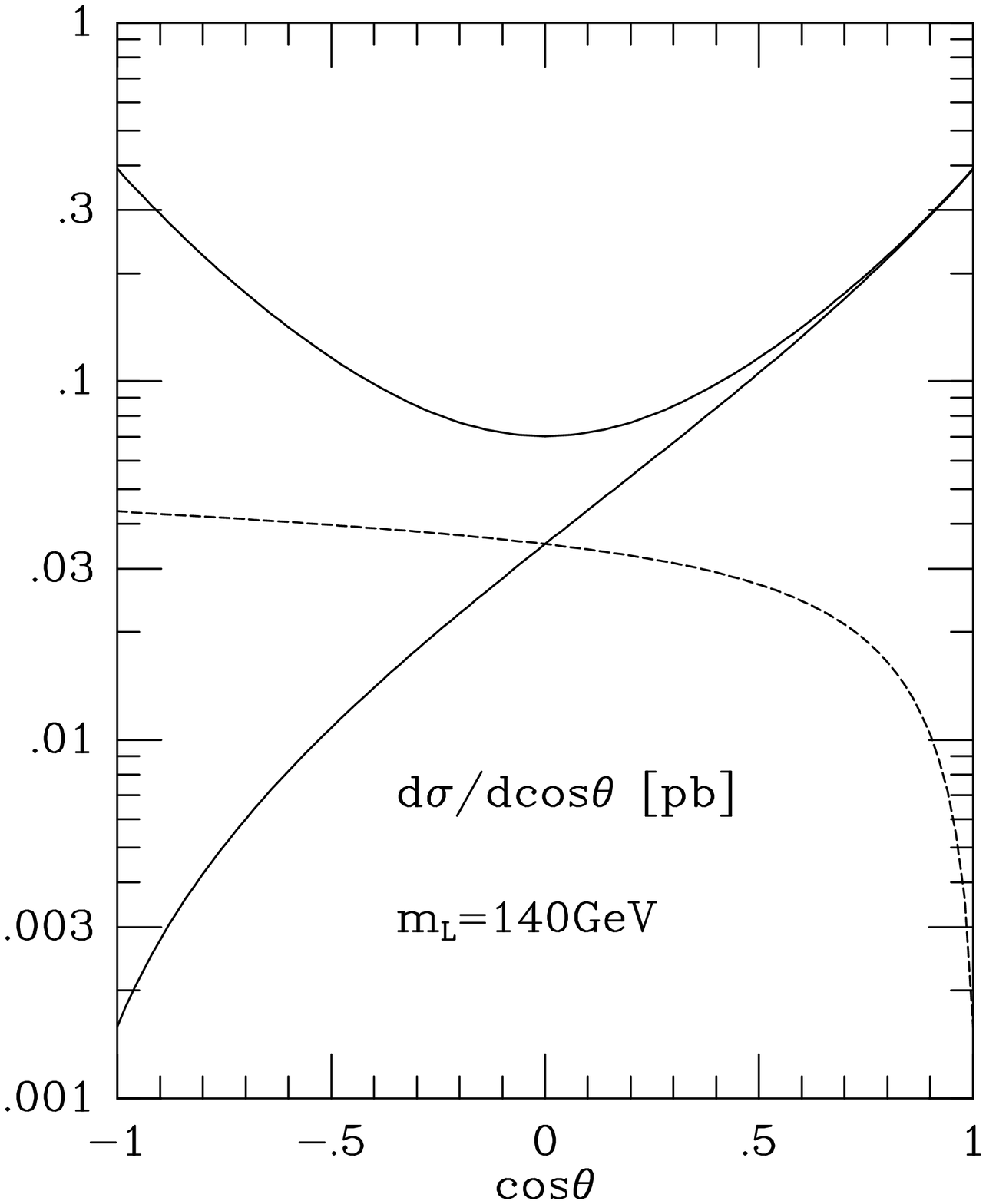,width=0.4\textwidth}}
\ccaption{}{\label{fig:exotic_sing}                       
Total cross section (a) and angular distribution (b) for the 
single production of exotic leptons in association with their ordinary 
light partners at LEP2 with $\sqrt{s}=190~\gev$. The solid (dashed) lines
are for first generation neutral leptons with a left--handed (right-handed) 
mixing, the long-dashed-dotted line is for second/third generation leptons 
with a left--handed or right-handed mixing, and the dotted (dot-dashed) lines
are for first generation charged leptons with a left--handed (right-handed) 
mixing. For the angular distribution the symmetric solid curve is for a 
Majorana neutrino.}
\end{figure} 

\subsubsection{Single Production} \label{elem:sp}
Introduction of exotic fermions, on the other hand, opens up the possibility
of tree-level FCNCs. This has the immediate consequence of allowing 
single production of a heavy fermion in association with a SM particle. 
While most such productions proceed through a
$s$-channel $\Z$-mediated diagram, for heavy leptons that have a direct
coupling with the electron, there is the additional contribution from
$t$-channel diagrams ($\W$-mediated for $N \nu$ production and  
$\Z$-mediated for $eE$). For such leptons (to be called ``first generation
exotics'' henceforth), the production cross section can be significantly 
higher.
In fig.~\ref{fig:exotic_sing}, 
we exhibit the mass-dependence of the cross section for various choices 
of exotic leptons~\cite{R3}. To be concrete, the left- (right-) handed 
mixing parameters are assumed to have the most optimistic 
value of $\zeta_{L,R} = 0.1$. As we see from the 
figure, the cross sections for the ``second and third generation'' exotics 
are below observable levels. As can be expected, 
the cross sections for quark-production ($Q q$) are of similar magnitude. 
We thus reach the conclusion that irrespective of the quantum numbers and
the production mode, LEP2 is unsuitable for exploring possible heavy quarks. 

While it is obvious that the  $E l$ production cross section is very small and
with LEP2 luminosities we can hope to gain over pair production only by a few
GeVs, the signals, nonetheless, are interesting. The heavy lepton $E$ may now
have both charged current (final state  $l \nu + 2 \;{\mbox{\rm jets}}$) and
neutral current (final state $l^+ l^- + 2 \;{\mbox{\rm jets}}$) decays. The
latter channel is better suited for mass reconstruction. The main backgrounds
arise from $l \nu W$ and $l^+ l^- Z$ production respectively. While these may
be eliminated by simple cuts, the signal itself is reduced to uninteresting
levels. 

\begin{figure}
\centerline{\epsfig{figure=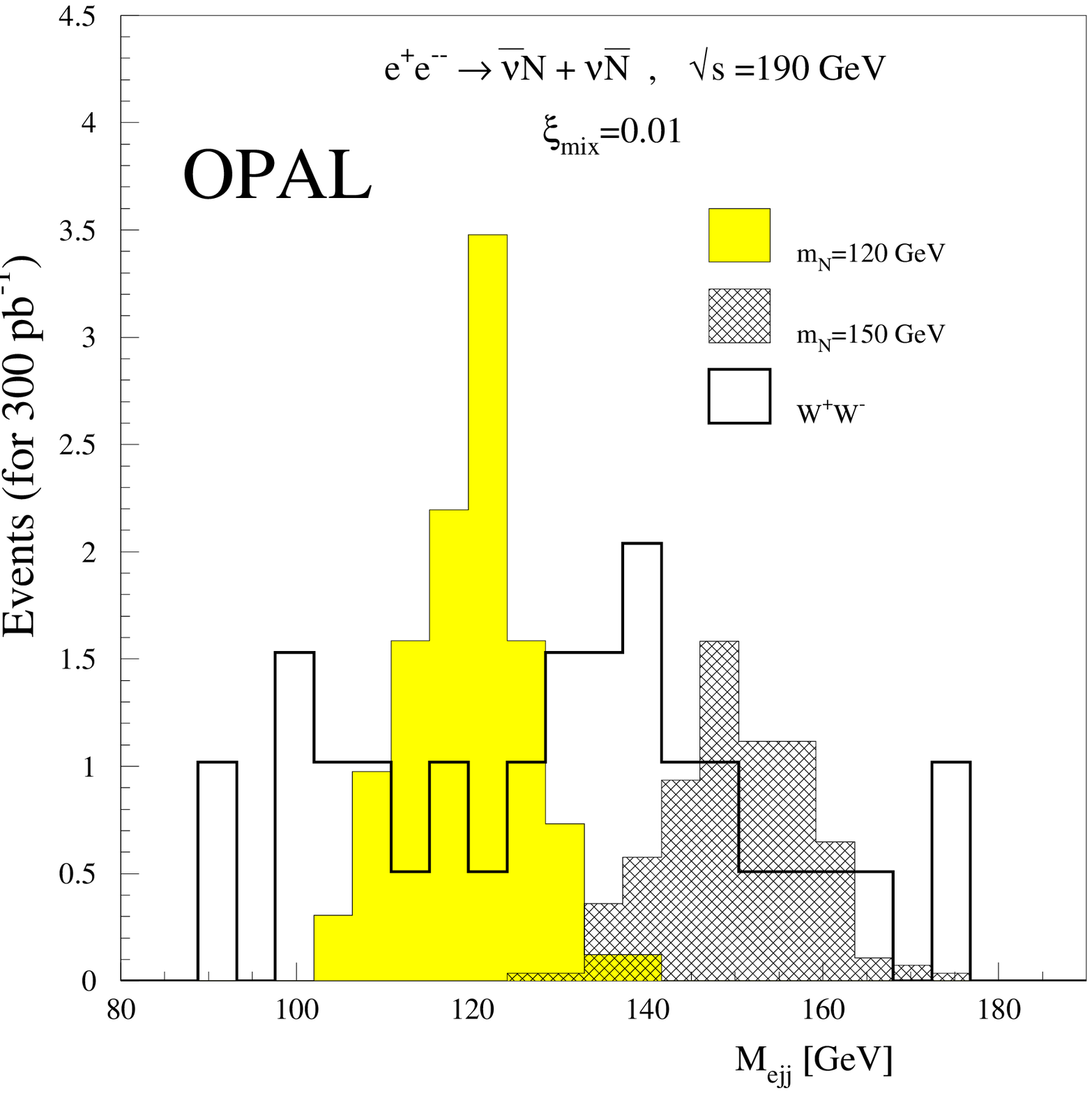,width=0.4\textwidth}
            \epsfig{figure=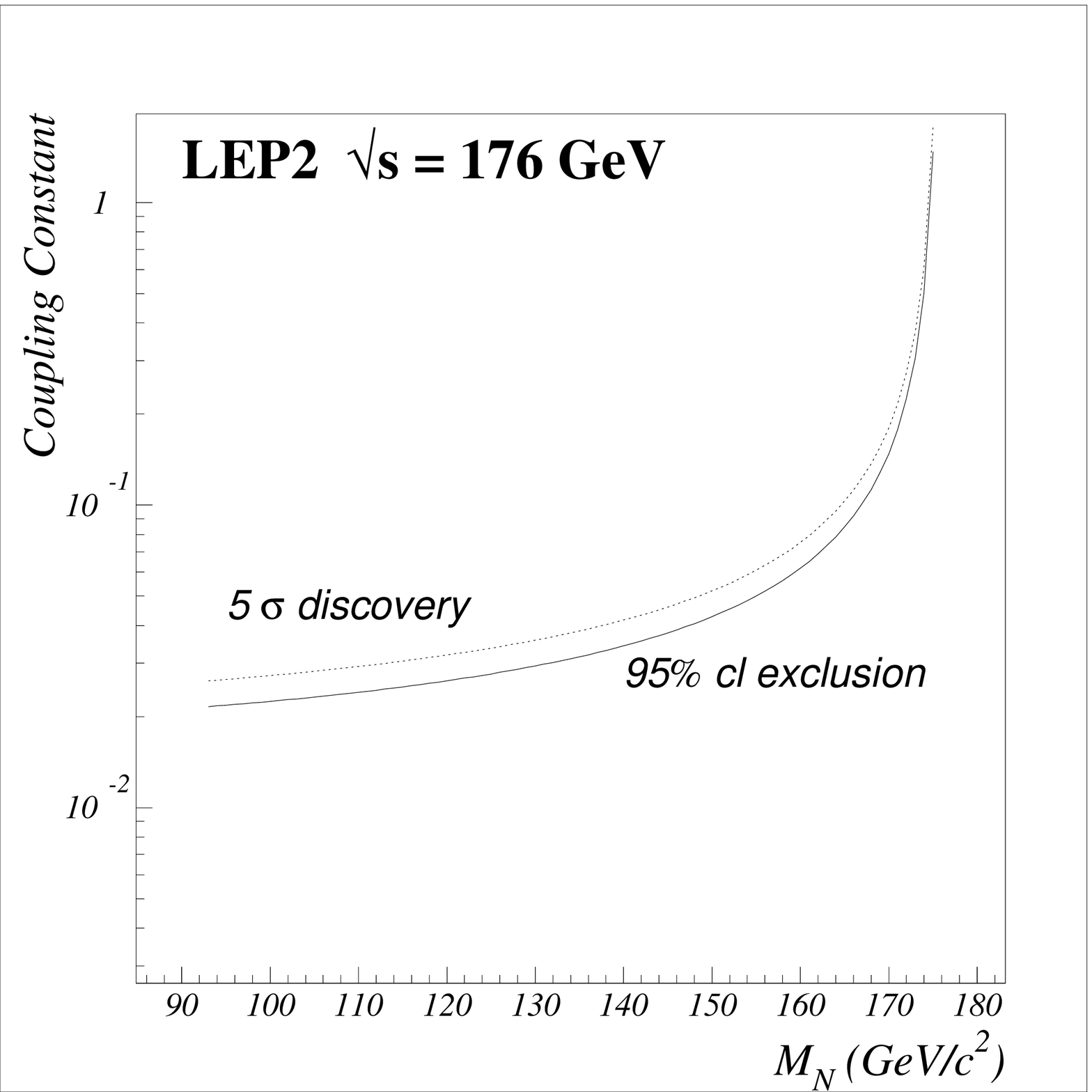,width=0.4\textwidth}}
\ccaption{}{\label{LBRYANT}
The singly produced first generation heavy neutrino: reconstructed
invariant masses (left) with $\WW$ background at $190~\gev$, and (right) limits
for $176~\gev$}
\end{figure}

Returning to the first generation case, we may be more optimistic.
Again $N$ can decay both through charged current (final state $l \nu +
2\;{\mbox{\rm jets}}$) and neutral 
current (final state $\nu \bar\nu + 2\;{\mbox{\rm jets}}$) 
channels, the former
being  more suitable for mass reconstruction.
Events are selected by requiring an isolated identified charged lepton
and missing
momentum. Then the hadronic activity (jets) is required to be consistent with
an on-shell $\W$, while the invariant mass of the electron and $\not p$ is
inconsistent with an on-shell $\W$. The corresponding SM background (mostly
$\WW$) is
reduced to a very low level. The heavy neutrino mass is found from the
invariant mass of all the visible particles, see fig.~\ref{LBRYANT}(left),
obtained from
a full simulation in the OPAL detector Monte Carlo~\cite{N:opal}. 
Limits derived from
a full simulation in the ALEPH detector Monte Carlo, using a similar analysis,
are shown in fig~\ref{LBRYANT}(right).

\subsection{Excited Leptons} \label{excited}
As the spectrum of the excited fermions ($F^\ast$) depends 
crucially on the dynamics of the particular  theory in question, the
lowest lying excited states can have various spin and isospin quantum
numbers. 
In this study we shall restrict ourselves to 
spin 1/2 and isospin 0 or 1/2 (other cases have been  discussed in
ref.~\cite{R3}). 
If we assume that the excited states acquire  mass 
above the EW breaking 
scale (so as to motivate the rather large mass gap), they necessarily 
would have vector-like couplings~\cite{BOUDJ}. The alternative is to 
assume a chirality structure identical to the ground state fermions. 
This is a more conservative choice from the experimentalist's 
point of view as it leads to lower production cross sections.
We shall examine both cases in some detail. 

Apart from the gauge couplings (governed by the quantum numbers), these 
particles would also couple to the ground states. 
As the relevant piece of the interaction Lagrangian 
has the structure of a magnetic form factor, 
(and hence is of dimension 5), its 
strength is determined by $f_i / \Lambda$,
where $\Lambda$ is the compositeness scale and $f_i$ are dimensionless 
coupling constants that may differ for the different gauge bosons. The 
presence of such couplings allow the excited fermions to decay 
to their ground state partners and a photon (or gluon in case 
of quarks). 

\subsubsection{Pair Production} \label{exc:pp}
The pair production cross sections are dominated by $s$-channel diagrams 
involving $\gamma$ and $Z$ exchange. While $e^\ast$ or $\nu_e^\ast$ production 
can receive contributions from the magnetic piece of the Lagrangian, 
for the allowed range of $f_i$ \cite{pdg}, these terms 
are non-negligible only for $\sqrt{s} \sim \Lambda$. 
The production rates are thus similar to those for the heavy 
fermions. 
The presence of the radiative decay modes thus provides us with 
tell-tale signatures, especially in the context of LEP2. The SM 
background may be further reduced by imposing isolation cuts on the 
photons. We examine below the specific case of $\lsls$ production. 

Events with two photons above $10~\gev$ and either two or four tracks are
required, all isolated from
each other by a cone of at least $25^{\circ}$, with the exception of the track
triplet, that must be a $\tau$ candidate.
Signal events were simulated by smearing Monte Carlo 4-vectors in accordance
with the ALEPH detector resolution for tracks and photons, after appropriate
`losses', such as photon conversion. Background events were looked for, using
standard EW generators, and none were found, but the generators are
not fully adequate to estimate the rate, and the program of 
ref.~\cite{MANEL} was used to confirm this prediction.
Because of the predominance of $\gamma/\z$ at masses below the nominal centre
of mass energy, it is essential to use events in which the visible energy
is deficient; however, with energy- and momentum-conserving reconstruction
techniques~\cite{SEARCHES}, 
the energy, $E_0$, of the most energetic photon below the polar
angular acceptance threshold can be estimated, and used in reconstructing the
event.

\begin{figure}
\centerline{\psfig{figure=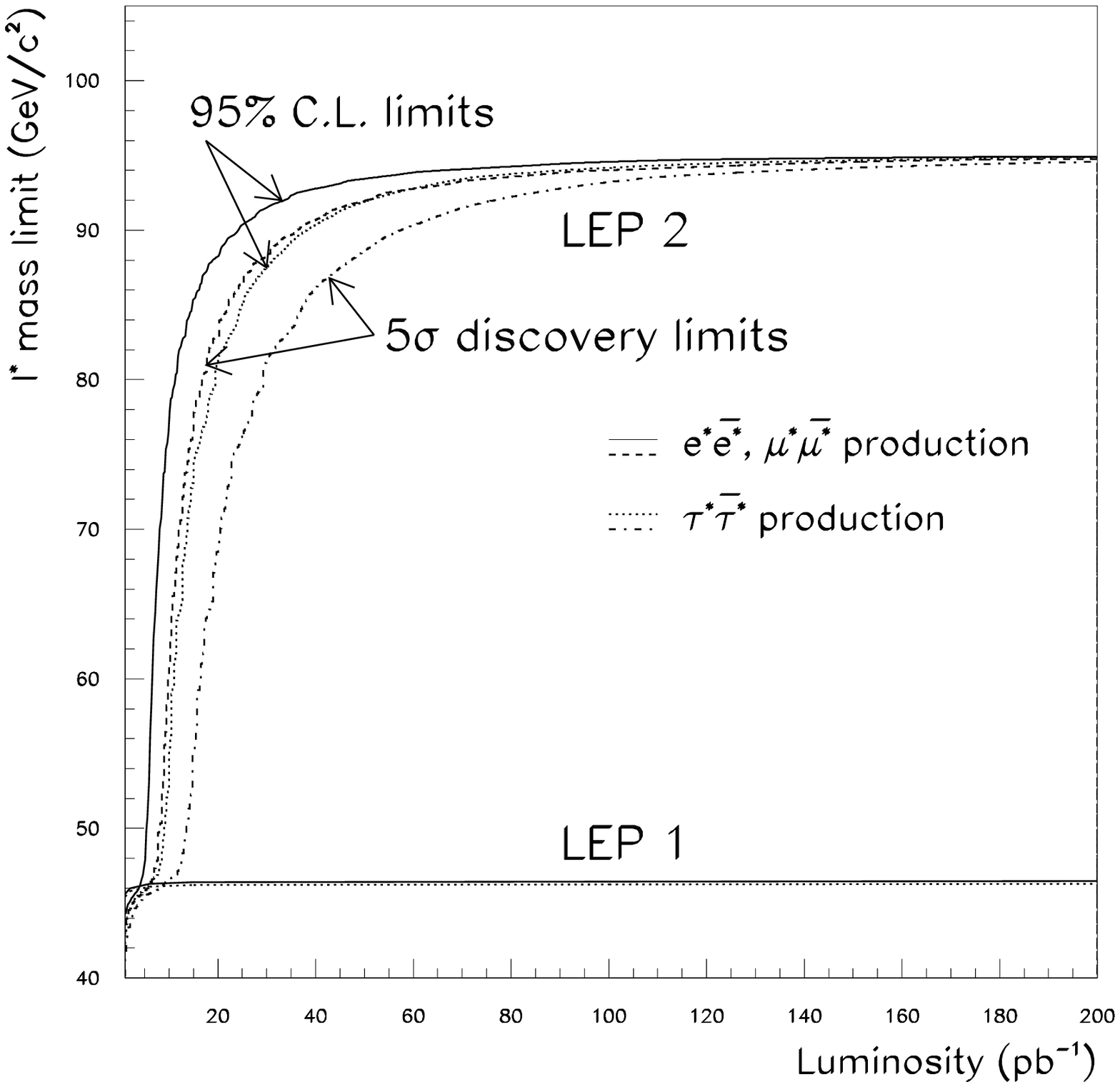,height=8cm}} 
\ccaption{}{\label{JWT1}                            
Limits for excited lepton pairs, decaying radiatively at LEP2
(190 GeV): 205 GeV is very similar, with more mass reach.}
\end{figure}

Additionally, when events with two energetic photons 
and two leptons are found,
it is required that there is one way of combining lepton and photon candidates
such that the invariant masses of the two lepton-photon pairs have a difference
which is smaller than $2\sigma$ of the resolution (improved by rescaling
events, using energy- and momentum-conservation) as the $\ls$s produced
should be identical. There is significant background prior to this last cut,
but after it the background is very small, and
the signal efficiency remains high: 67\% for $\eses$ and $\msms$, and
50\% for
$\tsts$, which has poorer resolution.
Limits for exclusion and discovery at $\sqrt{s}=190$
are shown in fig.~\ref{JWT1}.

\subsubsection{Single Production} \label{exc:sp}
Occurring  on account of the $f_i$ terms, these channels would allow us 
to probe for $F^\ast$ masses close to the c.m. energy. 
\begin{figure}
\centerline{\epsfig{figure=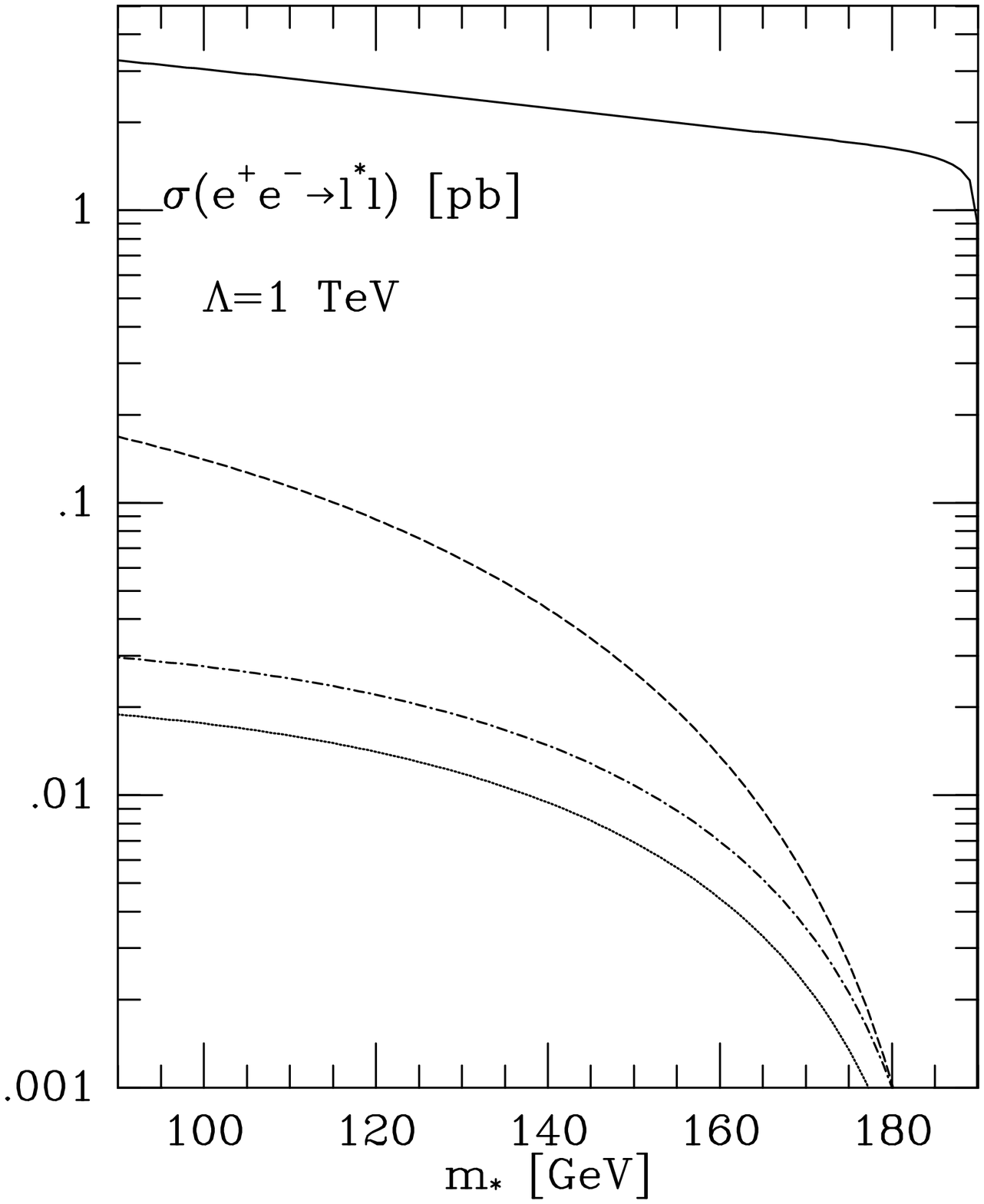,width=0.35\textwidth}
            \epsfig{figure=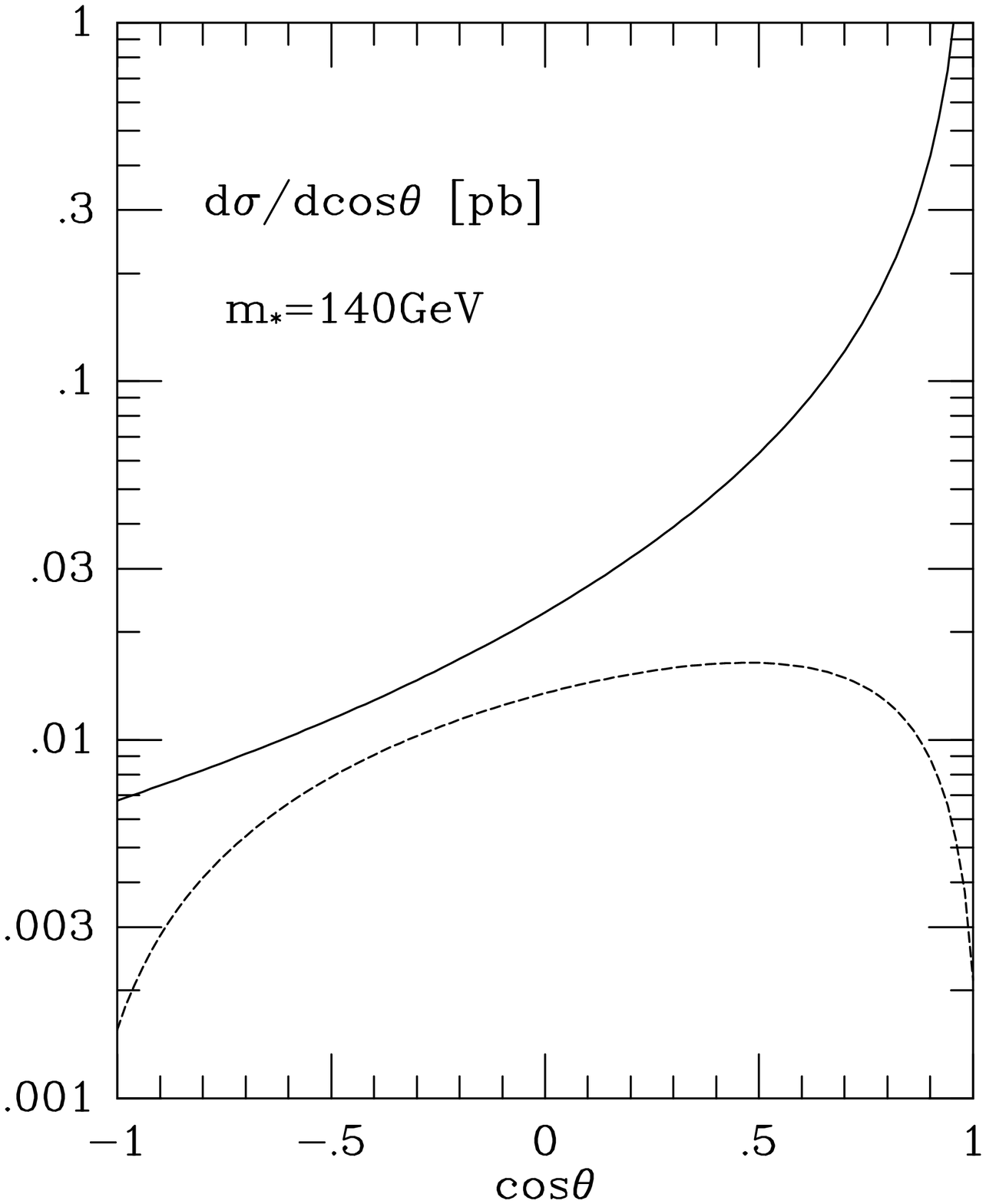,width=0.35\textwidth}}
\vspace{-1cm}
\ccaption{}{\label{fig:excited_sing}
Total cross section (a) and angular distribution (b) for the 
single production of vector-like
excited leptons in association with their ordinary light 
partners at LEP2 with $\sqrt{s}=190~\gev$ and $\Lambda=1$ TeV. The solid 
(dashed) lines are for first generation charged (neutral) excited leptons
and the dotted (dash--dotted) lines are 
for the second/third generation charged (neutral) excited leptons.}
\end{figure}
In fig.~\ref{fig:excited_sing}, we show the total cross sections 
for different choices of vector-like excited leptons 
at a c.m. energy of 190 GeV. The
largest production rate  occurs for the $e^\ast$ due to the $t$-channel photon
exchange. Compared to the other charged leptons this has a two-order of
magnitude enhancement. An analogous statement may be made about 
$\nu_e^\ast$ (as a consequence of the $\W$ exchange contribution) when 
compared to the other excited neutrinos. Similar statements also 
hold for the chiral $F^\ast$, and, henceforth, we shall concentrate on this 
possibility.

\begin{figure}
\centerline{\epsfig{figure=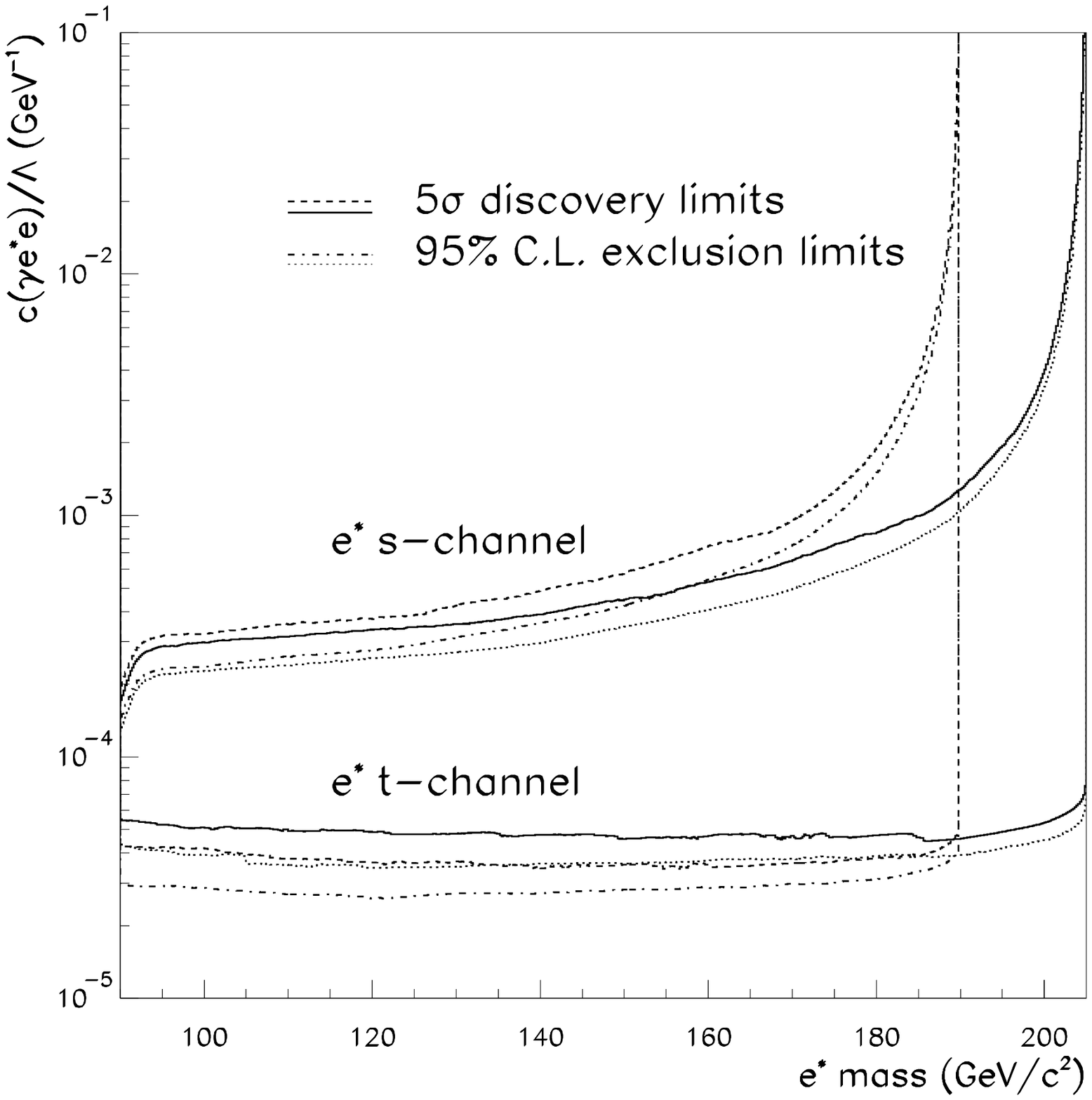,width=9cm}}
\ccaption{}{\label{JWT2}             
Exclusion and discovery limits (at 190 and 205 GeV) 
in the mass-coupling plane for singly produced chiral $\es$.  
The two sets derive from 2--track (``$s$--channel) and
1--track (``$t$--channel) signals respectively. 
$\msm$ limits are almost identical to $\ese$ (2--track), and
$\tst$ only a little worse.}
\end{figure}

Charged excited leptons\footnote{For excited neutrinos, 
the situation is similar to that of exotic neutrinos
(although the angular distributions are different, as shown in 
fig.~\protect\ref{fig:excited_sing}) and one
has to look for $e \nu jj$ events. A detailed analysis has
not been performed here. Note that the single photon signal 
has a large associated background.} 
should be looked for by exploiting their electromagnetic decays leading to a
signal 
\be
e^+ e^- \rightarrow l l^* \rightarrow l^+ l^- \gamma
\ee
All signal and background (EW) events were fully simulated in the
ALEPH detector Monte Carlo. To qualify, an event had to be comprised of a
dilepton pair in addition to  one and only one photon above 10 GeV. Isolation
criteria were held to be the same as in the previous  section, and a possible
low angle photon was reconstructed. Examples of the results are shown in
fig.~\ref{JWT2}. It must be borne in mind that the couplings $f_\gamma /
\Lambda$ and $f_Z / \Lambda$ are {\em a priori} different. To eliminate one
variable, the relationship used in ref.~\cite{hagiwara} was assumed. The lower
sensitivity for the $\tst$ case is due to the greater difficulty in
reconstructing the events. 

For the particular case of $\ese$ production, the photon and $Z$ contribution
can be unraveled by exploiting the difference in their angular distributions.
Since the $t$-channel $\gamma$--exchange diagram, unlike the others, gives a
very forward peaked contribution, it dominates the configurations wherein one
electron essentially disappears down the beam pipe. The final state is thus of
the form $(e^+) e^- \gamma$ and the main background is from higher order
processes in QED and misidentifications. Following the analysis of
ref.~\cite{SEARCHES}, the plane defined by the track and the photon is 
required to be nearly normal to the beam axis and the polar angle corresponding
to the missing momentum is required to be less than $2^\circ$. Possibility
exists though that the missed electron had a polar angle of greater than
$2^\circ$ but was too slow to get into the main tracking detector. This  is
taken care of by vetoing events with a large excess of hits in the inner
tracker (since this corresponds to a ``spiralling'' electron). The exclusion
limit after the imposition of these cuts is shown in fig.~\ref{JWT2}. As was
indicated earlier, this particular configuration is much more sensitive to the
presence of $\es$ than those where both the electrons are visible in the main
detector. 

\subsubsection{Virtual Effects} \label{exc:ve}
While the single production channels raise the mass reach of the machine to
nearly the kinematical limit, it is interesting to ask the question if one
could explore masses beyond the c.m. energy, albeit at the cost of
accommodating a somewhat larger coupling constant. The only existing
constraints for $m_{e^\ast} > 130~ \gev$ are derived from one-loop corrections
to low-energy observables such as $g - 2$ of the electron (see references
quoted in \cite{pdg}) and are subject to theoretical uncertainties. AT LEP2, a
particular window is provided by the possible $t$-channel exchange of a virtual
$\es$. The signal consists of a distribution of 2-photon final states that has
an excess at high polar angle (on account of the large mass of the $\es$), when
compared to the EW expectation.  In practice, 3-photon final states
(first order initial state radiative correction) have to be taken into account
as well, and the analysis uses the two most energetic photons and the angle
between them in their centre of mass frame, $\theta^*$. To perform a
quantitative analysis, we adopt the the maximum likelihood method~\cite{pdg}.
The cross section is parametrized~\cite{LITKE} in terms of the $e^\ast$ mass
and a dimensionless  effective coupling $\lambda$ and the resultant expression
compared to the actual (simulated) data. Assuming an integrated luminosity of
500 pb$^{-1}$, we find that for $\lambda = 1$ one can achieve 95\% C.L. mass
exclusion limits of $\sim 250 (300)~\gev$ at $\sqrt{s}=190 (205) \gev$. 
\section{Leptoquarks}
Leptoquarks carry, simultaneously, lepton and quark quantum numbers. They
naturally appear in unified and composite models \cite{models} and mediate
lepton--quark transitions. For this study we adopt the phenomenological
framework suggested in ref. \cite{buch} which involves a minimal number of model
assumptions. It is designed for spin 0 and 1 leptoquarks with the most general
dimensionless couplings to lepton-quark pairs which are $SU(3)_c\times
SU(2)_L\times 
U(1)$ symmetric, family--diagonal, and baryon and lepton number conserving.
The corresponding couplings to the EW gauge bosons are given in ref.
\cite{blum}. One can distinguish two classes of leptoquarks: color--triplets
with fermion number 2 decaying into lepton--quark final states and
color--antitriplets with fermion number 0 decaying into lepton--antiquark final
states. The weak isospin varies from 0 to 1. A complete list of the possible
species and their quantum numbers can be found in refs. \cite{buch,blum}. 

Experimentally, the existence of leptoquarks is constrained indirectly by 
low-energy data \cite{low} and 
precision measurements of the $Z$ widths \cite{z},
and by direct searches at high energies \cite{HERA,LEP,D0,CDF}. In particular,
rare processes and processes which are forbidden in the Standard Model provide
stringent bounds on $\lambda \over m_{LQ}$, where $\lambda$ denotes the
leptoquark--fermion Yukawa couplings. Essentially, below the TeV mass range and
for $\lambda$ of the order of the electromagnetic coupling, the only
allowed leptoquarks are those which 
couple only to a single fermion generation and, for the two lighter
generations, either only to the left- or to the right-handed components. HERA
experiments exclude first generation scalar leptoquarks up to $m_{LQ} = 250$
GeV for $\lambda = e$ ~$(\alpha = {e^2 \over 4\pi})$ ~\cite{HERA}. Bounds which
are independent of $\lambda$ and therefore unescapable come from LEP and
TEVATRON. The LEP experiments set a mass limit of 45.6 GeV for any leptoquark
species \cite{LEP}. Searches with the D0 detector exclude first generation
scalar leptoquarks below $m_{LQ} = 133$ GeV ~\cite{D0}, while the CDF
experiment puts a bound on second generation scalar leptoquarks at $m_{LQ} =
131$~GeV provided these leptoquarks decay to electron (muon) plus jet with
$100\%$ branching ratio~\cite{CDF}. 
Since there is at least one state in each isospin
multiplet with $Br(LQ \rightarrow l + jet) \ge 0.5$ where $l = e$ or $\mu$, the
absolute lower mass limit for first (second) generation scalar 
leptoquarks is  120~GeV (96~GeV), at least if the
members of a multiplet are nearly mass--degenerate. Otherwise, states decaying
exclusively into $\nu +$ jet final states are only required to respect the less
severe LEP1 mass limit. For sufficiently small $\lambda$, the LEP1 bound is
also the only one applying to third generation leptoquarks. 
                                                            
At LEP2, leptoquarks can be pair--produced via s--channel $\gamma$ and $Z$
exchange and, in the case of first generation leptoquarks carrying the electron
number, also via t--channel exchange of a $u$ or $d$--quark. The latter process
involves the unknown Yukawa coupling $\lambda$. The production cross section
depends very much on the leptoquark quantum numbers \cite{blum}. At $\sqrt{s} =
190$ GeV and for $m_{LQ} = 80$ GeV it varies from 0.04 pb for the scalar
isosinglet $S_1$ with charge $-{1 \over 3}$ to 12 pb for the vector isotriplet
$U_3$ with charge $+{5 \over 3}$ (possible t--channel contributions are
disregarded). Here, we follow the notation introduced in \cite{buch}.
Furthermore, leptoquarks in the LEP2 mass range are very narrow. The partial
width for a massless decay channel is expected to be of the order of 100 MeV
for $\lambda = e$.
 
The signatures for leptoquark pair--production and decay are: {\it i)}
two electrons
(muons) plus two hadronic jets; {\it ii)} one electron(muon), two hadronic 
jets, and
missing energy; {\it iii)} two hadronic jets and missing energy; 
{\it iv)} two tau leptons plus two
hadronic jets. Given the Tevatron bounds pointed out above, the focus is on the
last two signals. To estimate the acceptance for leptoquark final states we
have written a Monte Carlo program and generated events for 
$m_{LQ} = 50-80$~GeV, $0 < \lambda_{L,R}/e < 1$, and $\sqrt{s}$ = 150, 175 and
190 GeV. The generator is based on the analytical formulae for cross sections
and angular distributions given in ref. \cite{blum}. Initial state radiation is
included. Fragmentation is implemented according to JETSET 7.3~\cite{jetset}.
                                          
The following background processes have been considered and generated for
$\sqrt{s} =$ 150, 175 and 190 GeV using Pythia 5.7~\cite{jetset}: $e^+e^-
\rightarrow W^+W^-$, $We\nu$, $ZZ$, $Zee$, and two-photon 
processes. The numbers of
generated events correspond to an integrated luminosity of 500 pb$^{-1}$,
except for the two-photon processes, where they correspond to only 20
pb$^{-1}$. The events were passed through the standard L3 simulation and
reconstruction program \cite{L3s}. 
 
{\it Two electrons (muons) and two jets}
 
The signal events are required to consist of two energetic electrons (muons)
with energies greater than 20 GeV, and two hadronic jets. Furthermore, we
require the total visible energy to be greater than $0.9\times \sqrt s$. The
$ZZ$ background is suppressed by demanding that the invariant mass of the two
leptons is outside the $Z$ mass window. Since this cut is not efficient if one
$Z$ bosons is produced off-shell, we require in addition that the difference
between the two reconstructed leptoquark mass values is less than 4 GeV. The
masses are determined using energy--momentum conservation. We find that the
mass of an 80 GeV leptoquark of the first generation can be reconstructed with
a resolution of about 0.5 GeV. 

{\it One electron (muon) and two jets}
 
In this channel, we require the events to consist of one energetic electron
(muon) with energy greater than 20 GeV, and two hadronic jets. Furthermore, we
request the total visible energy to be greater than $0.5\times \sqrt s$ and
less than $0.9\times \sqrt s$. The $WW$ and $We\nu$ backgrounds are suppressed
by demanding that the invariant masses of the two jets and of the
lepton-neutrino pair is outside the $W$ mass window, and that the difference
between the two reconstructed leptoquark mass values is less than 8 GeV. In
order to calculate the leptoquark masses, we determine the missing momentum
from energy--momentum conservation. The leptoquark mass resolution turns out to
be about 2 GeV. 

{\it Two neutrinos and two jets}

This is a particularly important channel. The signal events are selected by
requiring that the event consists of two hadronic jets with a total energy less
than $0.8\times \sqrt s$ and that it
contains no electrons with energy greater than 20
GeV. Furthermore, the event multiplicity has to be greater than four, and the
polar angle of the missing momentum has to be larger than 30$^\circ$ or less
than 150$^\circ$. The background in this channel is composed of $ZZ$ events
with one $Z$ boson decaying into two neutrinos, $WW$ and $We\nu$ events with
one $W$ boson decaying hadronically, and two-photon events. To remove the
two-photon events, we require the total energy in the event to be greater than
$0.4\times \sqrt s$, and the minimal jet energy to be greater than 10 GeV. The
background from events in which some particles are not correctly reconstructed
in the detector is removed by demanding the calorimetric energy in a 30$^\circ$
cone around the missing momentum to be less than 1 GeV. No charged track must
be found in this cone. We find about ten $ZZ$, $WW$ and $We\nu$ events which
survive the above cuts. These can be removed by a cut on the invariant jet--jet
mass, ${ M_{jj} \frac{\sqrt s }{E_{vis}}} < {min}(\sqrt s - \mz,~\mw)-10$ GeV. 
 
{\it Two tau leptons and two jets}
 
This signal is the most difficult one to recover from the $WW$, $ZZ$, 
and $We\nu$ background.
We require that the event consists of four jets,
with a minimal jet energy greater than 8 GeV 
and two jets having multiplicity one.
Furthermore, the total energy in the event must be
greater than $0.5\times \sqrt s$ and less than $0.9\times \sqrt s$.
Finally,
we look for the most isolated track with momentum greater than
$ 3$ GeV, and require that there is
no other track inside a ${20^\circ}$ cone around this track.
The calorimetric energy inside that cone has to be more than 5 GeV,
and the difference of the calorimetric energies in the ${30^\circ}$ and
${20^\circ}$ cones
around the track has to be less than 1 GeV. 
 
\begin{figure}
\leavevmode   
\begin{center}
\vspace*{-1.2cm}\begin{tabular}{l l}
\hspace*{-1cm}\mbox{\epsfxsize=9.cm\epsfysize=8.cm\epsffile{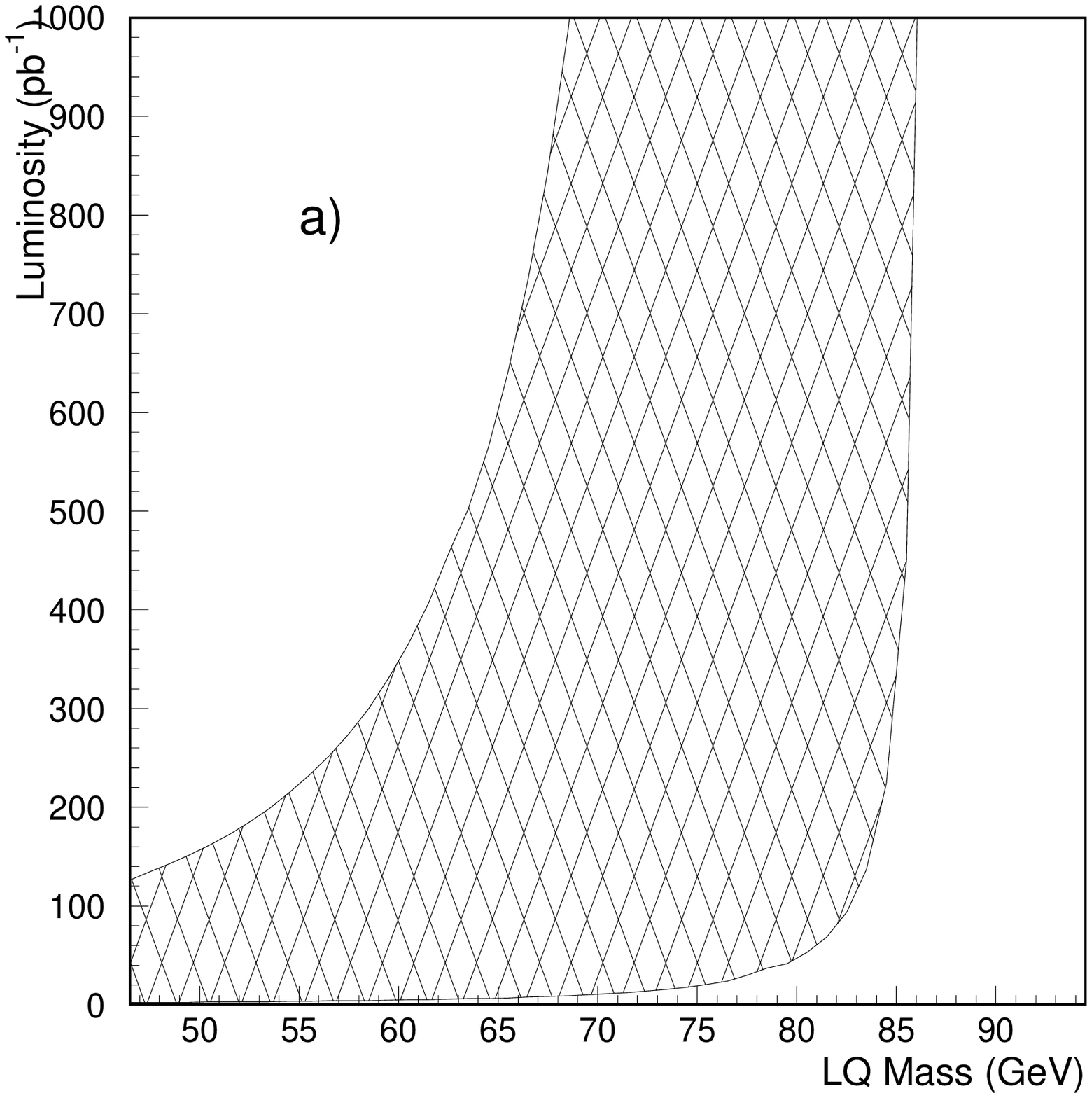}}  &
\hspace*{-1cm}\mbox{\epsfxsize=9.cm\epsfysize=8.cm\epsffile{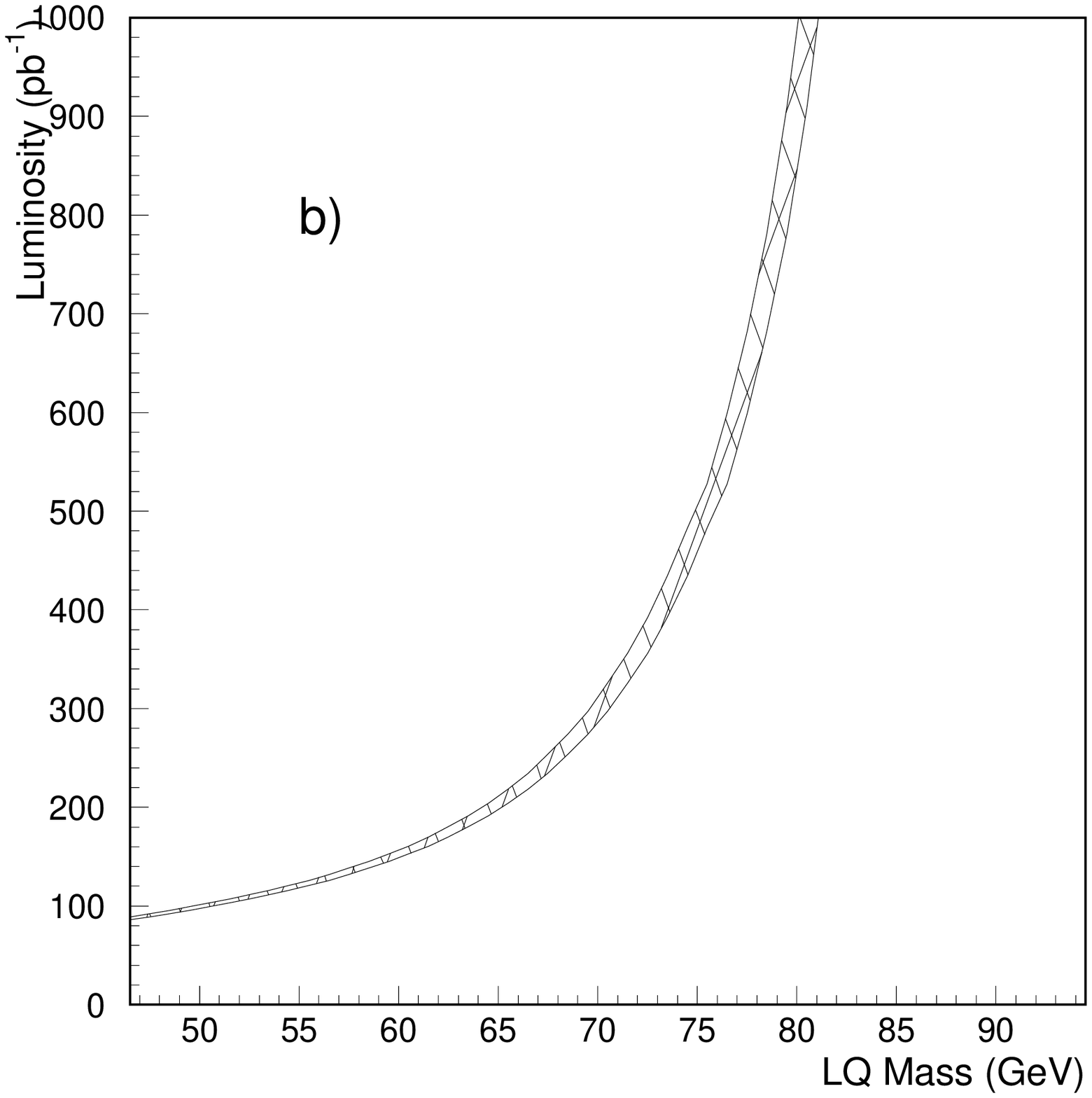}}  \\
\hspace*{-1cm}\mbox{\epsfxsize=9.cm\epsfysize=8.cm\epsffile{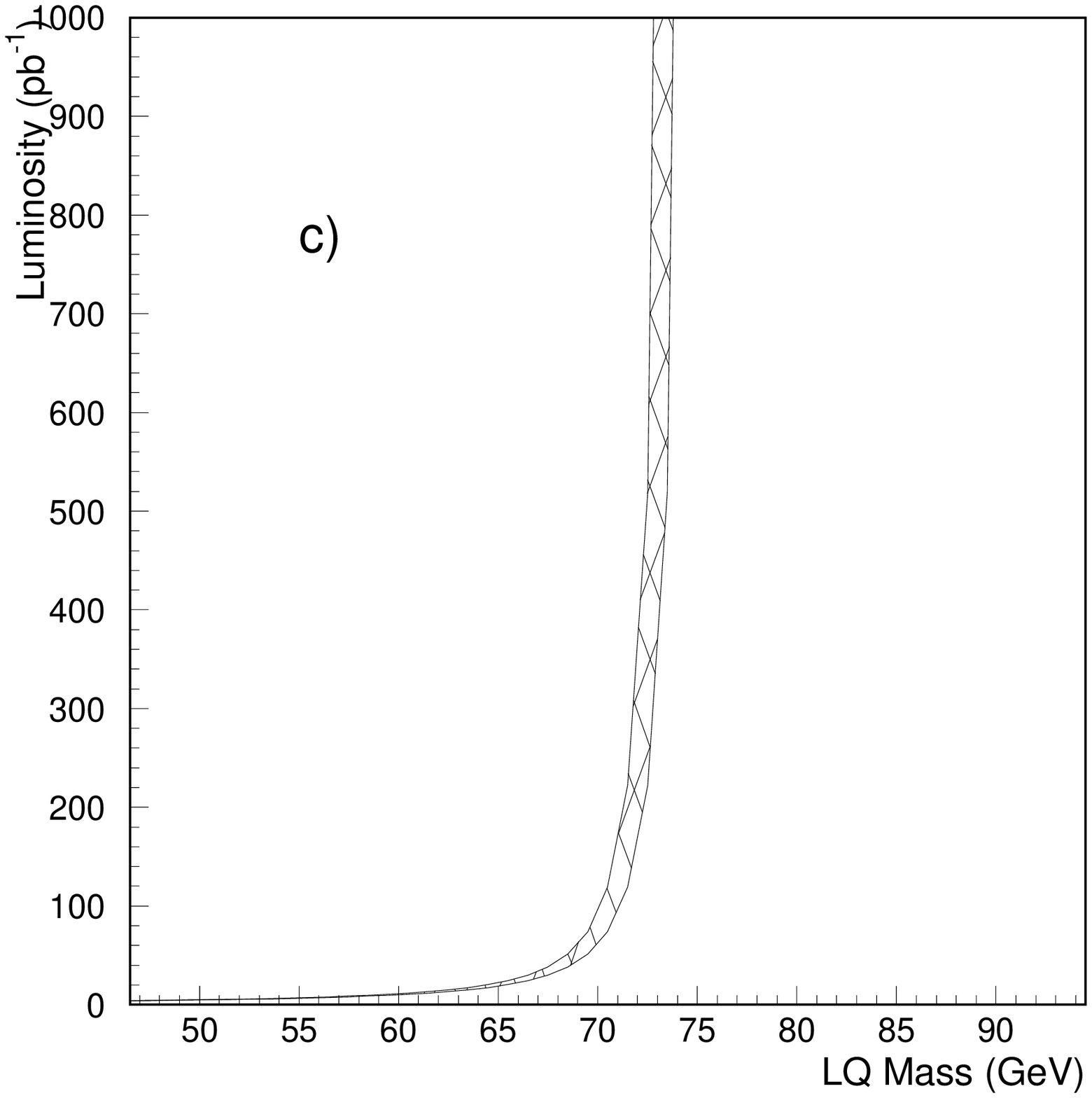}}  &
\hspace*{-1cm}\mbox{\epsfxsize=9.cm\epsfysize=8.cm\epsffile{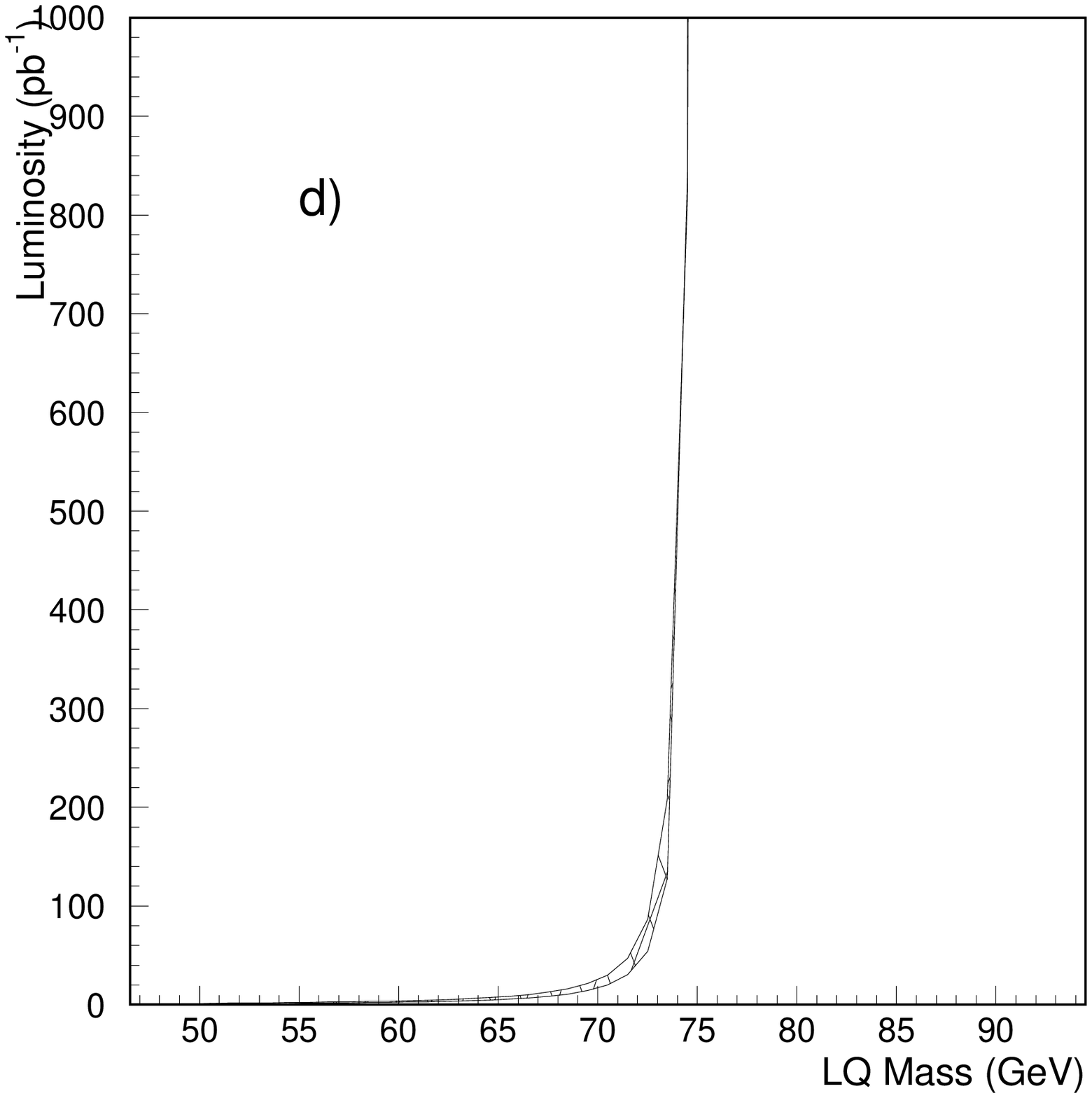}}  \\
\end{tabular}
\end{center}
\ccaption{}{\label{fig1}$5\sigma$ discovery limits for
a) $S_1^e$ at 175 GeV,  
b) $S_1^\tau$ at 190 GeV taking $Br(S_1^\tau \rightarrow \nu_{\tau} b) = 1$,
c) $S_3^\tau$ (4/3) at 150 GeV, and
d) $U_1^\mu$ at 150 GeV.
The shaded areas
show the effect of varying the Yukawa couplings
$\lambda_{R,L}/e$ from zero to unity, and the acceptance within
the estimated range given in table~\ref{eftab1}.}
\end{figure}                                   

Table~\ref{eftab1} shows the estimated acceptance and number of surviving
background events for different leptoquark
decay channels.
\begin{table}[htb]
\begin{center}
\begin{tabular}{|c|c|c|}
\hline
Decay & Acceptance (\%) for &   Background  \\
Channel        & $50<M_{LQ}<80$ GeV &  at 150-190 GeV         \\
\hline
$e^+e^- q \bar q$  & 39 to 45 &  0-1 $ZZ$   \\
\hline
$e\nu q \bar q$  & 22 to 25  & 3-5 $WW$  \\
\hline
$\mu^+\mu^- q \bar q$  & 42 to 48 & 0 \\
\hline
$\mu \nu q \bar q$  & 22 to 25  & 4-6 $WW$   \\
\hline
$\nu \bar \nu q \bar q$  & 20 to 23& 1 2$\gamma$, 1-3 $ZZ$,$WW$,$We\nu$  \\
\hline
$ \tau^+ \tau^- q \bar q$  & 11 to 12  &  10-13 $WW$,$ZZ$ \\
\hline
\end{tabular}
\end{center}
\ccaption{}{\label{eftab1}
Acceptance and number of background events for different leptoquark
signals assuming an integrated luminosity of 500 pb$^{-1}$.}
\end{table} 
Using these numbers and the leptoquark production cross sections,
we estimated the 5$\sigma$ discovery limits 
as a function of the integrated luminosity. Some illustrative results
are shown in fig.~\ref{fig1}.
Note that the large variation of the discovery
limit for the first generation leptoquark $S_1^e$ 
is mainly due to the variation of the Yukawa coupling $\lambda$.
  
%
To summarize,
a feasibility study on the search for leptoquarks at LEP2
with the L3 detector has been carried out. 
We have considered different leptoquark species and signatures.
The main result is that after applying the appropriate cuts
all signals except the one in the $\tau \tau$ plus two-jet channel
are practically background--free.
In the latter case, 
we expect 10 to 20 background events for 500~pb$^{-1}$.
Since the existing mass limits for leptoquarks decaying into electron (muon)
plus jet are already beyond the LEP2 energy range, the leptoquark
species relevant for LEP2 searches are those
which decay into $\tau$ plus jet and
$\nu$ plus jet. 
 
\section{The BESS Model for Dynamical EW Symmetry Breaking}
\label{besssection}
\begin{figure}
\centerline{\epsfig{figure=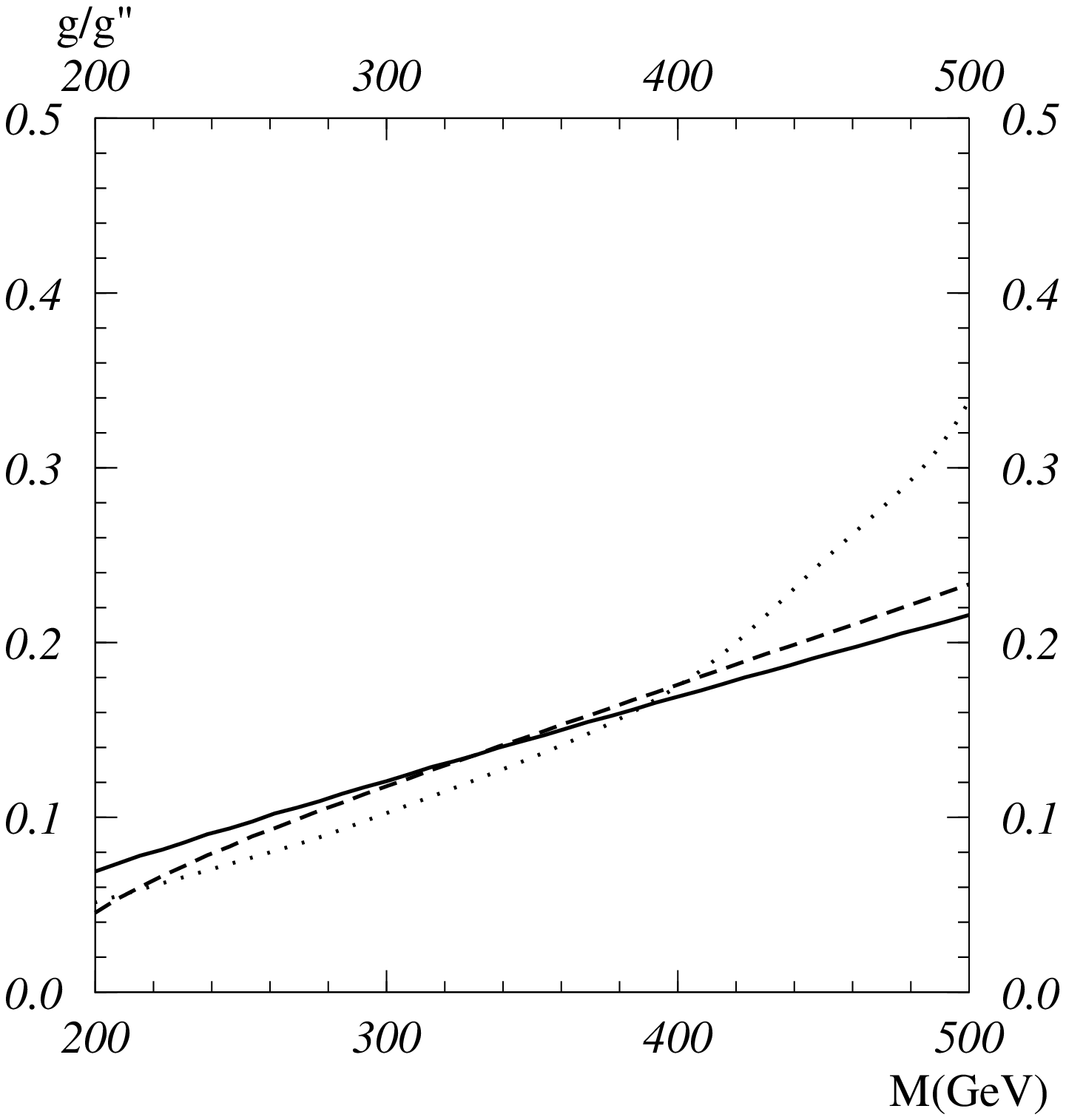,height=8cm,angle=0}}
\ccaption{}{\label{bessf}
95\% C.L. upper bounds on $g/g''$ vs. $M$ from LEP1 data 
(continuous line) and CDF data (dotted line) compared with
the expected bounds from LEP2 (dashed line).}
\end{figure}
It is a common theoretical idea that the parameterization of the mechanism of
EW mass generation in terms of scalar couplings may be the effective
low-energy manifestation of more fundamental dynamics. The new dynamics may,
for instance, have the form of a new strong interaction \cite{techni},
\cite{chanowitz}. The idea which leads to the formulation of the BESS model
(Breaking Electroweak Symmetry Strongly) \cite{bess} assumes the existence of a
strongly interacting longitudinal-scalar sector, and enables one to parametrize
the most relevant phenomenological effects, without assuming an explicit
dynamical realization. 

We discuss the sensitivity of LEP2 to possible new physics from a strong
EW symmetry breaking sector. The model proposed in ref. \cite{bess} is
an effective lagrangian description of Goldstone bosons and new vector
resonances as the expectedly most visible manifestations at low energy of the
strongly interacting sector. In particular we will focus on the case in which
the new resonances are degenerate in mass (neglecting the weak corrections)
\cite{axvec}. This degenerate model has the interesting feature of allowing for
a strong EW resonant sector at relatively low energies, while
satisfying the severe constraints from existing LEP, SLC and CDF data. This
type of realization corresponds to a maximal symmetry $[SU(2)\otimes SU(2)]^3$.
After gauging the standard $SU(2)\otimes U(1)$, the model describes the
ordinary standard gauge bosons $W^\pm$, $Z$ and $\gamma$ and, in addition, two
new triplets of massive gauge bosons, $L^\pm$, $L_3$, $R^\pm$, $R_3$,
self-interacting with gauge coupling constant $\gs$. These heavy resonances, as
a consequence of the chiral symmetry,  are degenerate in mass $M$ in the 
$\gs\to\infty$ limit. The main property of the degenerate BESS lagrangian is
that it becomes identical to that of the Standard Model (taken in the formal
limit of infinite Higgs mass) for $M\to\infty$. Therefore, differently from
ordinary BESS, the heavy mass decoupling occurs. Concerning the couplings to
fermions, we do not introduce extra parameters. In this way the new bosons are
coupled to fermions only through the mixings with the ordinary gauge bosons.
The peculiar feature of degenerate BESS is that for any $M$ the new bosons are
not coupled to the Goldstone bosons which are absorbed to give mass to $W^\pm$
and $Z$. As a consequence, the channels $W_L Z_L$ and $W_L W_L$ are not
strongly enhanced as it usually happens in models with a strongly interacting
symmetry breaking sector and this implies larger branching ratios of the new
resonances into fermion pairs. Degenerate BESS would thus be much more evident
experimentally than ordinary BESS, at the appropriate energy, but at the same
time much less constrained by existing EW limits. 

Let us now discuss how new resonance effects modify the observables which are
relevant for the physics at LEP and Tevatron. For LEP physics the modifications
due to heavy particles are contained in the so-called oblique corrections. In
the low-energy limit, one can expand the vacuum polarization amplitudes in
$q^2/M^2$ where $M$ is the heavy mass, and they can be parametrized, for
example, in terms of  the $\eps$ parameters \cite{FITS,ABC}. At the leading
order in $q^2/M^2$  the new contribution of the model to all these parameters
is equal to zero. This is due to the fact that in the $M\to\infty$ limit, the
new states decouple \cite{axvec}. We can perform the low-energy limit at the
next-to-leading order and study the virtual effects of the heavy particles.
Working at the first order in $1/\gs^2$ we get
$\epsilon_1=-(\cthe^4+\s^4)/(\cthe^2)~
X$, $\epsilon_2=-\cthe^2~ X$, $\epsilon_3=-X$ with $X=2({\mz^2}/{M^2})( g/\gs)^2$.
All these deviations are of order $X$ and therefore contain a suppression both
from $\mz^2/M^2$ and $(g/\gs)^2$. The sum of the SM contributions, functions of
the top and Higgs masses, and of these deviations has to be compared with the
experimental values for the $\epsilon$ parameters, determined from the all
available LEP data and the $\mw$ measurement at Tevatron \cite{alt}:
$\epsilon_1=(3.8 \pm 1.5)\cdot 10^{-3}$, $\epsilon_2=(-6.4\pm 4.2)\cdot
10^{-3}$, $\epsilon_3=(4.6\pm 1.5)\cdot 10^{-3}$. Taking into account the SM
values $(\epsilon_1)_{SM}=4.4\cdot 10^{-3}$, $(\epsilon_2)_{SM}=-7.1\cdot
10^{-3}$, $(\epsilon_3)_{SM}=6.5\cdot 10^{-3}$  for $\mt =180~GeV$ and
$m_H=1000~GeV$, we find, from the combinations of the previous experimental
results, the $95\%$ CL limit on $g/\gs$ versus the mass M given by the solid
line in fig. \ref{bessf}. 
The excluded region is above the curve. We see that there is room for
relatively light resonances beyond the usual SM spectrum. The same analysis
done for the ordinary BESS model gives bounds which are much more stringent
($g/\gs < 0.03$ for any value of $M$). In fig. \ref{bessf} the limits on the
degenerate model parameter space derived from CDF data are also shown (dotted
line). The curve was obtained using the CDF 95\% C.L. limit on the $W'$ cross
section times the branching ratio and comparing this limit with the predictions
of our model. The limit from CDF is more restrictive for low resonance masses
with respect to the LEP limit. 

Concerning the bounds which would come from LEP2 assuming no deviations from
the SM within the estimated errors, we have analyzed cross sections and
asymmetries for the channel $e^+e^-\rightarrow f^+f^-$ in the SM and in
degenerate BESS at tree level. We have first considered the following
observables: total hadronic and muonic cross sections $\sigma^h$, $\sigma^\mu$;
the forward-backward asymmetries $A_{FB}^{e^+e^- \to \mu^+ \mu^-}$,
$A_{FB}^{e^+e^- \to {\bar b} b}$; $\mw$ measurement. In fig. \ref{bessf} we
show LEP2 limits obtained considering $\sqrt{s}=175$ GeV and an integrated
luminosity of $500 pb^{-1}$, combining the deviations coming from the previous
observables (dashed line). For $\mw$ we assume a total error (statistical and
systematic) $\Delta \mw=50$ MeV. For $\sigma^h$ the total error assumed is 2\%.
For all the other observable quantities we assume only statistical errors. The
comparison with the bounds from LEP and CDF shows that LEP2 will not improve
the existing limits on degenerate BESS (the same conclusions hold for the
ordinary BESS model). We have also considered the possibility of having
polarized beams at LEP2 by analyzing the left-right asymmetries: 
$A_{LR}^{e^+e^- \to \mu^+ \mu^-}$, $A_{LR}^{e^+e^- \to {\bar b} b}$,
$A_{LR}^{e^+e^- \to {had}}$. However
the improvement with respect to the unpolarized case is only
marginal. Moreover  considering the option of LEP2
at $\sqrt{s}=190$ GeV does not modify the result in a significative way. 
\section{Virtual Effects}
\label{virtualeffects}
This section is devoted to the study of virtual effects from
new physics in the cross section for $e^+e^-\to W^+W^-$.
We have focused on two simple SM extensions: $i)$ the SM plus an extra doublet
of heavy quarks, exact replica of the SM counterparts as far as their
EW and strong quantum numbers are concerned; $ii)$  the MSSM with
relatively heavy (above the production threshold) EW gauginos and
higgsinos, and very heavy (decoupled) squarks, sleptons and additional higgses.
In both models the 1-loop corrections to the process in question are
concentrated in vector-boson self-energies and three-point functions, which
makes it possible to analyze the interplay between the two contributions. Box
corrections are obviously absent in the first model, while they are negligibly
small in the second case, provided that the scalar masses are sufficiently
large. Then the only relevant contributions remain those of gauginos and
higgsinos to vertices and self-energies. 

\begin{figure}[htb]
\vspace{0.1cm}
\centerline{
\epsfig{figure=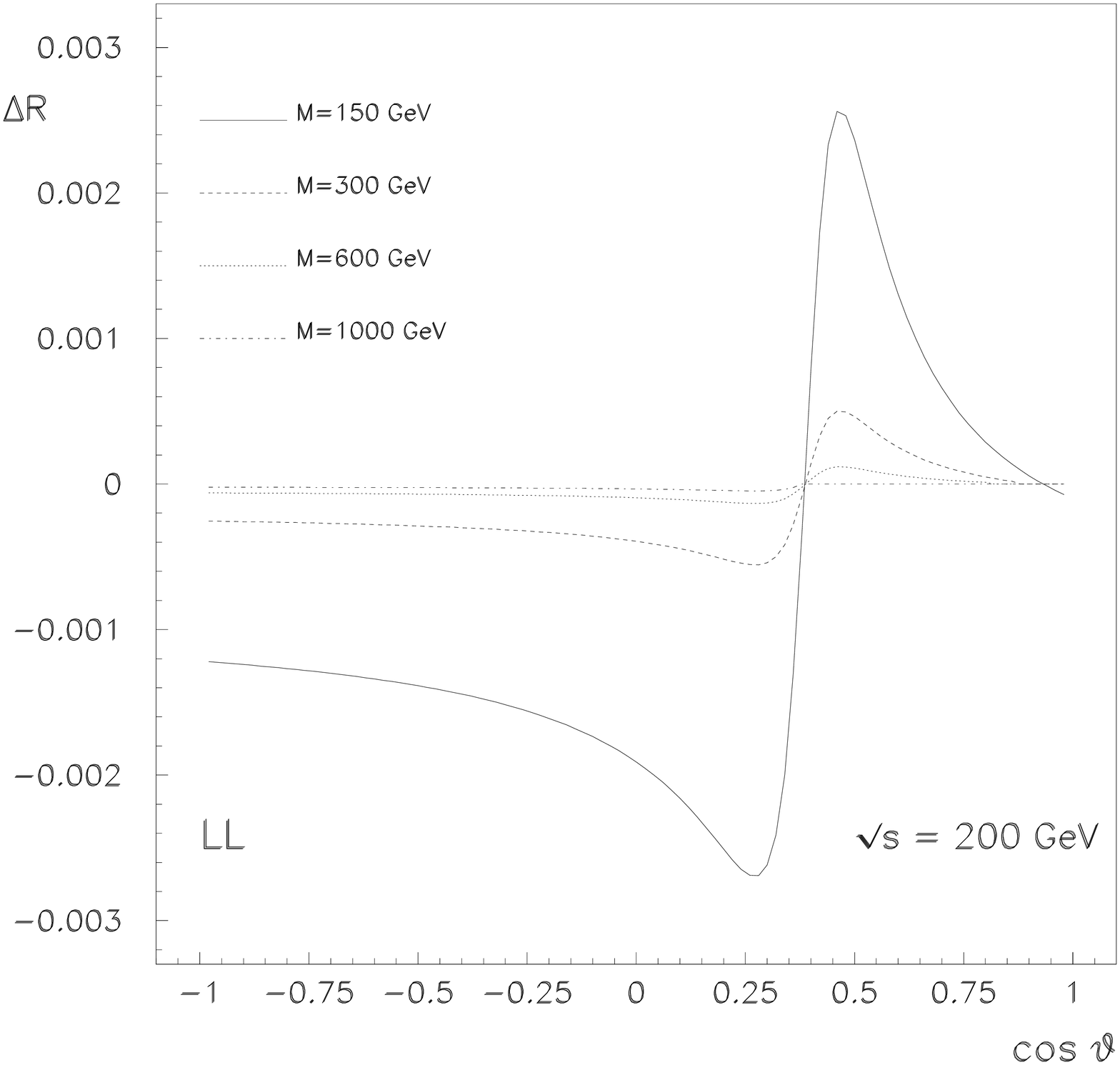,height=10cm,angle=0}
}
\ccaption{}{\label{pdfig}
Relative deviations in the differential cross section, due to gauginos 
contribution (LL channel) for $M_2 ~=~ 150,~300,~600,~1000$ {\rm GeV}
}
\end{figure}

To avoid conflicts with the LEP1 constraints on the $\epsilon$ parameters
\cite{FITS,ABC}, we consider the new doublet in $i)$ as mass
degenerate \cite{fmrs}. 
Model $ii)$ respects LEP1 bounds for arbitrary masses of the new particles, 
except for the case when they are very light and close to the production 
threshold of LEP1 \cite{bfc}.
However for practical purposes we have considered the case of degenerate
EW gauginos of mass $M_2$ and degenerate higgsinos of
mass $\mu$, in which the mixing terms among gauginos and 
higgsinos are negligible ($M_2,\mu >>m_W$).

After the inclusion of the 1-loop corrections due to the new particles
and of the appropriate counterterms, the reduced amplitude for the 
process at hand is (following the conventions of ref.\cite{hpz}):

\noindent
$\bullet \Delta\lambda=\pm2$
\be
{\tilde{\cal M}}~=~-\frac{\sqrt{2}}{\sin^2\thetab}
\delta_{\Delta\sigma,-1} \left[1-\frac{\sin^2\thetab}{\cos 2\thetab} 
\Delta r_W-e_6\right]
\frac{1}{1+\beta^2-2 \beta \cos\Theta}
\label{v2}
\ee

\noindent
$\bullet |\Delta\lambda| \le 1$
\bea
\Mg &=&-\beta \delta_{|\Delta\sigma|,1} \left[1+\Delta\alpha(s)\right]
\left[\Ag +\dAg (s)\right]\nn\\
\Mz &=&\beta \left[1+\Delta\rho(s)+\frac{\cos2\thetab}{\cos^2\thetab} 
       \Delta k(s)\right]\left[\delta_{|\Delta\sigma|,1}-
       \frac{\delta_{\Delta\sigma,-1}}{2\sin^2\thetab
(1+\Delta k(s))}\right] \frac{s}{s-m_Z^2} \left[\Az +\dAz (s)\right]\nn\\
\Mnu &=&\frac{1}{2\sin^2\thetab~\beta}
\delta_{\Delta\sigma,-1} \left[1-\frac{\sin^2\thetab}{\cos 2\thetab} 
\Delta r_W-e_6\right]
\left[B_{\lambda{\bar\lambda}}-
\frac{1}{1+\beta^2-2 \beta \cos\Theta} C_{\lambda{\bar \lambda}}\right]
\label{v3}
\eea
with $\beta=(1-4 m_W^2/s)^{1/2}$, $\Theta$ is the scattering angle of $W^-$
with respect to $e^-$ in the $e^+e^-$ c.m. frame; $\sigma$, $\bar\sigma$,
$\lambda$, $\bar\lambda$ are the helicities for $e^-$, $e^+$, $W^-$ and $W^+$,
respectively; $\Delta\sigma=\sigma-{\bar\sigma}$; $\Ag$, $\Az$,
$B_{\lambda{\bar\lambda}}$ and $C_{\lambda{\bar\lambda}}$ are tree-level SM
coefficients listed in table 3 of ref. \cite{hpz}; $\Delta\alpha(s)$, $\Delta
k(s)$, $\Delta\rho(s)$, $\Delta r_W$  are finite self-energy corrections,
appropriate extensions to the center of mass energy $\sqrt{s}$, of the
corresponding EW parameters defined in \cite{bfc,FITS,ABC} 
and $e_6$ is the
wave function renormalization of the external W' s lines. Finally, $\dAg$ and
$\dAz$ represent the corrections to the trilinear gauge boson vertices. For the
models considered, they can be expressed in terms of the $CP$-invariant form
factors $\delta f_i^{\gamma,Z}~~~ (i=1,...,4)$ according to the relations: 
\bea
\delta A_{++}^V & = & \delta A_{--}^V = \delta f_1^V\nn\\
\delta A_{+0}^V & = & \delta A_{-0}^V = \gamma(\delta f_3^V-i\delta f_4^V)\nn\\
\delta A_{0-}^V & = & \delta A_{0+}^V = \gamma(\delta f_3^V+i\delta f_4^V)\nn\\
\delta A_{00}^V & = & \gamma^2\left[-(1+\beta^2) \delta f_1^V + 4 \gamma^2 
                       \beta^2 \delta f_2^V + 2 \delta f_3^V\right]
\label{v4}
\eea
with $\gamma=\sqrt{s}/2 m_W$. Here $\delta f_i^V~~(i=1,...,4)~~(V=\gamma,Z)$
includes both the contribution coming from the 1-loop correction to the vertex
$VWW$ and the wave-function renormalization of the external $W$ legs, taken on
the mass-shell. This makes the terms $\delta f_i^V$ finite. 

The tree-level SM amplitudes are recovered from the above formulae by taking
$\Delta\alpha(s)=\Delta k(s)=\Delta\rho(s)= \Delta r_W=e_6=\dAg=\dAz=0$. In the
high-energy limit, the individual SM amplitudes from photon, $Z$ and $\nu$
exchange are proportional to $\gamma^2$ when both the $W$ are longitudinally
polarized ($LL$) and proportional to $\gamma$ when one $W$ is longitudinal and
the other is transverse ($TL$). The cancellation of the $\gamma^2$ and $\gamma$
terms in the overall amplitude is guaranteed by the tree-level, asymptotic
relation: $~\Ag~=~\Az~=B_{\lambda{\bar\lambda}}$. 

When one-loop contributions are included, one has new terms proportional to
$\gamma^2$ and $\gamma$ (see $\dAg$ and $\dAz$ in eq. (\ref{v4})) and the
cancellation of those terms in the high-energy limit entails relations among
oblique and vertex corrections. Omitting, for instance, the gauge boson
self-energies such cancellation does not occur any longer and the resulting
amplitudes violate the requirement of perturbative unitarity. 

On the other hand, one of the possibilities to have appreciable deviations in
the cross section is to delay the behaviour required by unitarity. This may
happen if in the energy window $m_W << \sqrt{s}\le 2 M$ ($M$ denoting the mass
of the new particles) the above cancellation is less efficient and terms
proportional to positive powers of $\gamma$ survive in the total amplitude.
Only if $\gamma$ were sufficiently large, beyond the LEP2 value, could sizeable
deviations from the SM be expected. 

Figure~\ref{pdfig} shows the relative deviation from SM results \cite{arg}
\be                          
\Delta R=\frac{\dd\left(\frac{d\sigma}{d\cos\Theta}\right)-
\dd\left(\frac{d\sigma}{d\cos\Theta}\right)_{SM}}
{\dd\left(\frac{d\sigma}{d\cos\Theta}\right)_{SM}}
\label{v9}
\ee
in model $ii)$, relative to the $LL$ channel, as function of $\cos \Theta$ at 
$\sqrt{s} = 200$ GeV for several values of the gaugino mass $M_2$ 
and negligible 
higgsino contribution ($\mu \gg M_2 \gg m_W$).

The deviations, even in the most favourable case $M_2=150$ GeV, are 
unobservable, being smaller than $3.0\cdot 10^{-3}$ \cite{bk}.
Similar magnitudes have been found for the channels $TL$ and $TT$ and when the 
higgsino contribution is singled out. Only when one considers
SUSY particles very close to the production threshold
(e.g. $M_2=105~\gev$ at $\sqrt{s}=200~\gev$), 
deviations of the order $1\%$ are obtained.
When model $i)$ is considered, the deviations at LEP2 are at the percent level
in the $LL$ and $TL$ channels, even smaller in the $TT$ one and in any case
well below the observability level.

Finally we would like to mention that at higher energies ( $\sqrt{s}=500$ GeV 
or even more) the deviations $\Delta R$ in model $ii)$ remain at the percent 
level, making questionable the possibility of observing such effect 
even in next generation $e^+ e^-$ colliders \cite{ls}.
While more interesting is the case of model $i)$, in which a delay of 
unitarity in the LL, LT channels, due to $\gamma$ enhancement factor, gives 
deviations from SM of the order 10-50 {\%}, for a wide range of new 
particles masses.
\section{CP-odd Correlations at LEP2}
\label{cpodd}
Unpolarized (and transversely polarized) $e^+e^-$ collisions can be used for
genuine tests of CP invariance at high energies (see, for example \cite{BBNO}).
We envi\-sage here statistical tests with observables which change sign under a
CP transformation. Measurement of a non-zero mean value of such an observable
would  signal violation of this symmetry. At LEP2 energies heavy fermion
production, {\it i.e.} tau pair and $b\bar{b}$ production appear to be
interesting channels for such tests (for the case of $W^+W^-$ production, see
ref. \cite{kneur}). It is possible to conceive that non-standard CP-violating
interactions induce substantially larger effects for heavy fermions than for
light flavours. Moreover, a study of various non-standard models of CP
violation shows that models with leptoquark mediated  CP violation may induce
form factors $d^{Z,\gamma}_{\tau}$ as large as $10^{-18} e \mbox{cm}$ at LEP2
energies \cite{BBO}. 

The first CP symmetry tests at high energies were made by the OPAL and ALEPH
collaborations for tau pair production at LEP1. These experiments also obtained
an upper bound on the CP-violating weak dipole form factor of the tau lepton
\cite{Opal2,Aleph2}. The combined OPAL and ALEPH measurements at the Z
resonance yield \cite{Stahl} 
$|\mbox{\cal Re}\, d^{Z}_{\tau}(\sqrt s=m_Z)| <
6.7 \times 10^{-18} e \mbox{cm} $ (95 $\%$ C.L.).
There is a direct bound on the tau electric dipole moment (EDM),
$|\mbox{\cal Re}\, d^{\gamma}_{\tau}(s=0)| <
4 \times 10^{-16} e \mbox{cm} $ (95 $\%$ C.L.),
from $\tau\tau\gamma$ events at LEP1 \cite{Venturi}.
The most stringent bound on the electric dipole moment (EDM) form factor,
$|d^{\gamma}_{\tau}(\sqrt s=m_Z)| < 5 \times 10^{-17} e \mbox{cm}$ (2$\sigma$),
has been derived indirectly \cite{EM}. However, one can imagine models
\cite{BBNO} which invalidate the arguments used for deriving indirect bounds on
CP-violating form factors. 

Here we wish to point out that, in spite of the limited statistics, one can
perform CP tests in tau pair production also at LEP2. These tests are useful
because  they would provide $direct$ information on the weak $and$ EDM form
factors at about twice the energy of LEP1. 

We consider tau pair production with unpolarized $e^+e^-$ collisions at LEP2
energies and their subsequent decays into the following channels: 
\begin{eqnarray}
e^+({\bf p_+}) + e^-({\bf p_-})
\to \tau^+({\bf k_+}) + \tau^-({\bf k_-})
\to A({\bf q_-}) + \bar B({\bf q_+}) +X,
\label{tau}
\end{eqnarray}
where the 3-momenta refer to the overall c.m. frame and $A,B=\ell,\pi,\rho,a_1$
($\rho$ and $a_1$ also denote the $2\pi$ and $3\pi$ states, respectively). CP
symmetry tests which are experimentally rather straightforward involve CP-odd
observables $\cal O({\bf q_+},{\bf q_-})$ = $ -\cal O(-{\bf q_-},-{\bf q_+})$
depending on the momenta of the charged particles in the final state. A
detailed analysis with simple observables of this form has been performed in
ref. \cite{BNO} for the channels (\ref{tau}). Repeating this analysis at $\sqrt
s =$ 190 GeV and assuming an integrated luminosity of 500 $(pb)^{-1}$ we find
that with these observables one can test the real parts of the weak and EDM
dipole form factors with the following accuracies: 
\begin{eqnarray}
\delta \mbox{\cal Re}\, d^{Z}_{\tau}(\sqrt s=190~ \mbox{GeV}) 
& \simeq & 3.1 \times
10^{-17} e \mbox{cm} \ \ (2\sigma) , \\
\delta\mbox{\cal Re}\, d^{\gamma}_{\tau}(\sqrt s=190~ \mbox{GeV}
) & \simeq & 5.1 \times
10^{-16} e \mbox{cm} \ \ (2\sigma) .
\end{eqnarray}
The considerably smaller statistics at LEP2 is only partially compensated by
the increase in sensitivity to the form factors at higher c.m. energies. 

However,  one can do better. For the channels with only two neutrinos in the
final state the tau direction of flight can be reconstructed up to a two-fold
ambiguity. This ambiguity can in principle be resolved \cite{Kuhn}; its
resolution is, however, not absolutely necessary (for details, see ref.
\cite{Opal2}). In the following we assume that the $\tau^{\pm}$ directions of
flight are known. The $\tau^+$ momentum direction in the overall c.m. frame is
denoted by $\hat{\bf k}$. Then one finds that the following CP- and T-odd
observables involving the $\tau^{\pm}$ spins ($\sigma_{i\pm}$ are the Pauli
matrices with $\pm$ refering to the respective spin spaces, and $\hat{\bf p}$
is the direction of the incoming positron), 
\begin{eqnarray}
{\cal O}_1 & = &
(\hat {\bf p}\times \hat{\bf k})\cdot(\bfsp-\bfsm),\\
{\cal O}_2 & = & \hat{\bf k}\cdot \bfsp
(\hat {\bf p}\times \hat{\bf k})\cdot \bfsm -
\hat{\bf k}\cdot \bfsm
(\hat {\bf p}\times \hat{\bf k})\cdot \bfsp,
\label{sigma}
\end{eqnarray}
are suitable for tracing the above form factors.  We find that ${\cal O}_1$ is
mainly sensitive to  $\mbox{\cal Re}\, d^{Z}_{\tau}$, whereas ${\cal O}_2$ is
mainly sensitive to  $\mbox{\cal Re}\, d^{\gamma}_{\tau}$. Needless to say the
tau spins are analysed by the decay distributions of the charged prongs. (The
distributions are given, for example, in ref. \cite{BNO}.) Moreover one can use
optimized observables ${\cal O}_i \to {\cal O}_i^{opt}$  with maximal
signal-to-noise ratio. At $\sqrt s$=190 GeV and with an integrated luminosity
of 500 $(pb)^{-1}$ we find that with these observables, using the channels
$A,B=\pi,\rho,a_1$, the following sensitivities can be reached: 
\begin{eqnarray}
\delta \mbox{\cal Re}\, d^{Z}_{\tau}(\sqrt s=190 \mbox{GeV}) 
& \simeq & 2.7 \times
10^{-17} e \mbox{cm} \    \ (2 \sigma) , \\
\delta \mbox{\cal Re}\, d^{\gamma}_{\tau}(\sqrt s= 190 \mbox{GeV}) 
& \simeq & 4.9 \times
10^{-17} e \mbox{cm} \   \ (2 \sigma) .
\label{sens}
\end{eqnarray}
We mention that one can find spin-momentum observables whose expectation values
are proportional to the imaginary parts of the form factors. 

Also of interest are searches of non-standard CP violation  in the reactions 
$e^+e^-\to b \bar{b}\  \ gluon(s) \to 3 jets$. (For details, see ref.
\cite{MMS}.) Here one would be sensitive to CP-violating but
chirality-conserving form factors of the $\gamma b \bar{b} g$ and $Z b \bar{b}
g$ vertices. However, due to the small number of events at LEP2, assuming an
integrated luminosity of 500 $(pb)^{-1}$, the sensitivity to these form factors
at $\sqrt s$ = 190 GeV is smaller by a factor of four as compared to the
sensitivity at  the $Z$ peak. 

In conclusion, measurements of CP-odd correlations in tau pair production at
LEP2 would test the weak dipole form factor $d_{\tau}^Z$ with a sensitivity
slightly below the accuracy reached at LEP1, but at higher energy. In addition,
one can probe the EDM form factor $d_{\tau}^{\gamma}$ directly, and with better
sensitivity than the direct test at LEP1. Essentially the same conclusions are
reached at $\sqrt s$ = 175  GeV assuming the  same integrated luminosity. 


\begin{thebibliography}{99}
\bibitem{wessz}
     Y. Gol'fand and E. Likhtam, \jetp{13}{71}{323};\\
     D. Volkov and V. Akulov, \pl{B46}{73}{109};\\
     J. Wess and B. Zumino, \np{B70}{74}{39}.
\bibitem{susyrev}
        J. Wess and J. Bagger, {\it Supersymmetry and Supergravity}
        (Princeton University Press,1983);\\ 
     H.P. Nilles, \prep{110}{84}{1};\\
     H.E. Haber and G.L. Kane, \prep{117}{85}{75};\\
        R. Barbieri, \rnc{11}{88}{1};\\
        R. Arnowitt, A. Chamseddine, and P. Nath, 
        {\it Applied N=1 Supergravity} (World Scientific, 1984);\\
        P. West, {\it Introduction to Supersymmetry and Supergravity} (World 
        Scientific, 1986);\\
        R.N. Mohapatra, {\it Unification and Supersymmetry} (Springer-Verlag, 
        1986).
\bibitem{natur}
     K. Wilson, as quoted by L. Susskind, \pr{D20}{79}{2619};\\
     G. 't Hooft, in {\it Recent Developments in Gauge Theories},
     ed. by G. 't Hooft {\it et al.} (Plenum Press, 1980);\\
     L. Maiani, Proc. Summer School of Gif-sur-Yvette (1980);\\
     M. Veltman, \app{B12}{81}{437}.
\bibitem{lim}
     J. Ellis, K. Enqvist, D.V. Nanopoulos, and F. Zwirner, \mpl{A1}{86}{57};\\
     R. Barbieri and G.F. Giudice, \np{B306}{88}{63};\\
     S. Dimopoulos and G.F. Giudice, \pl{B357}{95}{573}.
\bibitem{jelli}
     J. Ellis, J.S. Hagelin, D.V. Nanopoulos, K. Olive, and M. Srednicki,
     \np{B238}{84}{453}.
\bibitem{pdg}
     Review of particle properties, L. Montanet et al., \pr{D50}{94}{1173}.
\bibitem{opalstop}
   OPAL Collaboration, R. Akers {\it et al.}, \pl{B337}{94}{207}.
\bibitem{alephstop}                                
   ALEPH Collaboration, Contribution \#0416,
   International Europhysics Conference on High
   Energy Physics, Brussels, Belgium, 27 July - 2 August (1995).
\bibitem{neutr} 
     M. Acciarri et al., L3 Coll., \pl{B350}{95}{109}.
\bibitem{teva} 
    J. Hauser, for the CDF Coll., FERMILAB-CONF-95/172-E,
    to appear in the Proceedings of the 10th Topical Workshop on 
    Proton-Antiproton Collider Physics, Batavia, IL, May 9-13 1995;\\
    D\O\ Coll., contribution D\O\ \#434.                  
    presented at the                                
   International Europhysics Conference on High
   Energy Physics, Brussels, Belgium, 27 July - 2 August (1995).
\bibitem{EWWG2}
     The LEP Electroweak Working Group, CERN preprint LEPEWWG/95-02.
\bibitem{FITS}
     G. Altarelli and R. Barbieri, \pl{B253}{91}{161};\\
     G. Altarelli, R. Barbieri, and S. Jadach, \np{B369}{92}{3}.
\bibitem{ELLIS}
     J. Ellis, G.L. Fogli, E. Lisi preprint CERN--TH/95--202,
     BARI--TH/211--95;\\
     P.H. Chankowski and S.Pokorski \pl{B356}{95}{307} and hep-ph/9509207;\\
     J. Ellis, G.L. Fogli, E. Lisi \pl{B292}{92}{427},                 
     \pl{B318}{93}{375},                   
     \pl{B324}{94}{173}, \pl{B333}{94}{118}.
\bibitem{LANER}
     J. Erler and P. Langacker, \pr{D52}{95}{441}.
\bibitem{ABC}                                                       
     G. Altarelli, R. Barbieri, and F. Caravaglios  \np{B405}{93}{3},
     \pl{314B} {93} 357, \pl{349}{95}{145}.
\bibitem{MY_MSSM}                          
     P.H. Chankowski and S. Pokorski  preprint MPI-PTh/95-49\\
     hep-ph/9505308, {\it Phys. Lett.}  {\bf B} in press;\\
     P.H. Chankowski talk at the International Europhysics
     Conference on High Energy Physics, Brussels,
     27 July -- 2 August, 1995, to appear in the Proceedings.
\bibitem{BRUSS}
     A. Olchevski, plenary talk at the International Europhysics
     Conference on High Energy Physics, Brussels,
     27 July -- 2 August, 1995, to appear in the Proceedings.
\bibitem{BF} 
     M. Boulware and D. Finnell, \pr{D44}{91}{2054};\\
     J. Rosiek, \pl{B252}{90}{135};\\
     A.Denner et al., \zp{C51}{91}{695}.
\bibitem{SOLA}
     D. Garcia, A. Jimen\'ez and J. Sol\`a, \pl{347B}{95}{309}, 
        E {\bf351B} (1995) 602.                                   
\bibitem{KANE}                                                  
     G.L. Kane R.G. Stuart and J.D. Wells,
     \pl{B338}{94}{219};\\
     J.D. Wells and G.L. Kane, preprint SLAC-PUB-7038 (1995).
\bibitem{BSG} 
     S. Bertolini, F. Borzumati, A. Masiero and G. Ridolfi, 
     \np{B353}{91}{591};\\
     R. Barbieri and G.F. Giudice, \pl{309B}{93}{86};\\
     A. Buras, M. Misiak, M. M\"unz and S. Pokorski, \np{B424}{94}{376}.
\bibitem{BCTN} P. Krawczyk, S. Pokorski {\sl Phys. Rev. Lett.} {\bf 60}
              (1988) 182,\\
              G. Isidori {\sl Phys. Lett.} {\bf 298B} (1993) 409.
\bibitem{IR}
     M. Carena, S. Pokorski, and C.E.M. Wagner \np{B406}{94}{59}; \\
     W. Bardeen, M. Carena, S. Pokorski,and C.E.M. Wagner,
     \pl{B320}{94}{110}.
\bibitem{LARTAN}
     M. Olechowski and S. Pokorski, \pl{B214}{88}{393};\\
     G.F. Giudice and G. Ridolfi, \zp{C41}{88}{447};\\
     B. Ananhtarayan, G. Lazarides and Q. Shafi, \pr{D44}{91}{1613};\\
     S. Dimopoulos, L.J. Hall, and S. Raby, \prl{68}{92}{1984},
     \pr{D45}{92}{4192}.
\bibitem{OP2}
     M. Olechowski and S. Pokorski, \pl{344B}{95}{201}.
\bibitem{bartl-feng}
        See for instance:\\
        J.L. Feng and M.J. Strassler, \pr{D51}{95}{4661};\\
        A. Bartl, H. Fraas, W. Majerotto, and B. M\"osslacher,
        \zp{C55}{92}{257}, and references therein.
\bibitem{experim}
        J.-F. Grivaz (ALEPH),
        {\it Prospects for Supersymmetry Discoveries at Future
        \epem\ Colliders}, in the Proceedings of the Workshop
        ``Properties of Supersymmetric Particles'', Erice, Italy, 1992;\\
        P. Rebecchi (DELPHI), presentation at the May Meeting 
        of the LEP2 Workshop, ``New Particles" subgroup;\\
        S. Navas (DELPHI), presentation at the November meeting;\\
        A. Trombini (DELPHI), private communication;\\
        S. Rosier (L3), L3 Note 1863, in preparation;\\
        S. Komamiya (OPAL), presentation at the September meeting;\\        
        R. Van Kooten (OPAL), private communication.
\bibitem{jetset}
     T. Sj\"ostrand and M. Bengtsson, Comput. Phys. Commun. 43 (1987) 367;\\
     T.Sj\"ostrand,in: Z Physics at LEP 1, eds. G. Altarelli et al.,
     CERN Report CERN-89-08, Vol. 3 (1989) 143;\\
     T. Sj\"ostrand, Comput. Phys. Commun. 82 (1994) 74.
\bibitem{twogam}
     TWOGAM, S. Nova, A. Olshevski, and T. Todorov, DELPHI 90-35 PROG 152.
\bibitem{SUSYGEN}                                                       
        See the ``Event Generators for New Physics'' Chapter, in vol II of this
        Report.
\bibitem{leptwo}
     H. Baer, M. Brhlik, R. Munroe and X. Tata, 
        \pr{D52}{95}{5031}.
\bibitem{r-selL3}
        F. Nessi-Tedaldi,
        ``Monte Carlo study of selectron searches with L3 at LEP2'',
        L3 Internal Note 1576 (1995), and private communication.
\bibitem{r-BABY} R. Becker,  PITHA 93/27,
(Aachen, 1993).
\bibitem{IIHE}
      C.Vander Velde, IIHE report, ULB/VUB-Brussels,
      ``Monte Carlo study of slepton searches with DELPHI at LEP2'',
      in preparation.        
\bibitem{Favart}
      L.Favart,
      presentation at the September 27 1995 Meeting of the LEP2
      workshop, ``New Particles'' subgroup and private communication.
\bibitem{Koji}
        Koji Yoshimura,
        presentation at the November 1 1995 Meeting of the LEP2
        workshop, ``New Particles'' subgroup and private communication.
\bibitem{wwien}
    A. Bartl, W. Majerotto, and W. Porod, \zp{C64}{94}{499}.
\bibitem{drees} 
    M. Drees and K. Hikasa, \pl{B252}{90}{127}.
\bibitem{zerwas}
        W. Beenakker, R, Hoepker, P.M.Zerwas,
        \pl{B349}{95}{463}.
\bibitem{hikasa}                               
    K. Hikasa and M. Kobayashi, \pr{D36}{87}{724}.
\bibitem{Lthree}
   A. Sopczak, L3 Note 1860 (1995)
\bibitem{OPAL}            
    S. Asai, S. Komamiya and S. Orito, UT-ICEPP 95-10 (1995)
\bibitem{DELPHI}
   M. Besan\c{c}on, DELPHI Note, in preparation.
\bibitem{D0stop}
    D. Claes, for the D\O\ Coll., contribution D\O\ \#435.
    presented at the                                
   International Europhysics Conference on High
   Energy Physics, Brussels, Belgium, 27 July - 2 August (1995).
\bibitem{ambr-mele1} 
        S.~Ambrosanio and B.~Mele, \pr{D52}{95}{3900}. 
\bibitem{ambr-mele2}  
        S.~Ambrosanio and B.~Mele,
        ``Neutralino Decays in the Minimal Supersymmetric Standard Model", 
        Preprint ROME1-1095/95,  hep-ph/9508237, 
        August 1995, to appear in {\it Phys.~Rev.}~{D}.
\bibitem{last} 
        S.~Ambrosanio, M.~Carena, B.~Mele, C.~E.~M.~Wagner, 
        Preprint \mbox{CERN-TH/95-286}, ROME1-1121/95, hep-ph/9511259, 
        (1995), submitted for publication in {\it Phys.~Lett.}~{B}. 
\bibitem{pomar}
        G.F.~Giudice and A.~Pomarol, preprint \mbox{CERN-TH/95-337}.
\bibitem{pavia} 
        S.~Ambrosanio, B.~Mele, G.~Montagna, O.~Nicrosini, F.~Piccinini, 
        preprint FNT/T-95/32, ROME1-1126/95 (1995).
\bibitem{L3neutr}
        M. Felcini and J. Toth, L3 Note 1874, Nov 1995.
\bibitem{nmssm}
    H.-P.~Nilles, M.~Srednicki, and D.~Wyler, \pl{B120}{83}{346};\\
    J.-P.~Derendinger and C.A.~Savoy, \np{B237}{84}{307};\\ 
    J.~Ellis, J.F.~Gunion, H.E.~Haber, L.~Roszkowski, and F.~Zwirner,
    \pr{D39}{89}{844};\\
    M.~Drees, \ijmp{A4}{89}{3635};\\
    U. Ellwanger, M. Rausch de Traubenberg, and C. A. Savoy, 
    \pl{B315}{93}{331};\\
    T.~Elliott, S.F.~King, and P.L.~White, \pl{B351}{95}{213}.
\bibitem{franke1}
     F. Franke, H. Fraas and A. Bartl, \pl{B336}{94}{415};\\
     F. Franke and H. Fraas, \pl{B353}{95}{234}.
\bibitem{franke3}
     F. Franke and H. Fraas, WUE-ITP-95-021, hep-ph/9511275.
\bibitem{ibanross}
     F. Zwirner, \pl{132B}{83}{103};\\
     L.J. Hall and M. Suzuki, \np{B231}{84}{419};\\
     L. Ibanez and G.G. Ross, \np{B368}{92}{3}.
\bibitem{ali1}
     H. Dreiner and A. Chamseddine, ETH-TH-95-04, hep-ph/9504337.
\bibitem{LSPdecay}
     H. Dreiner and G.G. Ross, \np{B365}{91}{597}.
\bibitem{tatavernon}
     R.M. Godbole, P. Roy, and X. Tata, \np{B401}{93}{67};\\
     V. Barger, W.-Y. Keung, and R.J.N. Phillips, \pl{B356}{95}{546}.
\bibitem{morawitz}
     H. Dreiner and P. Morawitz, \np{B428}{94}{31}.
\bibitem{bargerhan}    
     V. Barger, G.F. Giudice, and T. Han, \pr{D40}{89}{2987}.
\bibitem{isajet}
     H. Baer, F. Paige, S. Protopopescu and X. Tata, in
     {\it Proceedings of the Workshop on Physics at Current Accelerators
     and Supercolliders}, ed.\ J. Hewett, A. White and D. Zeppenfeld,
     (Argonne National Laboratory, 1993); see also contribution by 
     H. Baer, F. Paige and X. Tata in event generators section.
\bibitem{helas}
     HELAS: HELicity Amplitude Subroutines for Feynman Diagram
     Evaluations, H. Murayama, I. Watanabe and K. Hagiwara,
     KEK-91-11 (1992).
\bibitem{exotic}
     J.C.~Pati and A.~Salam, \pr{D10}{74}{275}; \\
     R.N.~Mohapatra and J.C.~Pati, \pr{D11}{75}{366,2588};\\
     G.~Senjanovi\'c and R.N.~Mohapatra, \pr{D12}{75}{1502}.
\bibitem{guts}
     J.~Hewett and T.~G.~Rizzo, \prep{183}{89}{193};\\
     J.~Maalampi and M.~Roos, \prep{186}{90}{53};\\
     W.~Buchm\"uller and  C.~Greub, \np{B363}{91}{345}; \np{B381}{92}{109}.
\bibitem{4th family}
     Proceedings of the International Symposium on the 4$^{\rm th}$
     Family of Quarks and Leptons, Santa Monica (1987), Ann. New York
     Accademy of Science, 518 (eds. D.B. Cline and A. Soni).
\bibitem{follia}
     M.~Chanowitz, M.~Furman and I.~Hinchliffe, \pl{B78}{78}{285};\\
     M.~Drees, \np{298}{88}{333};\\
     F.~Csikor and I.~Montvay, \pl{B231}{90}{503}.
\bibitem{hagiwara}      
     K.~Hagiwara, S.~Komamiya and D.~Zeppenfeld, \zp{C29}{85}{115}.
\bibitem{dpfef}
         A.~Djouadi {\it et. al.}, SLAC-PUB-95-6772 
        (To appear in {\it Electroweak Symmetry Breaking and
        Beyond the Standard Model}, eds. T.~Barklow, S.~Dawson, H.E.~Haber 
        and S.~Siegrist, World Scientific.)
\bibitem{dp_bm}
     B.~Mukhopadhyaya and D.P.~Roy, \pr{D48}{93}{2105}. 
\bibitem{nardi}
     E.~Nardi, E.~Roulet, and D.~Tommasini, \pl{B344}{95}{225};\\
     G.~Bhattacharyya et al., \mpl{A6}{91}{2921};\\
     G.~Bhattacharyya, \pl{B331}{94}{143}.
\bibitem{LEP-precision}
     P.~Langacker, Proc. of SUSY-95, hep-ph/9511207.
\bibitem{rho}
     J.J.~van~der~Bij and F.~Hoogeveen, \np{B283}{87}{477};\\
     M.~Consoli, W.~Hollik and F.~Jegerlehner, \pl{B227}{89}{167};\\
     Also see  A.~Sirlin, hep-ph/9411363 (To appear
     in `{\em Reports of the Working Group on Precision Calculations 
     for the $Z$ resonance}') and the references therein.
\bibitem{shev}
     S.~Shevchenko and A.~Shvorob, L3 internal note (in preparation). 
\bibitem{gb_dc}
     G.~Bhattacharyya and D.~Choudhury, CERN preprint CERN-TH/95-306.
\bibitem{R3}
     A.~Djouadi, \zp{C63}{94}{317}; \\
     A.~Djouadi and G.~Azuelos, \zp{C63}{94}{327}. 
\bibitem{N:opal}
     R.~Tafirout and G.~Azuelos, OPAL Internal note.
 \bibitem{BOUDJ}
     F.~Boudjema, A.~Djouadi and J.L.~Kneur, \zp{C57}{93}{425}.
\bibitem{MANEL}                              
     M.~Martinez, and R.~Miquel, \pl{B302}{93}{108}.
\bibitem{SEARCHES}
     {\sc ALEPH} Collaboration, \prep{216}{92}{1}.
\bibitem{LITKE}
     A.~Litke, Experiments with Electron-Positron Colliding
     beams, PhD Thesis, Harvard University (1970).
\bibitem{models}
     J. Pati and A. Salam, \pr{D10}{74}{275};\\
     P. Langacker, \prep{72}{81}{185};\\
     B. Schrempp and F. Schrempp, \pl{B153}{85}{101};\\
     J.L. Hewett and T.G. Rizzo, \prep{193}{89}{193};\\
     P.H. Frampton, \mpl{A7}{92}{559};\\
\bibitem{buch}
     W. Buchm\"uller, R. R\"uckl and D. Wyler, \pl{B191}{87}{442}.
\bibitem{blum}
     J. Bl\"umlein and R. R\"uckl, \pl{B304}{93}{337}.
\bibitem{low}
     O. Shanker, \np{B204}{82}{375};\\
     W. Buchm\"uller and D. Wyler, \pl{B177}{86}{377};\\
     J.L. Hewett and T.G. Rizzo, \pr{D36}{87}{3367};\\
     M. Leurer, \pr{D49}{94}{333};\\
     S. Davidson, D. Bailey and A. Campbell, \zp{C61}{94}{613}.\\
\bibitem{z}
     J.K. Mizukoshi, O.J.P. Eboli and M.C. Gonzalez-Garcia,
     CERN-TH.7508/94 (1994);\\
     G. Bhattacharyya, J. Ellis and K. Sridhar, CERN-TH.7280/94 (1994).
\bibitem{HERA}
     H1 Collaboration, T. Ahmed et. al., \zp{C64}{94}{545}; 
     DESY 95-079.
\bibitem{LEP}
     L3 Collaboration, B. Adeva et. al., \pl{B261}{91}{169};\\
     OPAL Collaboration, G. Alexander et. al., \pl{B263}{91}{123};\\
     DELPHI Collaboration, P. Abreu et. al., \pl{B316}{93}{620};\\
     ALEPH Collaboration, D. Decamp et al., CERN PPE/91-149.
\bibitem{D0}
     D0 Collaboration, S. Abachi et.al., \prl{72}{94}{965}.
\bibitem{CDF}
     CDF Collaboration, F. Abe et. al., \prl{75}{95}{1012}.
\bibitem{L3s}                                        
     L3 Collaboration, O. Adriani et al., CERN-PPE/93-31 (1993).
\bibitem{techni}
     S.~Weinberg, \pr{D19}{79}{1277};\\
     L.~Susskind, \pr{D20}{79}{2619};\\
     E.~Farhi and L.~Susskind, \prep{74}{81}{277}.
\bibitem{chanowitz}
     For reviews, see M.~Chanowitz, \arnps{38}{88}{323};\\
     T.~Appelquist, Lectures given at Mexican School of Particles and Fields, 
     Mexico City, December 1990 (preprint YCTP-P23-91).
\bibitem{bess}
     R.~Casalbuoni, S.~De Curtis, D.~Dominici and R.~Gatto,
     \pl{B155}{85}{95}; \np{B282}{87}{235}; \\
     R.~Casalbuoni, S.~De Curtis, D.~Dominici, F.~Feruglio  and R.~Gatto, 
     \ijmp{A4}{89}{1065}.
\bibitem{axvec}
     R.~Casalbuoni, A.~Deandrea, S.~De Curtis, D.~Dominici, 
     F.~Feruglio, R.~Gatto and M.~Grazzini, \pl{B349}{95}{533};\\
     R.~Casalbuoni, A.~Deandrea, S.~De Curtis, D.~Dominici, R.~Gatto and 
     M.~Grazzini, UGVA-DPT 1995/10-906 (hep-ph/9510431), (1995).
\bibitem{alt}
     F. Caravaglios, talk given at the International Europhysics
     Conference on High Energy Physics, Brussels,
     27 July -- 2 August, 1995, to appear in the Proceedings.
\bibitem{fmrs}
     F. Feruglio, A. Masiero, S. Rigolin, R. Strocchi, \pl{B355}{95}{329}.
\bibitem{bfc}
     R. Barbieri, F. Caravaglios and M. Frigeni, \pl{B279}{92}{169}.
\bibitem{hpz}
     K.J.F. Gaemers, G.J. Gounaris, \zp{C1}{79}{259};\\
     K. Hagiwara, K. Hikasa, R. Peccei and D. Zeppenfeld, \np{B282}{87}{253}.
\bibitem{arg}
     E.N. Argyres, C.G. Papadopoulos, \pl{B263}{91}{298};\\
     E.N. Argyres, G. Katsilieris, A.B. Lahanas, 
     C.G. Papadopoulos, V.C. Spanos, \np{B391}{93}{23};\\
     J. Fleischer, J.L. Kneur, K. Kolodziej, M. Kuroda, D. Schildknecht, 
     \np{B378}{92}{443}, \np{B426}{94}{246}.
\bibitem{bk}
     M. Bilenky, J.L. Kneur, F.M. Renard, D. Schildknecht, \np{B409}{93}{22};
      \np{B419}{94}{240}.
\bibitem{ls}
     A.B. Lahanas, V.C. Spanos, \pl{B334}{94}{378}; hep-ph 9504340;\\
     A. Culatti, \zp{C65}{95}{537}.                 
\bibitem{BBNO}
     W. Bernreuther, G. Botz, O. Nachtmann,
     and P. Overmann, \zp{C52}{91}{567}.
\bibitem{kneur}
     Triple Gauge Boson Couplings, in this report.
\bibitem{BBO}
     W. Bernreuther, A. Brandenburg and P. Overmann, to be published.
\bibitem{Opal2} 
     R. Akers et al. (OPAL collab.), \zp{C66}{95}{31}.
\bibitem{Aleph2}
     D. Buskulic et al. (ALEPH collab.), \pl{B346}{95}{371}.
\bibitem{Stahl} 
     A. Stahl, {\it Nucl. Phys} {\bf B40} (Proc. Suppl.) (1995) 505.
\bibitem{Venturi}
     A. Venturi, in Proc. of the XXVII Int. Conf. on High Energy Physics, 
     ed. by P.J. Bussey and I.G. Knowles, Bristol (1995) p. 771.
\bibitem{EM} 
     R. Escribano and E. Masso, \pl{B301}{93}{419}.
\bibitem{BNO} 
     W. Bernreuther, O. Nachtmann and P. Overmann, \pr{D48}{93}{78}.
\bibitem{Kuhn}
     J.H. K\"uhn, \pl{B313}{93}{458}.
\bibitem{MMS} 
     J. K\"orner et al., \zp{C49}{91}{447};\\
     W. Bernreuther et al., \zp{C68}{95}{73}.
\end{thebibliography}
\end{document}